\documentclass{emulateapj}

\usepackage{psfig}
\usepackage{amsmath}
\usepackage{amssymb}
\usepackage{graphicx}
\usepackage{appendix}
\usepackage{color}
\usepackage{multirow}
\usepackage[normalem]{ulem}

\newcommand{\lSect}[1]{{\label{sec:#1}}}
\newcommand{\lFig}[1]{{\label{fig:#1}}}
\newcommand{\lEq}[1]{{\label{eq:#1}}}
\newcommand{\lTab}[1]{{\label{tab:#1}}}
\def\gtaprx {\lower .1ex\hbox{\rlap{\raise .6ex\hbox{\hskip .3ex
	{\ifmmode{\scriptscriptstyle >}\else
		{$\scriptscriptstyle >$}\fi}}}
	\kern -.4ex{\ifmmode{\scriptscriptstyle \sim}\else
		{$\scriptscriptstyle\sim$}\fi}}}
\def\ltaprx {\lower .1ex\hbox{\rlap{\raise .6ex\hbox{\hskip .3ex
	{\ifmmode{\scriptscriptstyle <}\else
		{$\scriptscriptstyle <$}\fi}}}
	\kern -.4ex{\ifmmode{\scriptscriptstyle \sim}\else
		{$\scriptscriptstyle\sim$}\fi}}}
\newcommand{\FIGFF}[2]{{\ref{fig:#2}{#1}}}

\newcommand{\FIG}[2]{{Fig.~\FIGFF{#1}{#2}}}
\newcommand{\Fig}[1]{{\FIG{}{#1}}}

\newcommand{\Sectff}[1]{{\ref{sec:#1}}}
\newcommand{\Sect}[1]{{\S\Sectff{#1}}}

\newcommand{\Msun}{\ensuremath{\mathrm{M}_\odot}}

\newcommand{\Tab}[1]{{Table \ref{tab:#1}}}

\newcommand{\KEPLER}{\ensuremath{\mathrm{\texttt{KEPLER}}}}

\def\gtaprx {\lower .1ex\hbox{\rlap{\raise .6ex\hbox{\hskip .3ex
	{\ifmmode{\scriptscriptstyle >}\else
		{$\scriptscriptstyle >$}\fi}}}
	\kern -.4ex{\ifmmode{\scriptscriptstyle \sim}\else
		{$\scriptscriptstyle\sim$}\fi}}}
\def\ltaprx {\lower .1ex\hbox{\rlap{\raise .6ex\hbox{\hskip .3ex
	{\ifmmode{\scriptscriptstyle <}\else
		{$\scriptscriptstyle <$}\fi}}}
	\kern -.4ex{\ifmmode{\scriptscriptstyle \sim}\else
		{$\scriptscriptstyle\sim$}\fi}}}

\begin{document}

\submitted{-- Oct, 2017}
\accepted{-- ---, 2018}

\title{High Resolution Study of Presupernova Core Structure}

\author{Tuguldur Sukhbold\altaffilmark{1,2}, 
                      S.\ E.\ Woosley\altaffilmark{3}, 
                 and Alexander\ Heger\altaffilmark{4,5,6}} 
\altaffiltext{1}{Department of Astronomy, The Ohio State University, Columbus, OH 43210, USA, tuguldur.s@gmail.com}
\altaffiltext{2}{Center for Cosmology and AstroParticle Physics, The Ohio State University, Columbus, OH 43210, USA}
\altaffiltext{3}{Department of Astronomy and Astrophysics, University
  of California, Santa Cruz, CA 95064, woosley@ucolick.org}
\altaffiltext{4}{Monash Center for Astrophysics, Monash University,
  Vic 3800, Australia, alexander.heger@monash.edu}
\altaffiltext{5}{School of Physics \& Astronomy,
  University of Minnesota, Minneapolis, MN 55455, U.S.A.}
\altaffiltext{6}{Department of Astronomy, Shanghai Jiao-Tong University, Shanghai 200240, P. R. China.}

\begin{abstract} 
The density structure surrounding the iron core of a massive star when
it dies is known to have a major effect on whether or not the star
explodes. Here we repeat previous surveys of presupernova evolution
with some important corrections to code physics and four to ten times
better mass resolution in each star. The number of presupernova
masses considered is also much larger. Over 4,000 models are
calculated in the range from 12 to 60 \Msun\ with varying mass loss
rates. The core structure is not greatly affected by the increased
spatial resolution. The qualitative patterns of compactness measures
and their extrema are the same, but with the increased number of
models, the scatter seen in previous studies is replaced by several
localized branches.  More physics-based analyses by \citet{Ert16} and
\citet{Mue16} show these branches with less scatter than the single
parameter characterization of \citet{Oco11}.  These branches are
particularly apparent for stars in the mass ranges 14 - 19 \Msun \ and
22 - 24 \Msun.  The multi-valued solutions are a consequence of
interference between several carbon and oxygen burning shells during
the late stages of evolution. For a relevant range of masses, whether
a star explodes or not may reflect more the small, almost random
differences in its late evolution than its initial mass. The large
number of models allow statistically meaningful statements about the
radius, luminosity, and effective temperatures of presupernova stars,
their core structures, and their remnant mass distributions.
\end{abstract}


\section{INTRODUCTION}
\lSect{intro}

As a massive star below the pair-instability threshold
($\sim80\,\Msun$; \citealt{Woo17}) evolves through its final stages of
nuclear burning, its central regions cool by neutrino emission, become
degenerate, and tend to decouple from the overlying layers and evolve
as separate stars. Although never becoming completely detached except
for the lowest mass stars, the presupernova core takes on a structure
similar to that of a white dwarf, mostly composed of iron, surrounded
by a dense mantle of oxygen and intermediate mass elements. As
numerous studies have shown, the structure of this configuration, and
especially the rate at which the density declines outside the iron
core is strongly correlated with the difficulty of blowing the star up
\citep[e.g.,][]{Bur87,Fry99}.  Recent studies have sought to capture
this complex structure in just one or two parameters that might
predict, albeit approximately, whether a star with a given mass blows
up or collapses to a black hole simply from looking at one-dimensional
models for stellar evolution
\citep{Oco11,Oco13,Ugl12,Pej15,Ert16,Suk16,Mue16}.

\citet{Suk14} discussed how the advanced burning stages, especially
convective carbon and oxygen burning, sculpt this structure, and
provided a library of presupernova stars consisting of 503 models in
the mass range 12 to 65 \Msun \ to demonstrate the systematics and its
dependence on input physics. \citet{Mue16} prepared a larger grid
of over 2,000 models between 10 and 32.5 \Msun\ and corrected some
errors in \citet{Suk14}. Their results showed finer structure, but
similar global systematics (see especially their Fig. 6).

More recently, \citet{Far16} using the \texttt{MESA} code, found that
the choice of mass resolution, i.e., zoning, ``\textsl{dominates the
  variations in the structure of the intermediate convection zone and
  secondary convection zone during core and shell hydrogen burning,
  respectively\/}'' and greatly affects the structure of presupernova
stars. They found that a minimum mass resolution of $\sim 0.01$ \Msun
\ was necessary to achieve convergence in the final helium core mass at
the $\sim$5 \% level. They also found $\sim$30 \% variations in the
central electron fraction and mass locations of the main nuclear
burning shells, and that a minimum of $\sim$127 isotopes was needed to
attain convergence of these values at the $\sim$10 \% level.

\citet{Ren17} also used the \texttt{MESA} code to explore the
sensitivity of massive star evolution due to variations in the mass
loss rate for stars with initial masses between 15 and 35 \Msun. They
found variations in the presupernova core compactness parameter
(\Sect{compact}) of $\sim$30 \% depending upon the choice of the
algorithm. In a limited study of resolution in one model, they found
roughly 9\% variation in final core compactness, smaller than that
resulting from uncertainties in mass loss prescription.

In this paper, we present a new survey similar to 
\citet{Suk14} and \citet{Mue16}, but using much finer zoning, a 
greater number of models and several different mass loss rates. 
The key differences of the new survey are listed in \Sect{physics}, 
including the correction to an erroneous pair-neutrino loss rate 
used by \citet{Suk14}. \Sect{physics} also offers a detailed 
discussion on the effects due to zoning, network size and the 
boundary pressure. In general, we find that the pattern of final 
core compactness seen in these previous studies is unaltered by 
finer zoning or larger reaction network, though the mass limits 
for different behaviors are shifted by about 10 \% when the 
pair-neutrino loss rate is corrected. In \Sect{procedure} and 
\Sect{results} we discuss the general characteristics of the new 
survey, including its observable properties. With a much greater 
number of models, clear evidence emerges for multi-valued solutions 
to the presupernova structure of stars in the 14 - 19 \Msun\ and 
22 - 24 \Msun\ ranges \citep[see also][]{Mue16}. In \Sect{interpret} 
we offer an interpretation for this feature through the physics of 
the advanced stage evolution in the cores of massive stars. Large 
variations in core structure are expected for stars of these masses 
no matter what resolution, mass loss rate, or reaction network is 
employed in the calculation \citep[for the effects of such 
variations in individual models see][]{Far16,Ren17}. Future surveys 
of supernova, as well as presupernova models should thus take care 
to examine, when feasible, a large number of masses in order to 
sample an outcome that is essentially statistical in nature. More 
specfic results, and effects on explodability and remnant masses 
are discussed in \Sect{remnants}. Finally, in \Sect{conclude} we 
offer our conclusions.

\section{Code Physics and Assumptions}
\lSect{physics}

With several important exceptions the code, physics, and input
parameters used here are the same as in \citet{Suk14} and
\citet{Mue16}.  The (solar) initial composition is from \citet{Asp09},
as was used in \citet{Mue16}, but with an appropriate correction for
the ratio $\rm ^{15}N/^{14}N$ \citep{Mei07}. Convective and
semi-convective settings, nuclear reaction rates, opacities, and mass
loss rates are the same as in both of the earlier works.  The mass
range studied and lack of rotation are the same as well.

Nuclear burning was handled as usual, using a $19$ isotope
approximation network until the central oxygen mass fraction declines
below $0.03$ and a silicon quasi-equilibrium network with $121$
isotopes \citep{Wea78} thereafter.  Numerous studies that compare this
treatment with the energy generation and bulk nucleosynthesis obtained
by ``co-processing'' with a network of several hundred to over 1,000
isotopes have shown excellent agreement \citep[e.g.,][]{Woo02,Woo07},
at least up to central oxygen depletion where the switch to the
quasi-equilibrium network is made.  By co-processing, we mean carrying
a large network in each zone using the same time step, temperature,
and density, but not coupling the energy generation from that large
network directly to the iterative loop within a time step.  After
oxygen depletion, the quasi-equilibrium network is more stable,
contains the same weak interaction physics, and is roughly an order of
magnitude faster.  An exception is the treatment of oxygen and silicon
burning in stars lighter than about 11 \Msun\ \citep{Woo15}.  For such
light stars, it is important to follow neutronization in multiple
off-center shells where the quasi-equilibrium approximation has
questionable validity and can be unstable.  Such light stars are not
part of the present survey which starts at 12 \Msun. \Sect{network}
discusses the issue of network sensitivity in greater detail and
confirms the validity of the approximation network plus
quasi-equilibrium for representative cases.

Key differences in the new study are:

\begin{itemize}

\item The models in this paper employ much finer mass resolution
  (\Fig{zone15}, \Fig{rhograd}, and \Fig{zone}). For the
  lightest stars considered (12 \Msun), the increase is roughly a
  factor of four. Since an effort was made to keep fine resolution
  even in the larger stars, the factor for the highest masses
  considered, around 60 \Msun, was closer to fifteen.  In the most
  massive models the total number of mass shells was approximately
  16,000.  Additional studies of individual cases of zoning
  sensitivity in $15\,\Msun$ and $25\,\Msun$ models (\Sect{ultra})
  showed little systematic difference.  In all cases, zones contained
  less than about $0.01\,\Msun$ everywhere, but were about ten times
  smaller than that in the heavy element core (\Fig{zone}).  It is
  this core that is most critical in determining the final
  presupernova core properties, such as compactness.  Surface zoning
  was a few times $10^{-5}\,\Msun$, and all temperature and density
  gradients were well resolved (\Fig{rhograd}).

\item A major, long standing coding error in the {\KEPLER} code was
  repaired.  The error affected the axial vector component of the pair
  neutrino losses.  In particular, the error resulted in the accidental
  zeroing of the second term involving $C_\mathrm{A}^{'2}$ and 
 $Q_\mathrm{pair}^-$ in the expression \citep[][eq. 2.1]{Ito96}
\begin{align}
\rm Q_{pair} &= \rm\frac{1}{2} [(C_V^2+ C_A^2) + n (C_V^{'2}+ C_A^{'2})] Q_{pair}^+\\
&  \rm + \frac{1}{2} [(C_V^2- C_A^2) + n (C_V^{'2}- C_A^{'2})] Q_{pair}^-
\end{align}
See \citet{Ito96} for the definitions of quantities.  The consequence
of this error was that the pair neutrino loss rate was underestimated
by close to a factor of two during carbon burning and somewhat less
during later, higher temperature stages. The error, introduced through
an inadequate checking of a routine provided by email, was
included in 2001, and affected all {\KEPLER} calculations published
through 2014. In particular it affected the often cited calculations
of \citet{Woo02}, \citet{Woo07}, and \citet{Suk14}.  Because some of
the models from \citet{Suk14} were used by \citet{Suk16}, it also
affected the outcome of that work for masses larger than 14 \Msun.
The bug was repaired, however, in the works of \citet{Woo15} on 9
\Msun -- 11 \Msun\ stars and \citet{Woo17} on pulsational pair
instability supernovae.  Most relevant to this present work, the bug
was also repaired for the work of \citet{Mue16}. Because of the strong
sensitivity of neutrino and nuclear reaction rates to temperature, the
effect of the bug was a slight shift upwards in the burning
temperature for the late stages of stellar evolution.  For a given
model the change in presupernova structure was not great and, as we
shall see, within the ``noise'' of other uncertainties. For the
lightest models ($<14\,\Msun$) the difference is hardly noticeable,
but it did systematically shift the outcome for a higher mass stars
significantly downwards.  For example, the peak in compactness that
occurred for \citet{Suk14} at about $24\,\Msun$ is shifted downwards
in the work of \citet{Mue16} and in the present work (\Sect{results})
by about $3\,\Msun$.

\item The surface boundary pressure is much less than in
  \citet{Mue16}, which significantly affects the final red supergiant
  (RSG) properties, but does not significantly alter the core
  structure. Whereas the mass loss rates are varied in the present
  study, the new calculations do not include Wolf-Rayet models where
  the envelope is completely lost. All stars studied here retained a
  substantial hydrogen envelope when they died.

\end{itemize}

\begin{figure}
\includegraphics[width=0.48\textwidth]{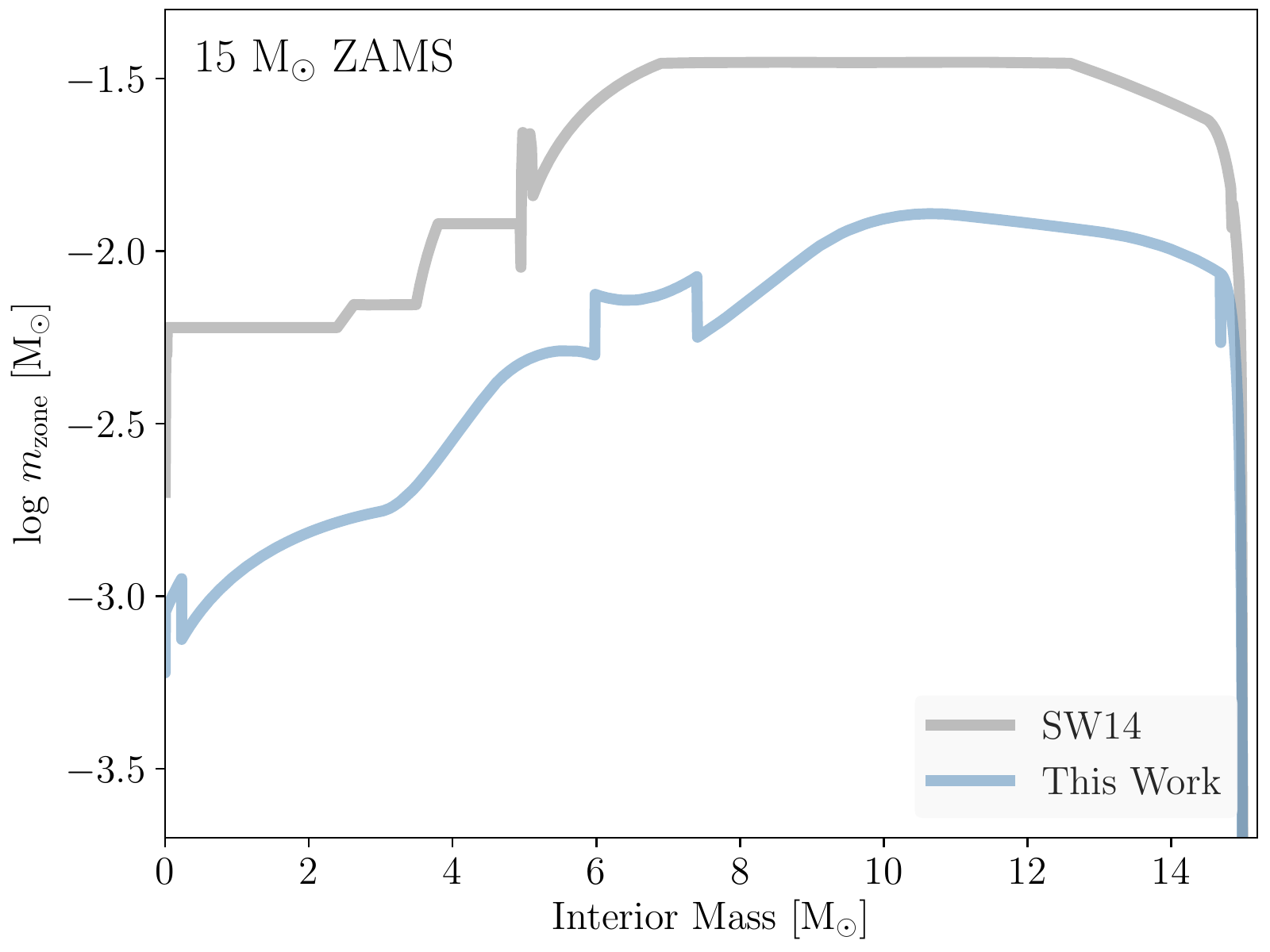}
\includegraphics[width=0.48\textwidth]{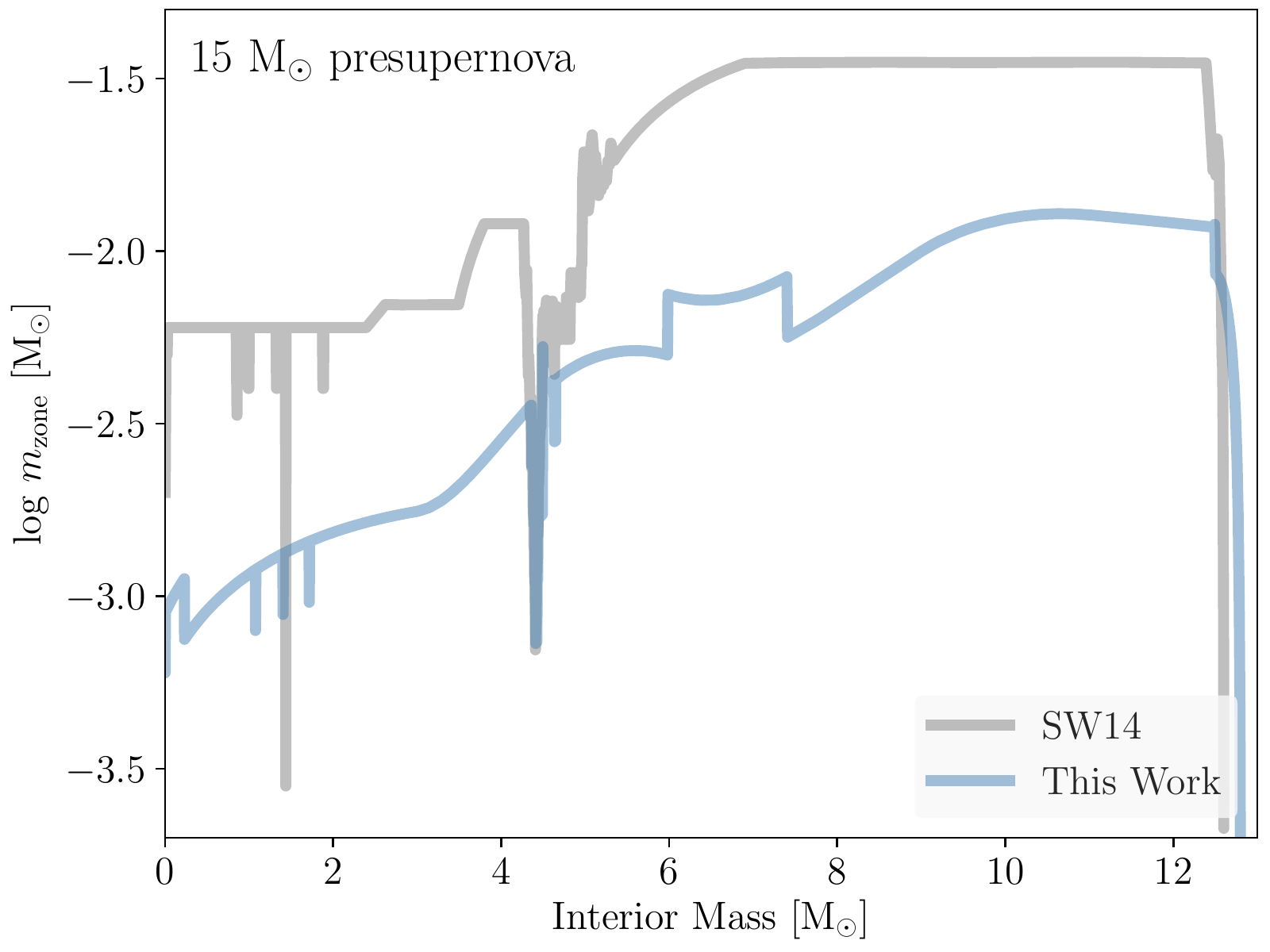}
\caption{The masses of individual zones as a function of interior mass
  for a 15 \Msun\ zero age main sequence star (top) and the
  corresponding presupernova star (bottom).  The gray curves are the
  zoning for the earlier model of \citet[][SW14]{Suk14} which was very
  similar to that of \citet{Woo07}, \citet{Suk16} and \citet{Mue16};
  the blue curves are for the new survey reported here. The previous
  ZAMS model had 1,068 zones, the new one, 4,257.  Stars are
  continually rezoned as they run to accommodate changing gradients in
  key quantities, but the zoning shown here did not vary greatly, with
  the exception of fine zones added at the base of the hydrogen
  envelope around 4.4 \Msun, and a few regions of finer zoning at what
  were once the boundaries of convective shells inside 2 \Msun.  The
  surface zoning was $\sim10^{28}$ g in all studies (\Sect{physics}). 
  \lFig{zone15}}
\end{figure}

\begin{figure}
\includegraphics[width=0.48\textwidth]{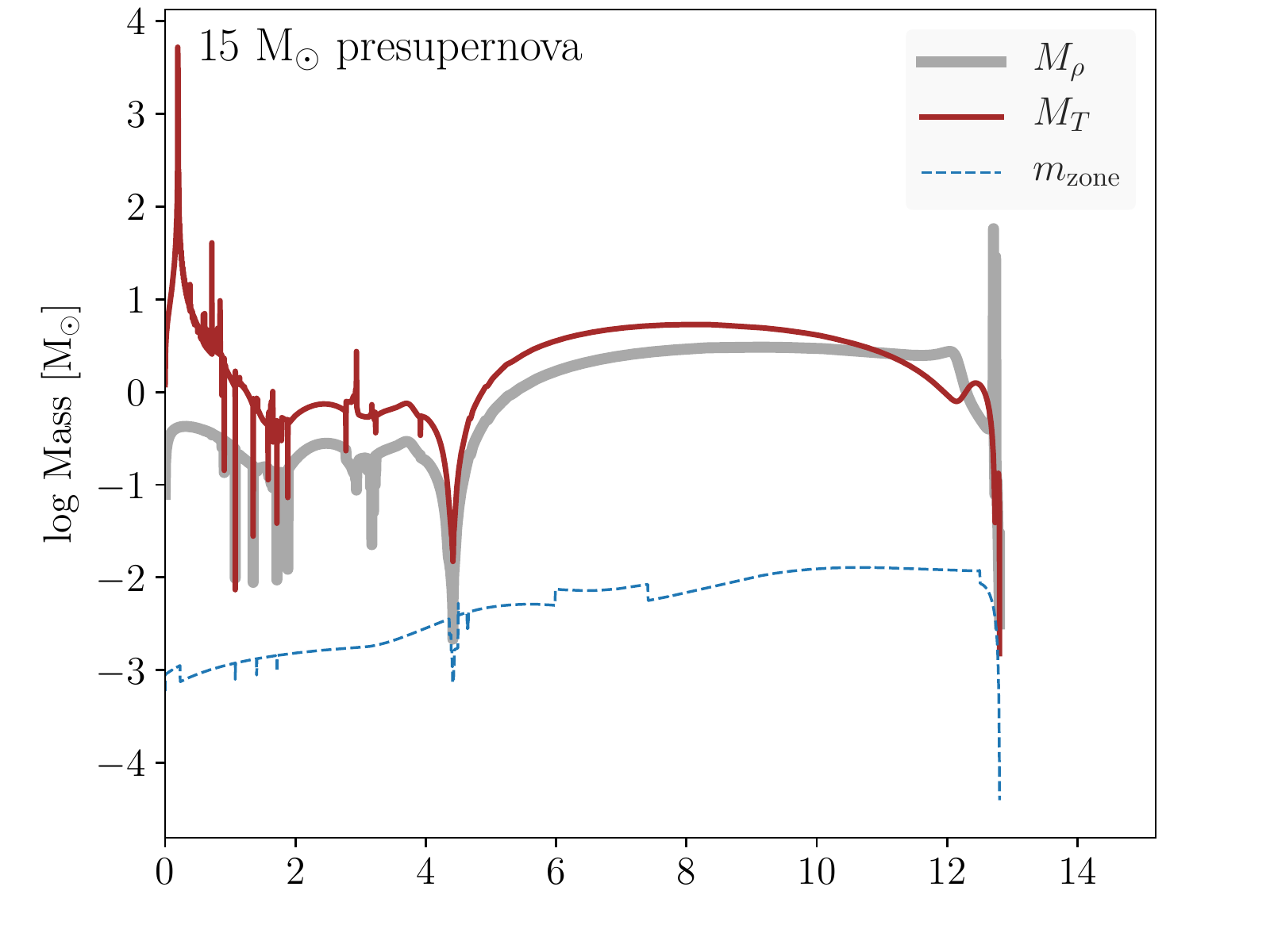}
\includegraphics[width=0.48\textwidth]{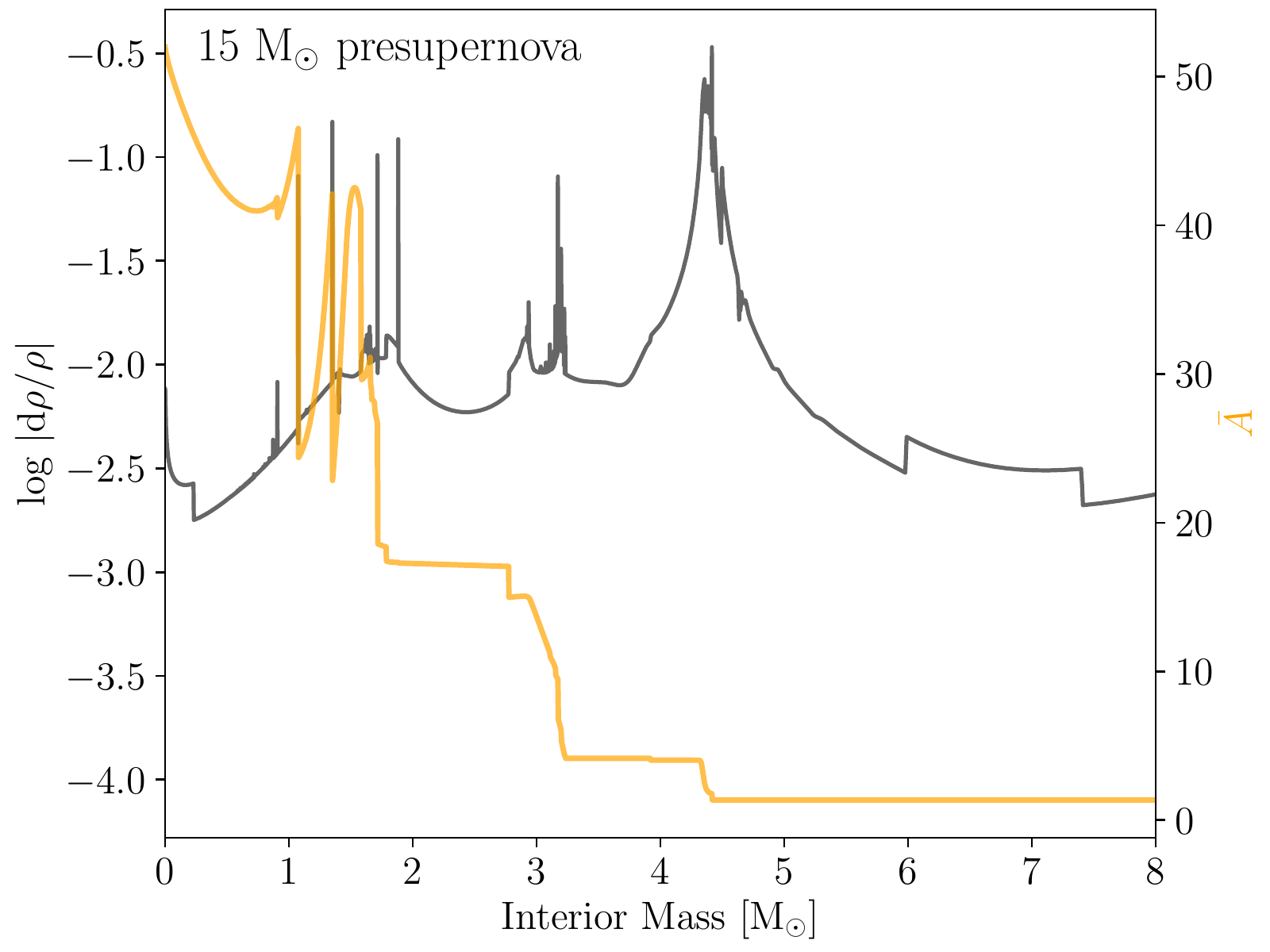}
\caption{Density and temperature resolution in the same 15 \Msun
    \ presupernova model shown in the lower panel of \Fig{zone15}. The
    top panel shows the scale heights ($M_x \equiv \mathrm{d}m/ 
    \mathrm{d}\ln x$) expressed in solar masses for the density ($M_{\rho}$;
    grey curve) and temperature ($M_T$; red curve). The blue dashed
    curve shows the zoning.  The bottom panel shows the actual
    variation in density between zones and mean atomic weight, $\bar
    A$, plotted also as a function of interior mass. Discontinuities
    in $\bar A$ occur at the boundries of active and fossil convective
    shells causing abrupt changes in density that the code attempts to
    resolve. The large spike in $M_T$ at 0.15 \Msun \ reflects a
    temperature inversion at the star's center (where $\mathrm{d}T/
    \mathrm{d}m$ is negative) and a small region of nearly constant 
    temperature bounding it. Both panels show that temperature and density 
    gradients are very well resolved throughout the star, except perhaps 
    in the very steep density gradient at the edge of the helium core 
    (4.33 \Msun), where changes of up to 30\% occur for the density in a 
    few zones.  \lFig{rhograd}}
\end{figure}

\subsection{The Effects of Zoning}
\lSect{ultra}

   One criterion for adequate zoning is that key variables like
   temperature and density be well resolved, that is, that they do not
   vary greatly in going from zone ``i'' to ``i+1''. Stellar evolution
   codes solve linear approximations to non-linear differential
   equations. \Fig{rhograd} shows the resolution for a standard
   $M_{\rm ZAMS}=15$ \Msun \ presupernova star with 4225 zones. The
   figure gives the scale heights in Lagrangian units (solar masses
   here) for the temperature and density. Pressure and other derived
   quantities, though not plotted, show similar variation. The scale
   height, e.g., for density is defined as 
   $M_\rho=\mathrm{d}m/\mathrm{d}\ln\rho$. In places, this quantity 
   can abruptly become artificially small because of discontinuous 
   changes in mean
   atomic weight at the edges of convective shells. Some of the
   prominent downward spikes in the iron core are where partial
   photodisintegration has changed $\bar A$ appreciably. These abrupt
   changes in $\bar A$ are responsible for most of the discontinuous
   spikes in the figure. Large upward spikes indicate regions of near
   constant temperature or density. An especially large spike near the
   center of the presupernova star reflects a temperature inversion
   (note that the scale height can be either positive or negative; the
   absolute value is plotted). Where the derivative changes sign and
   passes through zero, a spike results.

The figure shows that there are roughly 100 to 1000 zones per scale
height everywhere in the star, except in the steep gradient at the
edge of the helium core, where the actual density varies by six orders
of magnitude over an interval of $\sim$0.2 \Msun.  Here the zoning is
worst. In one location, the density changes by a maximum of 30\% from
one zone to the next. Zone masses here were $\sim 5 \times 10^{-4}$
\Msun. Because this location moves in mass as the star evolves, still
finer zoning would have significantly lengthened the calculation. We
conclude that the relevant physical quantities are well resolved in
the new study. In fact, except at the edge of the helium core, they
are over-resolved. This was done in order to explore the sensitivity
of outcome, e.g., the extent of convective shells, and to see if fine
zoning alone would lead to a ``converged'' answer.

Convergence assumes the existence of a well-defined solution to the
stellar structure equations.  As the numerical resolution is
increased, if the subgrid physics (e.g., convection) is coded in a
zoning independent way, the answer from a given calculation should
approach this solution and give a constant answer. Certainly it is
possible for a stellar model to be inadequately zoned. Envision
  \Fig{rhograd} with zoning of 0.1 \Msun. No one today would think of
trying to use just a few hundred zones in a presupernova model for a
25 \Msun \ star, though substantial progress was once made that way
\citep{Wea78}.  But given the power of modern machines, how many zones
are enough?

Typical \KEPLER\ models by \citet{Suk14} and \citet{Mue16} used
about 1,200 zones, roughly independent of the star's total mass,
though the zoning was by no means uniform. Zones were concentrated at
the center, where temperature-dependent burning required fine
resolution, in the steep density gradient at the base of the red giant
convective envelope, and at the surface. Continuous rezoning kept
gradients in density, temperature, and zone radius well resolved.  For
a large range of masses, characteristics of the presupernova star,
such as its helium, carbon, and iron core masses, its luminosity and
radius, and its ``compactness'' varied smoothly with mass and were
consistent with numerous past studies.  In some mass ranges, however,
especially in the range 14 \Msun\ to 19 \Msun, the compactness
parameter often varied wildly.  These variations were attributed by
\citet{Suk14} to the interaction of convective carbon and oxygen
burning shells. The characteristics of those burning shells were
irregular in location, extent, duration, and intensity, and so, for
some masses, was the final presupernova core structure. But might some
of that variation also be attributed to inadequate resolution?  Would
a set of finer-zoned models show less variation, or reveal structure in
the noise?

Most of this paper is about the results of repeating the survey of
\citet{Suk14} with finer zoning (\Fig{zone15}).  In this section,
however, we briefly examine, for just four masses, the question:
``What is enough?''  In part, the answer must have a pragmatic
aspect.  If running with 3,000 and 30,000 zones gives answers that
differ in some important quantity by 5 \%, but if changing the stellar
mass by 0.01 \Msun, or some bit of uncertain stellar physics by a
small fraction of its error bar alters the answer by 50 \%, perhaps
3,000 zones is ``good enough.''  Resources would be better spent
studying the variations that depend on these other variables.

There is a deeper issue though when the answer is deterministic, but
unpredictable. It is an inherent aspect of chaos that tiny changes
upstream produce large differences in outcome. As the noted chaos
theorist Edward Lorentz once said ``\textsl{Chaos exists when the
  present determines the future, but the approximate present does not
  approximately determine the
  future}''\footnote{mpe2013.org/2013/03/17/chaos-in-an-atmosphere-hanging-on-a-wall}. As
noted by \citet{Suk14} and \citet{Mue16} and, as will be explored in
greater depth here, for some ranges of mass, presupernova evolution is
like that. Seemingly minuscule changes in initial conditions may
determine whether a star in an important mass range explodes as a
supernova or collapses to a black hole. How different could two
outcomes be?

\begin{deluxetable*}{cllcccccccccc} 
\tablecaption{Presupernova properties with varying resolution and mass} 
\tablehead{ \colhead{${M_\mathrm{ZAMS}}$}         & 
            \colhead{${N_\mathrm{zones}}$}        & 
            \colhead{${N_\mathrm{steps}}$}        &
            \colhead{${M_\mathrm{presSN}}$}       & 
            \colhead{${L_\mathrm{preSN}}$}        & 
            \colhead{${R_\mathrm{preSN}}$}        &
            \colhead{${T_\mathrm{eff, preSN}}$}   &
            \colhead{${M_\mathrm{He}}$}           &
            \colhead{${M_\mathrm{Fe}}$}           &
            \colhead{${\xi_{2.5}}$}               &
            \colhead{${M_4}$}                     &
            \colhead{${\mu_4}$}
            \\
            \colhead{[\Msun]}                         &
            \colhead{}                                & 
            \colhead{}                                &
            \colhead{[\Msun]}                         &
            \colhead{[${\rm 10^{38}\ ergs\ s^{-1}}$]} &
            \colhead{[${10^{13}}$ cm]}                &
            \colhead{[K]}                             &
            \colhead{[\Msun]}                         &
            \colhead{[\Msun]}                         &
            \colhead{}                                &
            \colhead{[\Msun]}                         &
            \colhead{} 
            }\\
\startdata
\texttt{15.00B} &  1258  &  59992  &  12.935  &  3.57  &  6.04  &  3425  &  4.309  &  1.552  &  0.152  &  1.665  &  0.079  \\
\texttt{15.00C} &  2153  &  60162  &  12.866  &  3.58  &  5.97  &  3450  &  4.314  &  1.580  &  0.178  &  1.680  &  0.089  \\
\texttt{15.00D} &  4225  &  63435  &  12.804  &  3.59  &  6.06  &  3420  &  4.326  &  1.581  &  0.176  &  1.718  &  0.085  \\
\texttt{15.00E} &  8138  &  77578  &  12.485  &  3.69  &  6.19  &  3410  &  4.409  &  1.494  &  0.094  &  1.496  &  0.044  \\
\texttt{15.00F} &  15912 &  87612  &  11.717  &  4.02  &  6.64  &  3363  &  4.683  &  1.413  &  0.111  &  1.591  &  0.047  \\
\multicolumn{12}{c}{}\\
\texttt{25.00A} &  874    &  38558  &  16.047  &  7.88  &  8.56  &  3505  &  8.342  &  1.609  &  0.303  &  1.915  &  0.103  \\
\texttt{25.00B} &  1885   &  38202  &  16.026  &  9.59  &  9.78  &  3445  &  8.351  &  1.620  &  0.301  &  1.911  &  0.102  \\
\texttt{25.00C} &  3358   &  40799  &  15.690  &  9.72  &  9.79  &  3455  &  8.427  &  1.606  &  0.301  &  1.916  &  0.102  \\
\texttt{25.00D} &  6515   &  45929  &  15.902  &  9.59  &  9.77  &  3445  &  8.382  &  1.630  &  0.301  &  1.910  &  0.102  \\
\texttt{25.00E} &  12734  &  58975  &  15.816  &  9.70  &  9.85  &  3440  &  8.432  &  1.643  &  0.303  &  1.903  &  0.104  \\
\vspace{-0.2cm}\\
\hline
\vspace{-0.2cm}\\
\texttt{15.01B} &  1260  &  59794  &  12.922  &  3.59  &  6.05  &  3425  &  4.319  &  1.561  &  0.164  &  1.675  &  0.083  \\
\texttt{15.01C} &  2149  &  60641  &  12.701  &  3.64  &  6.05  &  3435  &  4.361  &  1.513  &  0.189  &  1.742  &  0.088  \\
\texttt{15.01D} &  4190  &  63675  &  12.574  &  3.67  &  6.16  &  3410  &  4.386  &  1.357  &  0.126  &  1.457  &  0.067  \\
\texttt{15.01E} &  8103  &  76827  &  12.157  &  3.83  &  6.37  &  3390  &  4.523  &  1.506  &  0.100  &  1.517  &  0.046  \\
\texttt{15.01F} &  15952 &  77854  &  11.845  &  3.95  &  6.55  &  3373  &  4.638  &  1.438  &  0.159  &  1.599  &  0.070 \\
\multicolumn{12}{c}{}\\
\texttt{25.01A} &  876   &  38416  &  16.138  &  7.88  &  8.56  &  3505  &  8.269  &  1.606  &  0.301  &  1.911  &  0.102  \\
\texttt{25.01B} &  1897  &  38165  &  16.319  &  9.57  &  9.80  &  3440  &  8.348  &  1.610  &  0.298  &  1.909  &  0.101  \\
\texttt{25.01C} &  3335  &  41211  &  15.141  &  9.85  &  9.82  &  3460  &  8.498  &  1.613  &  0.300  &  1.910  &  0.101  \\
\texttt{25.01D} &  6466  &  46494  &  15.235  &  9.81  &  9.89  &  3445  &  8.485  &  1.606  &  0.302  &  1.907  &  0.103  \\
\texttt{25.01E} &  12268 &  67699  &  13.264  &  10.34 &  9.86  &  3495  &  8.720  &  1.600  &  0.315  &  1.901  &  \ 0.110 
\enddata
\tablecomments{All quantities are measured at the presupernova stage. 
               $N_{\rm zones}$ is slightly larger at the beginning of calculation.}
\lTab{jmztable}
\end{deluxetable*}

\Tab{jmztable} gives the final presupernova properties of four sets of
models with varying resolution for initial masses of 15.00, 15.01,
25.00, and 25.01 \Msun. The spatial resolution ranges from about 1,200
to 16,000 mass shells. The most finely resolved models used roughly
twice as many timesteps to reach the presupernova stage as the least
resolved models.  The ``compactness'' parameter, $\xi_{2.5}$
\citep{Oco11} is inversely proportional to the radius enclosing
innermost 2.5 \Msun\ of the presupernova core, while $M_4$ and $\mu_4$
\citep[][ further discussed in \Sect{compact}]{Ert16} are, respectively, 
the lagrangian location of the mass shell with entropy per baryon 
$s = 4 \, k_B$ and the derivative of mass at that location evaluated 
over a mass interval of 0.3 \Msun. The motivation for using these 
parameters is discussed in \Sect{compact}.

The zoning for the typical 15 \Msun\ case in \Fig{zone15} corresponds
to the moderately zoned case (\texttt{15.00D}) in \Tab{jmztable}. Initial
zoning in other 15.00 \Msun\ and 15.01 \Msun\ cases was scaled at all
locations in the star by roughly factors of two. For the 25.00
\Msun\ and 25.01 \Msun\ models the number of zones was proportionately
larger. That is Model \texttt{25.00C} had roughly $25/15$ as many
zones as Model \texttt{15.00C}, etc. There was no Model
\texttt{25.00F}, instead there was a Model \texttt{25.00A} which had
about half the zoning of \texttt{25.00B}.  The continuous rezoning
parameters were set so as to preserve, within about 10 \%, the initial
zoning. The number of zones given in \Tab{jmztable} is for the
presupernova star.

Two closely adjacent masses were studied at 15 \Msun\ and 25
\Msun\ in order to compare the effects of fine zoning with slight
variations in any other parameter of the problem. One might have chosen
instead to vary time step, overshoot, the rate for
$^{12}$C($\alpha,\gamma)^{16}$O, semiconvection, mass loss, or
rotation.  Although no substitute for a full survey of such
dependencies, varying the mass slightly in a situation where the
solution is chaotic might be expected to send the calculation down one
path or another. Stated another way, if varying the mass by less than
0.1 \% results in an answer significantly different than one obtained
by increasing the zoning by a factor of two, perhaps there is no
reason to use much finer zoning for that mass.  On the other hand, if
all models give the same answer, one might have confidence in the
uniqueness of that model, at least for an assumed set of stellar
physics.

The two choices, 15 \Msun\ and 25 \Msun, illustrate these two
possibilities nicely.  In all cases the observable properties of the
presupernova star, its luminosity and radius, and hence its effective
temperature, are well determined.  Changing opacities, composition,
etc., would certainly cause variation, but the bulk observables of our
models are well determined.  There is a weak, but noticeable trend to
produce smaller presupernova stellar masses and larger helium core
masses when the zoning is finer. As will be discussed later, this is a
consequence of variation in the time spent as a red or blue
supergiant, which, in turn, is sensitive to an uncertain treatment of
semiconvection. A small number of episodic mixing events at the edge
of convective hydrogen burning core can temporarily ``rejuvinate'' the
core and by affecting the gravitational potential of hydrogen burning
shell, eventually causing the transition from blue to red to
occur at different times. This trend does not emerge because the
models are converging with higher resolution, but rather as a result
of enhanced mixing due to the uncertain treatment of convection
physics.

The measures of presupernova core structure on the other hand vary by
factors of two for the 15 \Msun\ models, but scarcely at all for the
25 \Msun\ models.  The reasons for the difference will be discussed in
\Sect{interpret}, but we note that any reasonable zoning suffices to
get the structure of the 25 \Msun\ correct, but no reasonable zoning
shows convergence for the 15 \Msun\ model.  Not knowing beforehand
what the outcome will be for an arbitrary mass, we used here the
maximum zoning that time and the desire to survey thousands of models
allowed. This is most like the ``D'' models in \Tab{jmztable}. There
is little motivation, however, for carrying out still finer resolution
studies when a trivial change in mass, 15.00 \Msun\ to 15.01 \Msun,
causes more variation than an order of magnitude increase in zoning.

\subsection{The Effects of Networks}
\lSect{network}

Uncertainty in nuclear energy generation and neutronization during the
various episodes of burning can be another source of
variability in presupernova models.  In this section the effect of
using either the standard 19-isotope approximation network plus
silicon quasiequilibrium (\Sect{physics}) or a much larger nuclear
reaction network is explored. It should be noted that while the
standard network carries the abundances of only $19$ species, it has
the power of a network roughly twice that size since reactions through
trace species like $^{23}$Na, $^{27}$Al, etc.\ are carried in a
steady-state approximation \citep{Wea78}.  Reactions coupling
$^{54}$Fe to $^{56}$Ni are also included, as are approximations to
energy generation by the CNO- and pp-cycles and to
$^{14}$N($\alpha,\gamma)^{18}$F($e^+
\nu)^{18}$O($\alpha,\gamma)^{22}$Ne\footnote{In the code, $^{14}$N is
  converted to $^{20}$Ne at an appropriate rate conserving mass.}.

The large network used for comparison here is drawn from an extensive
nuclear data base maintained and used for nucleosynthesis studies for
decades \citep[e.g.,][]{Woo02}.  Key reaction rates such as
$^{14}$N($p,\gamma)^{15}$O, $3 \alpha$,
$^{12}$C($\alpha,\gamma)^{16}$O, the $(\alpha,\gamma)$ and
($\alpha,p$) reactions on $\alpha-$nuclei and the weak interaction
rates are the same in both networks. The network is ``adaptive''
\citep{Rau02} in the sense that isotopes are added or subtracted at
each step to accommodate the reaction flow (e.g., there is no need to
include a detailed network for the iron group during hydrogen
burning). Usually the settings on the network size are quite liberal,
resulting in from 700 to 1,200 isotopes being carried in the early and
late stages of evolution respectively \citep{Woo07}.  Here, because we
are only interested in stellar structure and not, e.g., the
\textsl{s}-process, a smaller number was carried, typically $300$
isotopes extending up through the element Selenium. All of the
iron-group isotopes in the quasi-equilibrium network were included,
and many more, so that core neutronization during and after silicon
burning was equivalently calculated.

Network sensitivity was examined for four different masses of star,
15.00, 15.01, 25.00, and 25.01 \Msun.  Two studies were carried out to
test network sensitivity.  \Tab{neteffect} edits the evolution of the
15.00 star, similar to Model \texttt{15.00D} in \Tab{jmztable}, at
different times in its evolution using 3 approximations to the energy
generation and neutronization. The \textsc{Approx} case corresponds to
using the 19 isotope approximation network and quasiequilibrium
network, as is standard in the rest of this paper. The \textsc{Coproc}
case carries the large network of about 300 isotopes along with the
approximation network in ``co-processing mode''. That is, the
approximation and quasi-equilibrium networks are used for energy
generation in the stellar structure calculation, but the network is
also carried along in passive mode. The same time step is used at the
same temperature and density to evolve, zone by zone, a much larger
number of isotopes in parallel with the approximation
network. Sub-cycling is done within a given time step to follow the
abundances of trace isotopes accurately. Coprocessing is only
performed in the code up to the point where the transition to
quasiequilibrium occurs in a zone, typically at oxygen depletion
$X(^{16}O$ = 0.03).
The output from the large co-processing network is used to continually
update the electron mole number, $Y_{\rm e}$, which feeds back into
the structure calculation.  Because $Y_{\rm e}$ evolves slowly, no
iteration is required. The energy generation from the large network
can also be used to check the validity of the approximation network,
though no correction is applied. In the third case \textsc{Full}, the
big network is coupled directly to the structure calculation
throughout the entire life of the star, including silicon burning and
core collapse. This might be deemed the most accurate approach, but it
is slow, requiring an order of magnitude more computer resources, and
prone to instability during silicon burning because of strongly
coupled flows.

In \Tab{neteffect}, up until oxygen depletion, two different numbers
are given for the co-processing run (\textsc{Coproc}). The first is the 
energy generation from the approximation network, which is used to 
calculate the stellar model, and the second is the output from the 
passively carried big network.
Hydrogen burning and depletion correspond to central 
hydrogen mass fractions of 0.4 and 0.01. Helium burning and helium 
depletion are when the central helium mass fraction is 0.5 and 0.01. 
Carbon ``ignition'' is actually evaluated during the Kelvin-Helmholtz 
contraction between helium depletion and real carbon burning when the 
central temperature is $5 \times 10^8$ K, and carbon depletion is when 
the central carbon bass fraction is 1 \%. Similarly, ``oxygen ignition'' 
is when the central temperature during Kelvin-Helmholtz contraction is 
$1.5 \times 10^9$ K and oxygen has yet to burn, and oxygen depletion 
is when the central oxygen mass fraction is 0.03. Silicon ignition is 
at $3.0 \times 10^9$ K, and silicon depletion is when the silicon mass 
fraction is 1 \%.  Presupernova is  when the core collapse speed 
exceeds a maximum of 900 $\rm km\ s^{-1}$.

The table shows near exact agreement in energy generation and $Y_{\rm
  e}$ calculated using all three approaches up until at least oxygen
ignition. The slight differences in energy generation at carbon
ignition do not matter because neutrino losses (far right column)
dominate at the time examined.  By oxygen depletion though, things are
starting to mildly diverge. This divergence has three causes. One is
the electron capture that goes on in the late stages of oxygen burning
and starts to appreciably affect $Y_{\rm e}$. This change is not
followed by the \textsc{Approx} network. Carrying $Y_{\rm e}$ in the
co-processing run addresses most of this divergence; the values of
$Y_{\rm e}$ in the co-processing and full network runs are nearly the
same. Another effect, more difficult to disentangle, is the fact that,
at late times, the stellar structure in the different calculations
starts to diverge. It diverges, in part, of the different nuclear
physics, but even more because of the chaotic nature of the evolution.
As was seen in the zoning study (\Sect{ultra}), in certain initial
mass ranges two models with even slightly different physics will
differ appreciably in final outcome. The third effect is the
instantaneous difference due solely to the differing treatments of
nuclear physics.  This is what we are actually trying to study, but
difficult to separate from cumulative effects due to structural
changes. Even so, at silicon depletion the three values of $Y_{\rm e}$
agree very well and the energy generation differs by a factor of two,
which is mostly due to the higher temperature in the \textsc{Approx}
case and the very sensitive temperature dependence of the silicon
burning reactions.

These small differences probably cause very little change in
the presupernova model because the star has a certain amount of fuel to
burn and the total energy release is set by known nuclear binding
energies. The different burning rates only affects modestly the
temperature at which the silicon burns in steady state
\citep{Woo02}. We conclude that the approximation plus
quasiequilibrium approach is adequate for our survey and introduces
errors that are small compared with other variations that occur when
other uncertain quantities, or even the zoning change.

\begin{deluxetable}{lcclcc}
\tablecaption{Central properties for different network treatments for 15.00 \Msun\ model}
\tablehead{
\colhead{network}               &
\colhead{${\log T_{\rm c}}$}    &
\colhead{${\log \rho_{\rm c}}$} &
\colhead{${Y_{\rm e,c}}$}       &
\colhead{${\log S_{\rm n}}$}    &
\colhead{${\log |S_\nu|}$}\\ 
&
\colhead{[K]}                   &
\colhead{[$\rm{g\ cm^{-3}}$]}   &
\colhead{}                      &
\multicolumn{2}{c}{[$\rm{ergs\ g^{-1}\ s^{-1}}$]}
}\\
\startdata
\multicolumn{6}{c}{H burning}\\
\textsc{Approx} & 7.564 & 0.822 & 0.70042 & 5.104 & 3.910 \\
\textsc{Coproc} & 7.564 & 0.822 & 0.70196 & 5.104 & 3.910 \\
                &       &       & 0.70196 & 5.107 &       \\
\textsc{Full}   & 7.565 & 0.820 & 0.69991 & 5.109 & 3.914 \\
\multicolumn{6}{c}{H depletion}\\
\textsc{Approx} & 7.676 & 1.138 & 0.50581 & 5.376 & 4.114 \\
\textsc{Coproc} & 7.676 & 1.138 & 0.50616 & 5.376 & 4.114 \\
                &       &       & 0.50616 & 5.469 &       \\
\textsc{Full}   & 7.677 & 1.136 & 0.50489 & 5.387 & 4.129 \\
\multicolumn{6}{c}{He burning}\\
\textsc{Approx} & 8.249 & 3.127 & 0.49998 & 5.600 & 1.375 \\
\textsc{Coproc} & 8.250 & 3.123 & 0.49932 & 5.621 & 1.383 \\
                &       &       & 0.49932 & 5.631 &       \\
\textsc{Full}   & 8.249 & 3.122 & 0.49932 & 5.621 & 1.371 \\
\multicolumn{6}{c}{He depletion}\\
\textsc{Approx} & 8.406 & 3.426 & 0.49998 & 5.624 & 2.560 \\
\textsc{Coproc} & 8.406 & 3.426 & 0.49932 & 5.627 & 2.562 \\
                &       &       & 0.49932 & 5.667 &       \\
\textsc{Full}   & 8.407 & 3.422 & 0.49932 & 5.667 & 2.577 \\
\multicolumn{6}{c}{C ignition}\\
\textsc{Approx} & 8.703 & 4.544 & 0.49998 & 0.405 & 4.750 \\
\textsc{Coproc} & 8.704 & 4.546 & 0.49933 & 0.434 & 4.756 \\
                &       &       & 0.49933 & 0.368 &       \\
\textsc{Full}   & 8.700 & 4.513 & 0.49933 & 0.159 & 4.729 \\
\multicolumn{6}{c}{C depletion}\\
\textsc{Approx} & 9.081 & 6.602 & 0.49998 & 6.193 & 8.198 \\
\textsc{Coproc} & 9.080 & 6.598 & 0.49894 & 6.181 & 8.202 \\
                &       &       & 0.49894 & 6.108 &       \\
\textsc{Full}   & 9.079 & 6.580 & 0.49897 & 6.063 & 8.212 \\
\multicolumn{6}{c}{O ignition}\\
\textsc{Approx} & 9.176 & 6.895 & 0.49998 & 10.63 & 9.123 \\
\textsc{Coproc} & 9.176 & 6.899 & 0.49896 & 10.60 & 9.116 \\
                &       &       & 0.49896 & 10.58 &       \\
\textsc{Full}   & 9.176 & 6.944 & 0.49899 & 10.57 & 9.188 \\
\multicolumn{6}{c}{O depletion}\\
\textsc{Approx} & 9.343 & 7.056 & 0.49998 & 11.53 & 11.05 \\
\textsc{Coproc} & 9.346 & 7.037 & 0.49242 & 11.60 & 11.12 \\
                &       &       & 0.49242 & 11.64 &       \\
\textsc{Full}   & 9.352 & 6.990 & 0.49309 & 11.81 & 11.27 \\
\multicolumn{6}{c}{}\\
\multicolumn{6}{c}{Si ignition}\\
\textsc{Approx} & 9.477 & 8.068 & 0.48679 & 12.14 & 11.21 \\
\textsc{Coproc} & 9.477 & 8.087 & 0.48609 & 12.18 & 11.19 \\
\textsc{Full}   & 9.477 & 8.084 & 0.47945 & 11.51 & 11.29 \\
\multicolumn{6}{c}{Si depletion}\\
\textsc{Approx} & 9.583 & 7.699 & 0.47060 & 12.71 & 13.01 \\
\textsc{Coproc} & 9.571 & 7.706 & 0.46987 & 12.67 & 12.97 \\
\textsc{Full}   & 9.477 & 7.739 & 0.46809 & 12.40 & 12.92 \\
\multicolumn{6}{c}{presupernova}\\
\textsc{Approx} & 9.885 & 9.747 & 0.43848 & (16.11) & 16.24 \\
\textsc{Coproc} & 9.880 & 7.762 & 0.43747 & (16.09) & 16.20 \\
\textsc{Full}   & 9.869 & 10.10 & 0.43266 & (16.83) & 16.94
\enddata
\tablecomments{$S_{\rm n}$ is the energy generation rate excluding 
neutrino-losses, and is negative during the presupernova stage due 
to photodisintegration. Values in parentheses denote $\log |S_{\rm n}|$. 
The two values of $Y_{\rm e,c}$ and $S_{\rm n}$ given for co-processing 
calculations until core oxygen depletion (until quasi-equilibrium) are 
form approximation network (top) and co-processed big network (bottom).
\lTab{neteffect}}
\end{deluxetable}

\begin{deluxetable*}{clcccccccccc}
\tablecaption{Presupernova properties with varying network size}
\tablehead{ \colhead{${M_{\rm ZAMS}}$} &
\colhead{${\rm network}$} &
\colhead{${N_{\rm zones}}$} &
\colhead{${N_{\rm steps}}$} &
\colhead{${M_{\rm He}}$} &
\colhead{${M_{\rm CO}}$} &
\colhead{${\rm ^{12}C_{ign.}}$} &
\colhead{${M_{\rm Fe}}$} &
\colhead{${\xi_{2.5}}$} &
\colhead{${M_4}$} &
\colhead{${\mu_4}$} &
\colhead{${Y_{\rm e,c}}$}
\\
\colhead{[\Msun]} &
\colhead{} &
\colhead{} &
\colhead{} &
\colhead{[\Msun]} &
\colhead{[\Msun]} &
\colhead{} &
\colhead{[\Msun]} &
\colhead{} &
\colhead{[\Msun]} &
\colhead{} &
\colhead{}
}\\
\startdata
15.00 & \textsc{Approx} & 4233 & 32013 & 4.319 & 2.949 & 0.215 & 1.527 & 0.187 & 1.760 & 0.087 & 0.438 \\
15.00 & \textsc{Coproc} & 4227 & 33294 & 4.319 & 2.946 & 0.215 & 1.537 & 0.179 & 1.742 & 0.084 & 0.437 \\
\smallskip
15.00 & \textsc{Full}      & 4185 & 39361 & 4.383 & 3.014 & 0.202 & 1.407 & 0.139 & 1.469 & 0.068 & 0.433 \\
15.01 & \textsc{Approx} & 4235 & 33091 & 4.308 & 2.937 & 0.211 & 1.373 & 0.143 & 1.449 & 0.088 & 0.434 \\
15.01 & \textsc{Coproc} & 4248 & 35736 & 4.298 & 2.925 & 0.216 & 1.387 & 0.135 & 1.421 & 0.097 & 0.432 \\
15.01 & \textsc{Full}      & 4194 & 36083 & 4.390 & 2.998 & 0.214 & 1.511 & 0.195 & 1.761 & 0.089 & 0.437 \\
\multicolumn{12}{c}{}\\
25.00 & \textsc{Approx} & 4717 & 30080 & 8.663 & 6.711 & 0.179 & 1.604 & 0.311 & 1.913 & 0.107 & 0.444 \\
25.00 & \textsc{Coproc} & 4721 & 30143 & 8.670 & 6.722 & 0.180 & 1.580 & 0.306 & 1.891 & 0.106 & 0.443 \\
\smallskip
25.00 & \textsc{Full}      & 4589 & 42642 & 8.179 & 6.322 & 0.179 & 1.806 & 0.240 & 1.807 & 0.084 & 0.438 \\
25.01 & \textsc{Approx} & 4696 & 30317 & 8.633 & 6.720 & 0.176 & 1.617 & 0.312 & 1.912 & 0.108 & 0.444 \\
25.01 & \textsc{Coproc} & 4681 & 30120 & 8.640 & 6.670 & 0.176 & 1.564 & 0.289 & 1.854 & 0.110 & 0.443 \\
25.01 & \textsc{Full}      & 4531 & 43698 & 8.411 & 6.513 & 0.180 & 1.835 & 0.252 & 1.837 & 0.088 & 0.438
\enddata
\tablecomments{All quantities are measured at the presupernova stage, 
except $\rm ^{12}C_{ign}$ (\Sect{network}). ${N_{\rm zones}}$ is 
slightly larger at the beginning of calculation. $Y_{\rm e,c}$ is 
evaluated at the center.}
\lTab{nettable}
\end{deluxetable*}

\Tab{nettable} examines the effect of using either the large network
or the approximation network plus quasiequilibriumin to study stars of
15.00, 15.01, 25.00, and 25.01 \Msun, the same masses considered in
the zoning sensitivity study (\Sect{ultra}). Only one zoning was
considered here, but the resolution was about three times greater in
all 4 cases than in \citet{Suk14}. Fewer time steps were taken during
hydrogen burning than in the zoning sensitivity study. An overall
limit of $5 \times 10^{10}$ s was placed on the time step as opposed
to $1 \times 10^{10}$ s in the survey and zoning sensitivity
study. Here, the emphasis is on comparing runs with different
networks, not on the highest spatial and temporal resolution. 
Consequently the 15.00 model in \Tab{nettable} is not directly 
comparable with the one in \Tab{jmztable}.

\Tab{nettable} gives, in addition to the presupernova zoning and time
steps, the masses of the iron, carbon-oxygen, and helium cores, the
central mass fraction of carbon evaluated just prior to carbon 
ignition, the central value of the electron mole number, $Y_{\rm e}$ 
in the presupernova star, and the descriptors of presupernova core
structure, $\xi_{2.5}$, $\mu_4$, and $M_4$. Though not given, the
presupernova masses, radii, luminosities, and effective temperatures
are in the same range as for the same mass stars in \Tab{jmztable}.

Based solely on the changes in the core structure and iron core mass
for the 15.00 and 15.01 \Msun\ models, one might conclude that using a
large network made a substantial difference and all future runs should
take much greater care with the nuclear physics. Though the models do
indeed differ, this would not be a valid inference. As shown in
\Sect{ultra} and discussed elsewhere in the paper, changing {\sl
  anything} for models in some mass ranges can give qualitatively
different answers for core structure. The differences between runs
with large and small networks is within the range of differences
resulting from increased or decreased zoning (\Tab{jmztable}) or a
small change in the star's mass. More telling is the fact that the
central carbon abundances, the helium and carbon-oxygen core masses,
and most critically the central value of $Y_{\rm e}$ all agree very
well. The approximation network is doing a fine job representing the
nuclear physics. It is just that other, uncontrollable factors 
introduce large variations in the core structure of a 15 \Msun, and
to a lesser extent 25 \Msun\ star. We conclude that the 19-isotope
approximation network and quasiequilibrium hypothesis are quite
adequate for surveys like this. Instead, focus should be placed on 
studying the impact of stellar physics on the statistical properties 
over the entire mass range.

\subsection{Boundary Pressure}
\lSect{pbound}

Especially during its post-main sequence evolution, the radius of a
model star is sensitive to surface boundary conditions.  Different
codes treat the surface in different ways. Many use a boundary
condition on pressure or density that sometimes includes a reduction
of gravity by the Eddington factor and by the inertia term from wind
acceleration \citep[e.g.,][]{App70,Heg98,Pax15}. Others solve the wind
equation for conditions at the sonic radius (Grassitelli, private
communication, 2018). Still others fit the surface to a stellar
atmosphere of varying complexity (e.g. plane-parallel gray or tables)
\citep[e.g.][]{Pax11,Chi89}. \KEPLER\ uses a constant boundary
pressure, $P_{\rm bound}$ that does not vary during the evolution. The
advantage of such an approach is its simplicity, but care must be
taken that $P_{\rm bound}$ is not so large as to greatly alter the
solution.

Ideally, $P_{\rm bound}$ only influences the structure of a tiny mass
near the surface.  No reasonable value of $P_{\rm bound}$ affects
main sequence evolution, for example, because the the pressure
gradient there is very steep. The gradient becomes shallower during
the RSG phase though, and even slightly different radii there can
significantly alter the mass loss rate. The timing and development of
the surface convection zone is also affected. Both can have a
significant affect on the star's location in the HR-diagram and, to a
lesser extent, the helium core mass and presupernova structure.
Traditionally, studies with \KEPLER \ have used $P_{\rm bound}$ = 50
to 100 dyne cm$^{-2}$ \citep[e.g.,][]{Woo02,Woo07,Suk14}. An exception
is the work of \citet{Mue16}, where a much larger value, $\sim3500$
dyne cm$^{-2}$ was used.  

In the present study a value of 50 dyne cm$^{-2}$ is employed. Is this
low enough?  To explore the sensitivity of results to boundary
pressure, a standard 15.00 \Msun, Model S15.00D in \Tab{jmztable}, was
calculated using a variety of boundary pressures from 10 to 6400 dyne
cm$^{-2}$ maintained throughout the evolution. The resulting HR
diagrams and radial histories are shown in \Fig{pb}. For $P_{\rm
  bound} \ltaprx 500$ dyne cm$^{-2}$ the trajectory in the HR-diagram
is essentially identical for six different choices of $P_{\rm
  bound}$. The presupernova radius and luminosity do not vary.  For
larger values of $P_{\rm bound}$ though, especially $P_{\rm bound}
\gtaprx 1000$ dyne cm$^{-2}$ significant variation is found.

Even the models with small surface pressures show some variation in
final properties. For example, for $P_{\rm bound}$ between 10 and 400
dyne cm$^{-2}$ the final star mass varied between 12.59 \Msun \ and 12.61
\Msun, while the helium core mass varied between 3.61 \Msun \ and 3.67
\Msun. Some of this variation is a consequence of the irreducible
noise in running any model twice with even small variations in the
physics or mass, but part could be a residual sensitivity to $P_{\rm
  bound}$.

\begin{figure}
\begin{center}
\leavevmode
\includegraphics[width=0.48\textwidth]{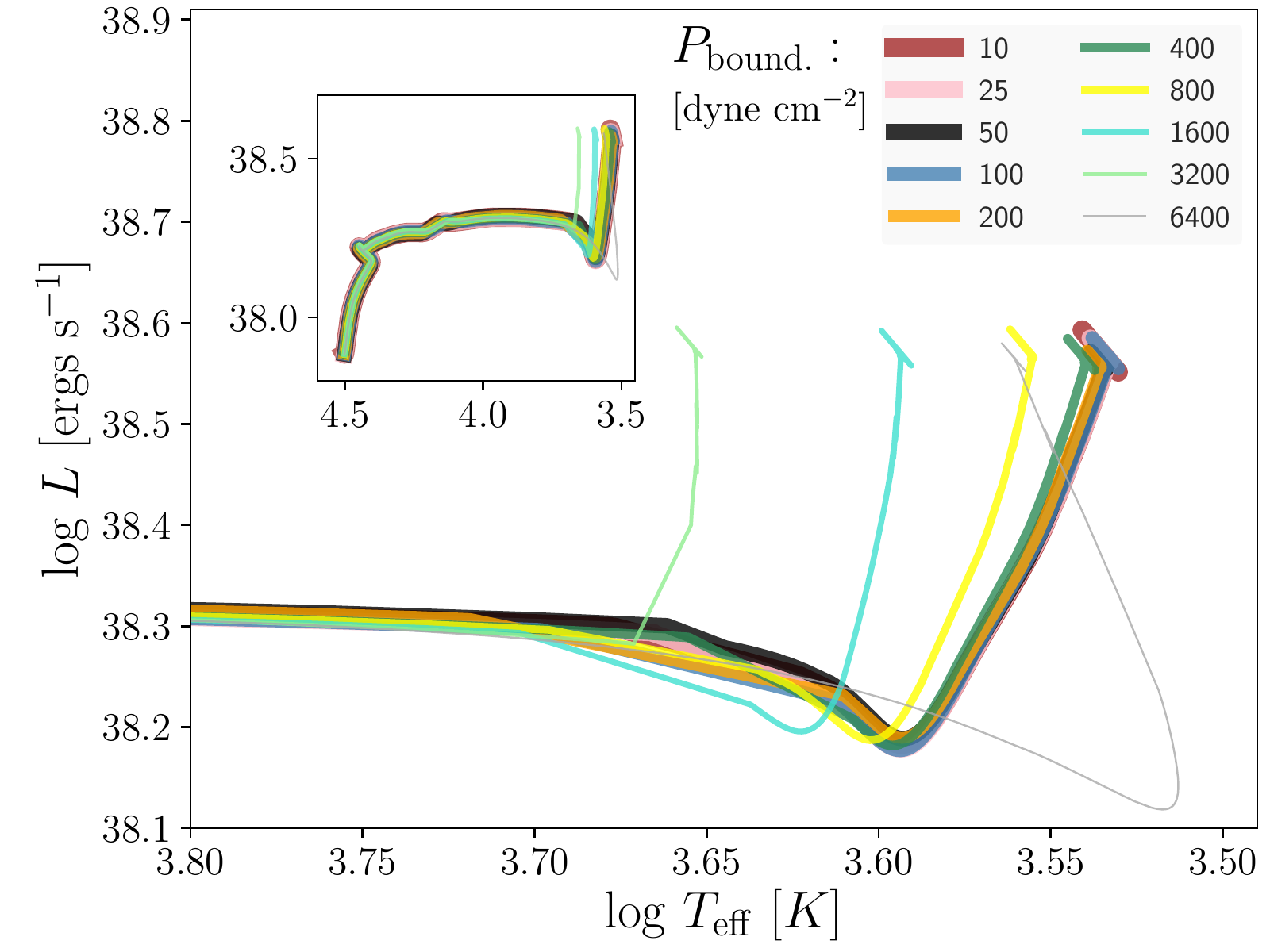}
\includegraphics[width=0.48\textwidth]{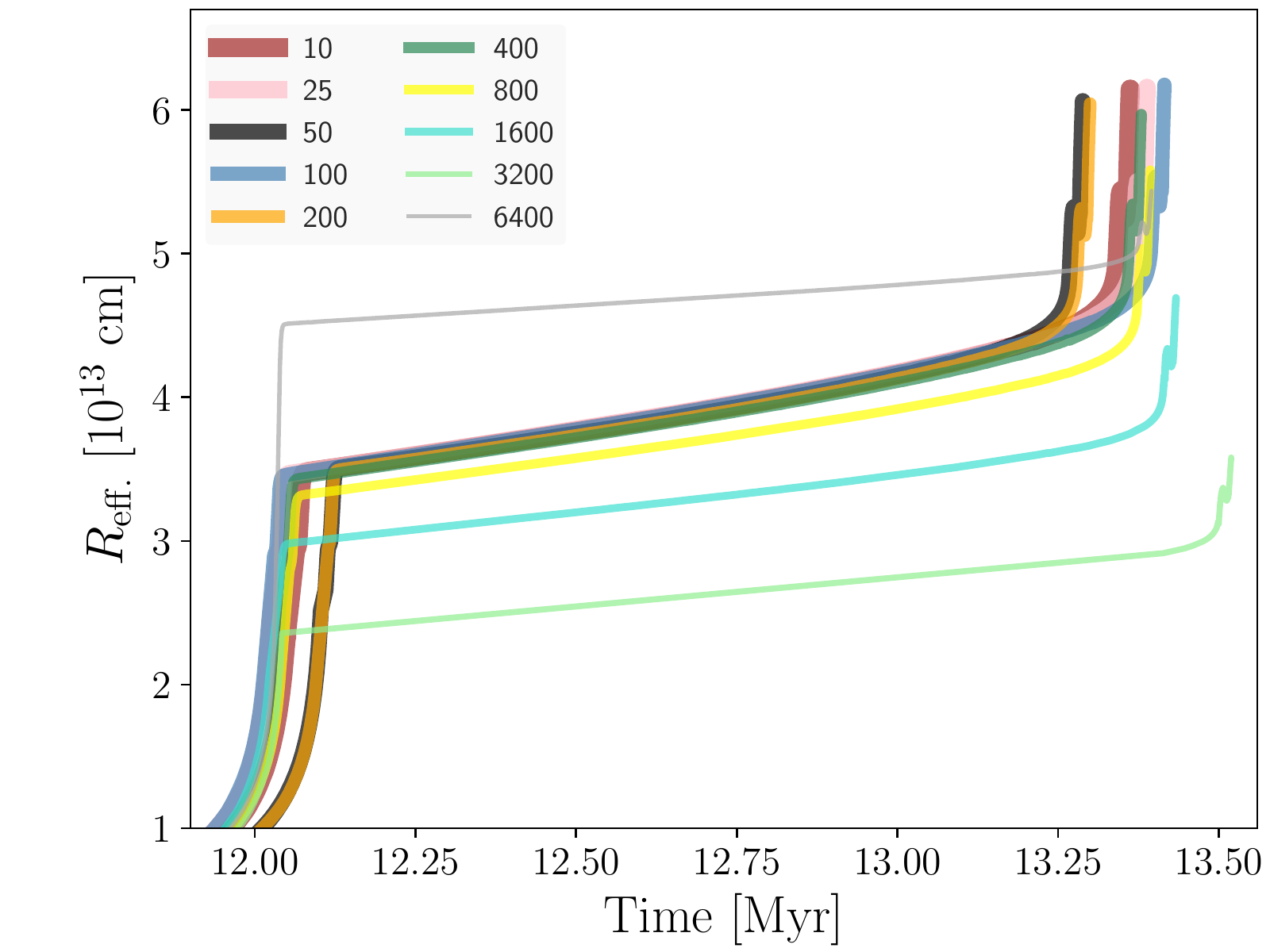}
\caption{(Top panel:) The HR-diagram for variations of Model 15.00D
  (\Tab{jmztable}) in which the surface boundary pressure is varied
  from 10 dyne cm$^{-2}$ to 6400 dyne cm$^{-2}$ in roughly factor of
  two steps. The inset shows that the evolution is insensitive to this
  boundary condition until the star becomes a red supergiant. The main
  figure shows nearly identical results for pressures $\ltaprx 500$
  dyne cm$^{-2}$, but significant variations for larger
  values. (Bottom panel:) Radius as a function of time during the
  final evolution of the same models. Only the last 1.4 My when the
  stars become RSGs is shown. Models with boundary
  pressures less than 500 dyne cm$^{-2}$ have nearly identical radii
  as supergiants until very close to the end while those with bigger
  pressures vary. The time spent as RSG's is similar. \lFig{pb}}
\end{center}
\end{figure}

Closely related to the treatment of surface boundary conditions is
surface zoning. In our recent studies \citep{Suk14,Mue16}, rezoning
was allowed to continue all the way to the surface of the star
throughout its evolution. This often resulted in coarse zoning near
the photosphere during the RSG stage. Fine zoning on the main sequence
was lost to dezoning when the temperature and density gradients were
shallow. See \Fig{zone15} for a sample comparison with the older
model. In the new models rezoning was not allowed in the outer 250
zones, typically 0.3 \Msun. Thus the fine surface zoning shown in
\Fig{zone15} was maintained throughout the evolution for all models.
Typical zoning near the surface is less than $5 \times 10^{-4}$ \Msun.

The use of finer surface zoning and a reduced boundary pressure
generally gave larger radii and increased mass loss during the RSG
stage, especially as compared with \citet{Mue16}. Because the helium
core mass was not appreciably affected though, the final core
compactness, nucleosynthesis, and remnant masses are not altered by
this change.

\section{The New Survey}
\lSect{procedure}

Three grids of models were calculated for initial main sequence masses
in the range 12 through 60 \Msun \ using the zoning described in
\Sect{physics} and three choices of mass loss rate. Since all stars
retained at least a small envelope mass as presupernova
stars, the relevant mass loss rate was that of \citet{Nie90}. The
first set of models uses that rate unmodified ($\dot{M}_{\rm N}$). 
This results in so much of the envelope being lost by stars
above about 27 \Msun, that the residual envelope, typically a few
solar masses, becomes unstable and develops density inversions of more
than an order of magnitude. We suspect that massive stars with
  extremely large radii and low mass envelopes become unstable in
nature at this point and lose their remaining hydrogen rapidly
\citep[e.g.,][]{San15,San17}. Mass loss is not the focus here
  though. In order to have a greater range of helium core masses, a
second set of models used one-half of the standard mass loss
($\dot{M}_{\rm N}/2$) appropriate for solar metallicity and
encountered no instability up to 40 \Msun.  Finally, a third set was
calculated with one-tenth the standard mass loss ($\dot{M}_{\rm
  N}/10$) to provide a sparse mass grid between 12 \Msun\ and 60
\Msun. Considering that mass loss prescriptions are probably uncertain
\citep{Ren17,Bea18}, the factors of two and ten could reflect a sensitivity
study, or they might be appropriate for stars with reduced
metallicity.

Altogether 1,499 models were calculated in the mass range 12 to 27
\Msun\ using the standard mass loss rate and 2,799 models from 12 to
40 \Msun\ using half that value. The grid spacing was 0.01 \Msun\ for
both sets. The third set with one-tenth the standard mass loss rate,
consisted of only 49 models between 12 and 60 \Msun\ with a mesh size
of 1 \Msun. Lower metallicity stars were not considered but, except
for zero, metallicity would have affected the answer chiefly by
changing the mass loss and hence hydrogen envelope mass of the
presupernova star.  As we shall see, the helium core mass is not
entirely independent of the remaining envelope mass, but it is
insensitive.

The code was compelled to spend a much greater number of time
steps during hydrogen and helium burning by imposing a limit on the
maximum time step of $10^{10}$ s. A minimum of roughly 10,000 steps
was thus spent burning hydrogen on the main sequence. This helped to
weaken the impact of random semiconvective mixing events that affected
the final envelope mass and, to a smaller extent, the helium core
mass. It also improved the accuracy of the treatment of convection and
burning as ``split'' operations (convection and burning are not
implicitly coupled in the code). Very little difference was noted in a
few test cases when this maximum step was increased to $5 \times
10^{10}$ s.

Minor changes to improve the stability of the code when transitioning
to silicon quasiequilibrium and the convergence criteria were also
incorporated.  These had insignificant effects on the outcome,
  but improved code performance. Models for the main survey were all
calculated on identical processors using the same version of the code
and compiler so as to reduce any possible noise introduced by machine
architecture \citep{osc}.

\begin{figure}
\includegraphics[width=0.48\textwidth]{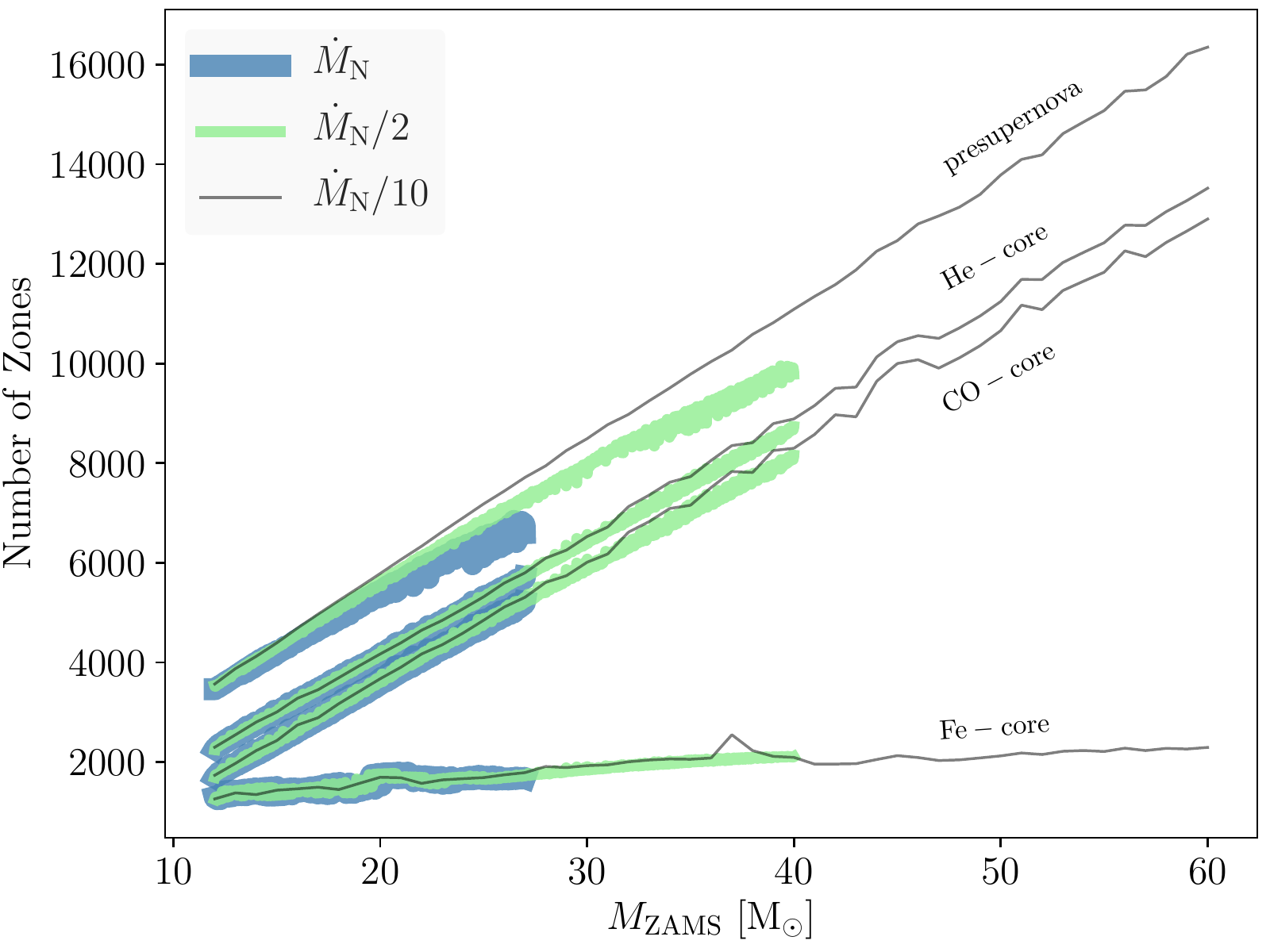}
\caption{Total number of zones in the presupernova star as a function
  of main sequence mass for the model sets with standard mass loss rate 
  $\dot{M}_{\rm N}$ (blue), with reduced rate of $\dot{M}_{\rm N}/2$ 
  (green) and reduced rate of $\dot{M}_{\rm N}/10$ (black). 
  In order to maintain fine resolution in stars of increasing mass, the 
  number of shells becomes greater. The lower curves show the zoning 
  inside the helium core, the carbon-oxygen core and iron-core. These 
  new models retain more zones only in the iron-core alone as the 
  earlier studies did in the whole star. \lFig{zone}}
\end{figure}

One major goal of this work was to explore the effect of resolution on
calculations of presupernova structure.  The number of zones
  used in the main survey was typically $4,000\times(M_{\rm
  ZAMS}/12\,\Msun)$. Stars more massive than 30 \Msun\ thus had over
10,000 initial mass shells. In all cases, the maximum mass zone
anywhere in the star was, at all times, $\lesssim\ 0.01$ \Msun, but
zoning was by no means uniform. The initial zoning (\Fig{zone}), which
continued to characterise most of the star until the presupernova
stage, had much finer zoning in the helium core, down to about 0.001
\Msun \ at the stars center. 

Except as noted above for hydrogen burning, the time step
criteria were not varied from the previous studies.  Nevertheless the
finer resolution of the new runs did result in taking more time steps
as smaller zones experienced mass loss or were cycled through the large
density contrast at the base of the hydrogen convective
shell. Abundances within a given zone convectively coupled to many
other zones also required more time steps to change.  The typical
number of steps for a new calculation is 45,000, about twice the
earlier studies $\sim$25,000.

The presupernova structure and composition tables for all models from
the main sets with $\dot{M}_{\rm N}$, $\dot{M}_{\rm N}/2$ and
$\dot{M}_{\rm N}/10$ are available through the Harvard-Dataverse
archive
\footnote{http://doi.org/10.7910/DVN/VOEXDE}. 

\section{Results}
\lSect{results}

\subsection{Observable Properties}
\lSect{stars}

The observable properties of the new models are summarized in
\Fig{mfin}, \Fig{HR}, and \Tab{bigtable}.  Final (presupernova) masses
are sensitive both to the uncertain mass loss rate prescription
\citep{Nie90}, and to the uncertain history of the stellar radius. The
results vary considerably. For lower mass stars, the spread in final
mass is smaller since the star only loses a small amount of mass. For
bigger stars, though, a large fraction of the hydrogen envelope may be
lost.  Most of the mass loss occurs during helium burning, and most of
that, during the late stages when the central helium mass fraction is
less than 0.5.  

Since the luminosity does not vary greatly during helium burning, the
amount of mass lost, and its uncertainty, depends mostly upon the
radial history of the star - $R^{0.8}$ in the \citet{Nie90} formalism,
i.e., the amount of time the helium burning star spends as a blue
supergiant (BSG) or RSG. It is well known that the semi-convective
mixing and overshoot mixing play a key role in determining the ratio
of time spent as each \citep{Lau71}. In the present calculations,
small changes in individual semiconvective mixing episodes affect the
timing of the development of a deep surface convection zone, and hence
the movement of the star to the red. With more semiconvection, the
star spends a greater fraction of its helium burning lifetime as a BSG
and hence loses less mass, ending its life with a larger hydrogen
envelope still intact. RSGs have the converse behaviour.  Less
semiconvection, e.g., strictly Ledoux convection, favors a longer time
as a RSG and thus reduces the threshold for a massive star to lose its
envelope. Though the spread in \Fig{mfin} looks large, the uncertainty
is usually a small fraction of the total mass lost. For example a 25
\Msun\ star with the standard mass loss rate might end up as a 12 or
15 \Msun\ star.  That is it might lose 10 to 13 \Msun, a range of
about 25 \%. Lower mass models have a smaller variation.

The final masses for stars calculated using $\dot{M}_{\rm N}$/2, of 
course, are larger. Bigger stars on the main sequence then retain more 
envelope by the time they die, and we were thus able to determine the 
core structure of more massive stars. For the most massive models 
considered the final hydrogen envelope mass was roughly 4 \Msun\ for the 
$\dot{M}_{\rm N}$ models and 5 \Msun\ for the $\dot{M}_{\rm N}/2$ models. 
For smaller envelope masses, the radius expanded beyond 
$1.5 \times 10^{14}$ cm and became unstable, in the \KEPLER\ code, due 
to recombination. In these cases, the envelope could only be retained 
on the star by the addition of a large, unrealistic boundary pressure 
that resulted in gross density inversions in the outer envelope. We 
suspect that this is a real instability in nature, that once the 
envelope mass decreases below a critical value in a very massive star 
and the radius extends beyond 10 AU, the remaining envelope is lost 
through non-steady processes \citep{San15,San17}. The luminosity in 
the hydrogen envelope is a substantial fraction of the Eddington value. 
This interesting possibility is deferred for a later study.

Since a larger pressure can give a smaller radius and reduced mass
loss, different choices of boundary pressure can lead to
  significant differences in the final mass.  The boundary pressure
used in \citet{Mue16}, for example, several thousand dyne cm$^{-2}$,
was larger than what we now believe reasonable and much larger than
used in the present study, 50 dyne cm$^{-2}$. This probably accounts,
at least partly, for the larger final masses found in the
\citet{Mue16} study and accounts for their ability to study higher
mass stars using a single mass loss prescription.

This boundary pressure does not appreciably affect the compactness of 
the cores for stars of given main sequence masses, however. This is 
because the large spread in final star masses (\Fig{mfin}) is 
not reflected in the helium core mass (\Sect{coremass}) or the 
luminosity of the star (\Fig{HR}). Since the core structure, which is 
the main focus of this paper, is chiefly sensitive to the helium core 
mass and not the envelope mass, the spread in final masses seen in 
\Fig{mfin} does not affect our major conclusions.

\begin{figure}
\includegraphics[width=0.48\textwidth]{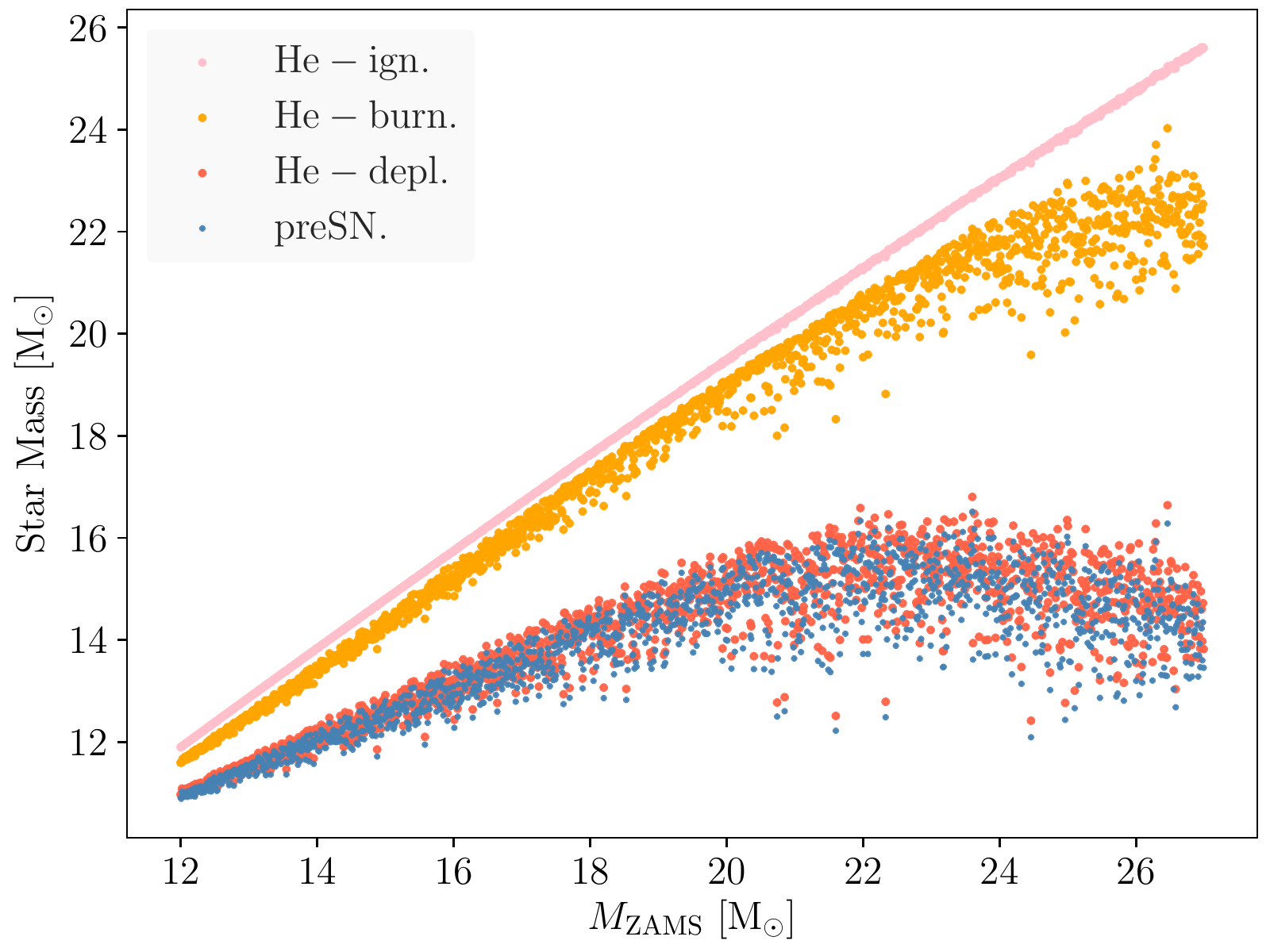}
\includegraphics[width=0.48\textwidth]{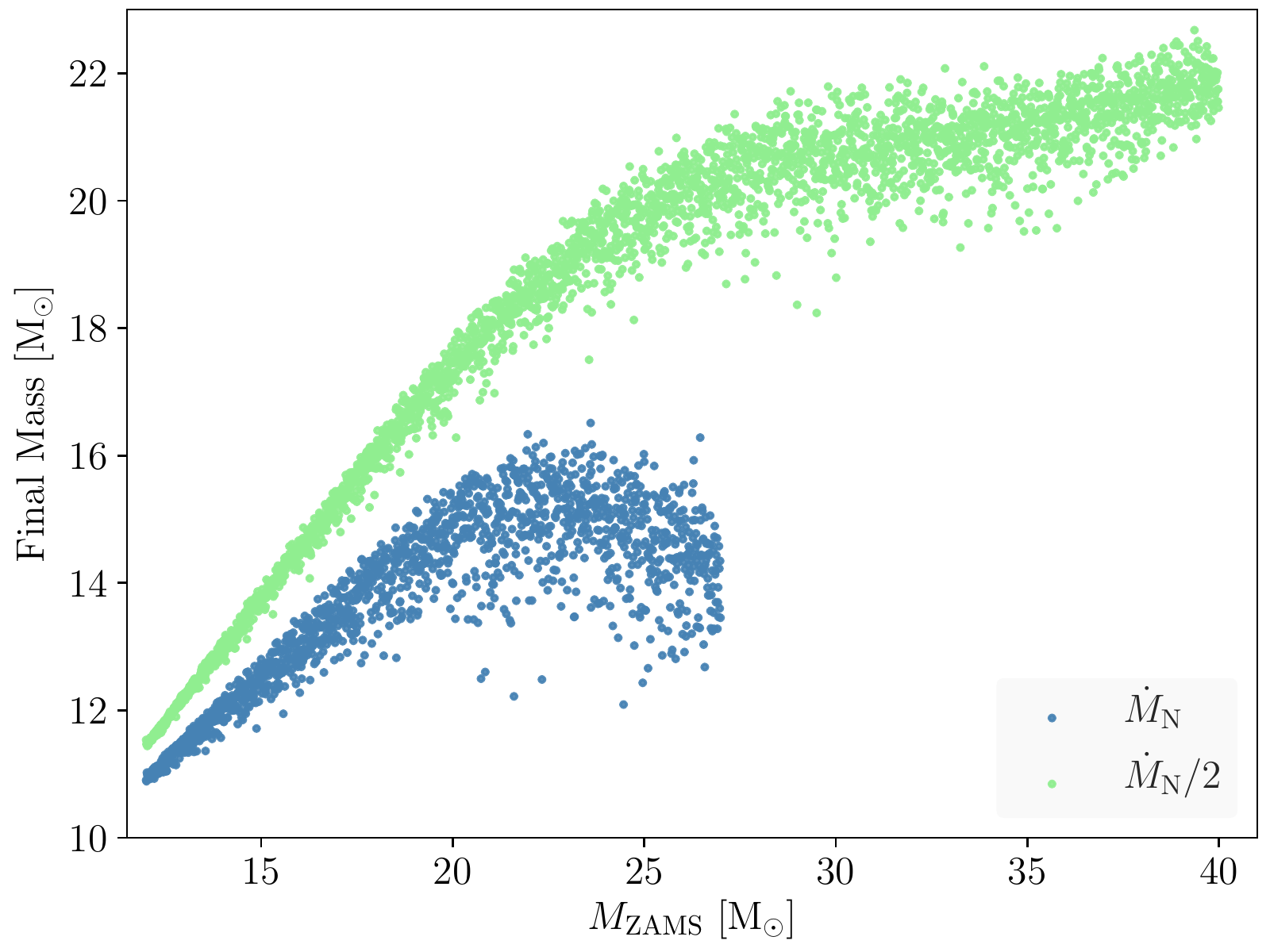}
\caption{Stellar masses during helium burning and for presupernova
  stars. The upper panel shows results for standard mass loss 
  ($\dot{M}_{\rm N}$) at helium ignition (pink) when 1 \% of the helium 
  has burned to carbon; helium burning (orange) when central helium mass 
  fraction is 50 \%; helium depletion (red) when the central helium 
  mass fraction is 1 \%; and the presupernova star (blue). The 
  presupernova masses and helium depletion masses are almost 
  indistinguishable. Most of the mass loss, and most of the dispersion 
  in the presupernova mass, arises due to radius expansion after 
  substantial helium has already burned. The lower panel shows the 
  larger presupernova masses expected when the mass loss rate is 
  reduced by a factor of two. Models with reduced mass loss generally 
  have lower dispersion, but both show the dispersion increase with 
  increasing initial mass (thus increasing amount of mass lost). The 
  presupernova helium core masses show much less variation
 (see \Fig{cores}). \lFig{mfin}}
\end{figure}

\begin{figure}
\begin{center}
\leavevmode
\includegraphics[width=0.48\textwidth]{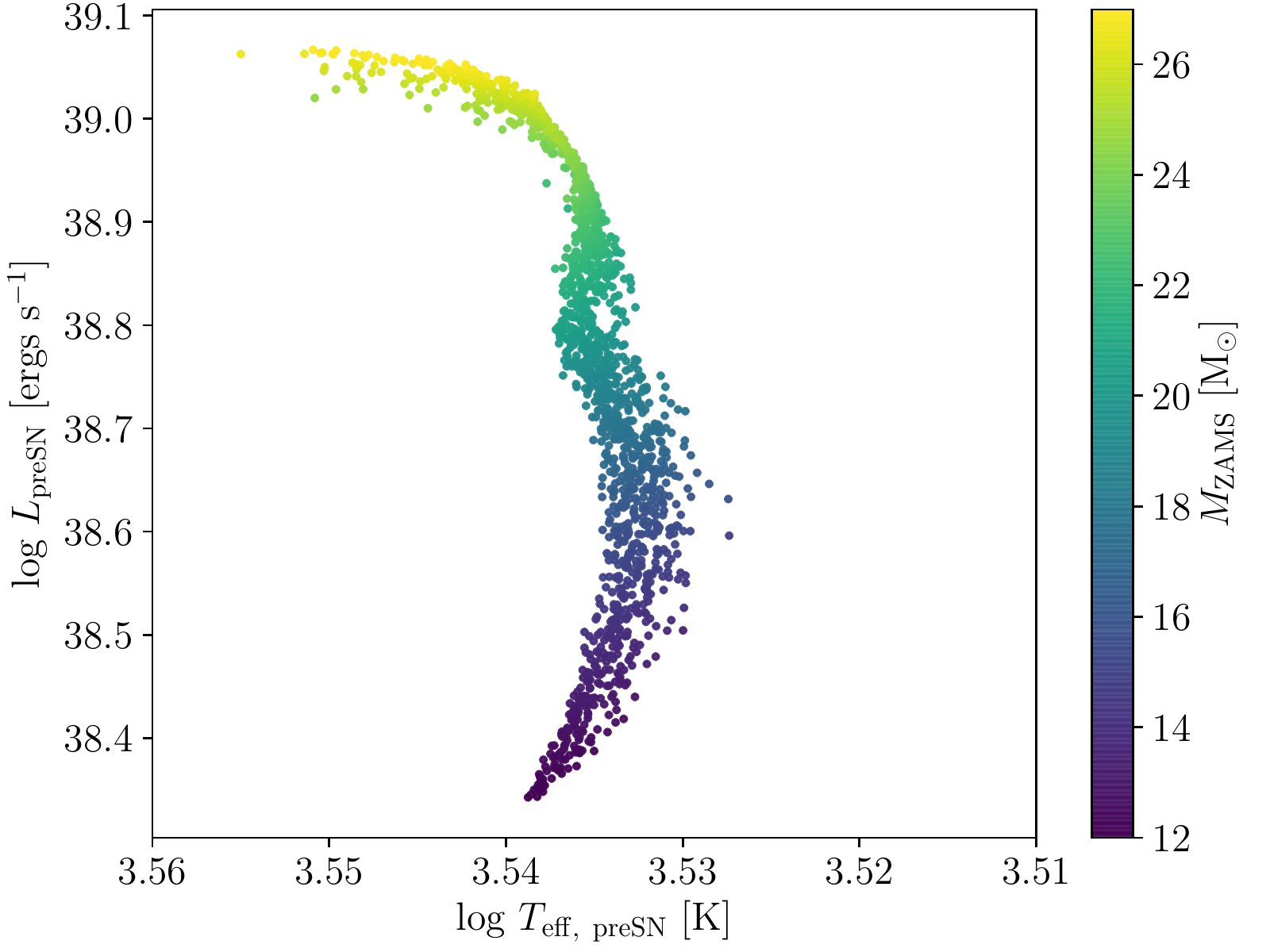}
\includegraphics[width=0.48\textwidth]{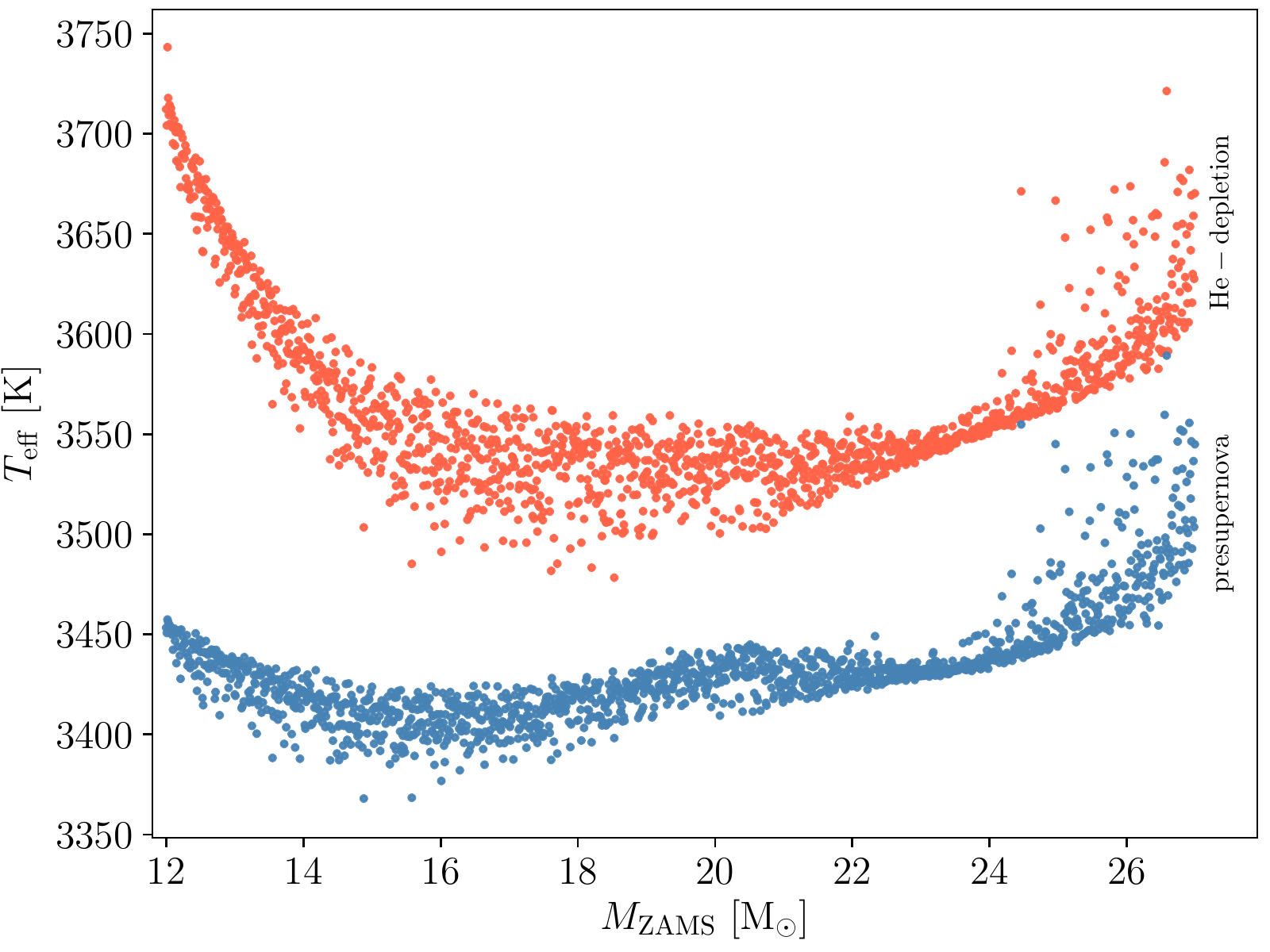}
\caption{(Top panel:) Location in the HR-diagram for the presupernova
  stars that used the standard mass loss rate ($\dot{M}_{\rm N}$).
  (Bottom panel:) $T_{\rm eff}$ at helium depletion (red) and for the
  presupernova stars (blue). The presupernova radii vary almost
  linearly from $5 \times 10^{13}$ cm to 10 $\times 10^{13}$ cm as the
  mass increases from 13 to 26 \Msun. The stars all die with a nearly
  constant effective temperature between 3400 and 3600 K. Observations
  would select RSGs prior to helium depletion, so the
  helium-depletion curve (red) in the lower panel is a lower bound to
  what is likely to be observed for $T_{\rm eff}$. \lFig{HR}}
\end{center}
\end{figure}

\Fig{HR} and \Tab{bigtable} give the final luminosities and effective
temperatures of our presupernova models. The luminosity of the
presupernova star depends chiefly on the helium core mass and is
approximately given by $L_{\rm preSN}\approx5.77\times10^{38}\ (M_{\rm
  He}/6\Msun)^{1.5}\ \rm ergs\ s^{-1}$.  The effective temperature,
essentially bounded by the Hayashi limit remains pegged at close to
$\sim$3500 K for all presupernova stars.  Thus the radius of the star
increases as $M_{\rm He}^{0.75}$. Except for a few presupernova stars,
most RSGs will be observed during their helium burning stage where
their effective temperatures will be hotter.  Our calculations show a
systematic $\sim$200 K offset between the effective temperatures at
helium depletion and presupernova stars.  Earlier in helium burning
the temperature will be even hotter and thus $T_{\rm eff}\approx3,550$
K should be a lower bound to what is seen. Larger values are also
expected for sub-solar metallicity and lower values for presupernova
stars. These numbers are in good accord with measurements of the
hottest RSGs \citep{Dav13,Lev05}.

\begin{deluxetable*}{ccccccccc} 
\tablecaption{Properties of three main model sets} 
\tablehead{ \colhead{${M_{\rm ZAMS}}$}          & 
            \colhead{${M_{\rm preSN}}$}         & 
            \colhead{${M_{\rm He}}$}            &
            \colhead{${M_{\rm CO}}$}            & 
            \colhead{${\log L_{\rm preSN}}$}    & 
            \colhead{${T_{\rm eff,\ preSN}}$}   &
            \colhead{${T_{\rm eff,\ He\ dep}}$} & 
            \colhead{${R_{\rm preSN}}$} 
            \\
            \colhead{[\Msun]}                &  
            \colhead{[\Msun]}                & 
            \colhead{[\Msun]}                &
            \colhead{[\Msun]}                &
            \colhead{[${\rm ergs\ s^{-1}}$]} & 
            \colhead{[K]}                    &
            \colhead{[K]}                    &
            \colhead{[${\rm 10^{13}}$ cm]} 
            }\\
\startdata
\multicolumn{8}{c}{${\dot{M}_{\rm N}}$ (binned by 1 \Msun)}\\
\multicolumn{8}{c}{}\\
12  &  11.19  &   3.35  &   2.20  &  38.39  &  3440  &  3670  &   4.97  \\ 
13  &  11.72  &   3.79  &   2.53  &  38.47  &  3420  &  3610  &   5.49  \\ 
14  &  12.22  &   4.20  &   2.85  &  38.54  &  3410  &  3570  &   5.97  \\ 
15  &  12.73  &   4.61  &   3.19  &  38.60  &  3405  &  3545  &   6.41  \\ 
16  &  13.23  &   5.01  &   3.53  &  38.65  &  3405  &  3535  &   6.80  \\ 
17  &  13.74  &   5.42  &   3.87  &  38.69  &  3410  &  3535  &   7.15  \\ 
18  &  14.12  &   5.84  &   4.24  &  38.73  &  3420  &  3530  &   7.45  \\ 
19  &  14.60  &   6.24  &   4.60  &  38.77  &  3425  &  3530  &   7.72  \\ 
20  &  14.87  &   6.66  &   4.98  &  38.81  &  3430  &  3535  &   8.07  \\ 
21  &  14.97  &   7.09  &   5.35  &  38.85  &  3430  &  3535  &   8.51  \\ 
22  &  15.09  &   7.49  &   5.71  &  38.90  &  3430  &  3540  &   8.95  \\ 
23  &  15.11  &   7.90  &   6.05  &  38.94  &  3435  &  3550  &   9.36  \\ 
24  &  14.72  &   8.34  &   6.44  &  38.98  &  3450  &  3565  &   9.73  \\ 
25  &  14.49  &   8.75  &   6.80  &  39.01  &  3470  &  3590  &  10.01  \\ 
26  &  14.27  &   9.15  &   7.15  &  39.04  &  3500  &  3620  &  10.20  \\
\multicolumn{8}{c}{}\\
\multicolumn{8}{c}{${\dot{M}_{\rm N}/2}$ (binned by 1 \Msun)}\\
\multicolumn{8}{c}{}\\
12  &  11.82  &   3.39  &   2.22  &  38.40  &  3460  &  3685  &   4.95  \\ 
13  &  12.59  &   3.82  &   2.55  &  38.47  &  3450  &  3630  &   5.44  \\ 
14  &  13.35  &   4.24  &   2.88  &  38.54  &  3440  &  3595  &   5.90  \\ 
15  &  14.12  &   4.65  &   3.22  &  38.60  &  3440  &  3580  &   6.31  \\ 
16  &  14.87  &   5.07  &   3.58  &  38.65  &  3440  &  3570  &   6.71  \\ 
17  &  15.62  &   5.49  &   3.93  &  38.70  &  3450  &  3565  &   7.04  \\ 
18  &  16.37  &   5.91  &   4.30  &  38.74  &  3460  &  3565  &   7.32  \\ 
19  &  17.01  &   6.36  &   4.69  &  38.78  &  3465  &  3560  &   7.65  \\ 
20  &  17.66  &   6.79  &   5.08  &  38.82  &  3465  &  3560  &   8.04  \\ 
21  &  18.27  &   7.22  &   5.45  &  38.87  &  3460  &  3560  &   8.49  \\ 
22  &  18.69  &   7.67  &   5.85  &  38.92  &  3455  &  3560  &   9.00  \\ 
23  &  19.25  &   8.07  &   6.20  &  38.96  &  3455  &  3565  &   9.42  \\ 
24  &  19.58  &   8.52  &   6.59  &  39.00  &  3455  &  3565  &   9.86  \\ 
25  &  19.95  &   8.94  &   6.94  &  39.03  &  3455  &  3570  &  10.23  \\ 
26  &  20.14  &   9.38  &   7.31  &  39.06  &  3460  &  3575  &  10.60  \\ 
27  &  20.36  &   9.80  &   7.67  &  39.09  &  3465  &  3585  &  10.92  \\ 
28  &  20.66  &  10.19  &   7.99  &  39.11  &  3470  &  3590  &  11.19  \\ 
29  &  20.61  &  10.65  &   8.38  &  39.14  &  3480  &  3600  &  11.50  \\ 
30  &  20.72  &  11.08  &   8.73  &  39.17  &  3495  &  3610  &  11.75  \\ 
31  &  20.83  &  11.49  &   9.10  &  39.19  &  3510  &  3625  &  11.97  \\ 
32  &  20.85  &  11.95  &   9.48  &  39.21  &  3525  &  3640  &  12.18  \\ 
33  &  20.98  &  12.37  &   9.86  &  39.23  &  3540  &  3655  &  12.37  \\ 
34  &  21.03  &  12.85  &  10.27  &  39.26  &  3560  &  3675  &  12.55  \\
35  &  21.08  &  13.33  &  10.70  &  39.28  &  3580  &  3690  &  12.74  \\
36  &  21.36  &  13.69  &  11.01  &  39.29  &  3590  &  3705  &  12.86  \\ 
37  &  21.40  &  14.18  &  11.46  &  39.31  &  3610  &  3725  &  12.99  \\ 
38  &  21.63  &  14.56  &  11.80  &  39.33  &  3625  &  3735  &  13.15  \\ 
39  &  21.80  &  14.98  &  12.18  &  39.34  &  3635  &  3750  &  13.29  \\
\multicolumn{8}{c}{}\\
\multicolumn{8}{c}{${\dot{M}_{\rm N}/10}$ (not binned)}\\
\multicolumn{8}{c}{}\\
12  &  11.88  &   3.21  &   2.09  &  38.36  &  3480  &  3725  &   4.69  \\ 
15  &  14.76  &   4.46  &   3.07  &  38.58  &  3465  &  3615  &   6.05  \\ 
20  &  19.48  &   6.66  &   4.96  &  38.81  &  3500  &  3590  &   7.76  \\ 
25  &  24.00  &   8.86  &   6.89  &  39.02  &  3495  &  3605  &   9.88  \\ 
30  &  27.75  &  11.39  &   8.95  &  39.17  &  3510  &  3615  &  11.73  \\ 
35  &  31.55  &  13.89  &  11.05  &  39.29  &  3535  &  3625  &  13.17  \\ 
40  &  35.25  &  16.23  &  13.19  &  39.37  &  3575  &  3660  &  14.23  \\ 
45  &  36.65  &  21.41  &  18.22  &  39.53  &  3610  &  3650  &  16.73  \\ 
50  &  41.40  &  21.21  &  17.87  &  39.52  &  3635  &  3710  &  16.28  \\ 
55  &  44.27  &  23.73  &  20.20  &  39.58  &  3655  &  3715  &  17.26  \\ 
60  &  47.31  &  25.64  &  22.00  &  39.62  &  3680  &  3740  &  17.79  \\
\enddata
\tablecomments{All quantities are measured at the presupernova stage, 
except $T_{\rm eff,\ He\ dep}$, which was measured when the helium mass 
fraction in the core was depleted to 1 \%. All $T_{\rm eff}$ values were 
rounded to multiples of 5.}
\lTab{bigtable}
\end{deluxetable*}

\subsection{Core Masses}
\lSect{coremass}

\Fig{cores} shows the helium, carbon-oxygen (CO), and iron-core masses
for all our presupernova models with $\dot{M}_{\rm N}$ and 
$\dot{M}_{\rm N}/2$. The dispersion in helium and CO core masses for a 
given main sequence mass is small, much less than the hydrogen envelope 
masses inferred in \Fig{mfin}. This implies that the helium and CO core 
masses are only slightly affected by the assumed mass loss rate and are 
probably not very sensitive to metallicity. The helium core mass as a function of main 
sequence mass is approximately 
$M_{\rm He}\approx6.46\ \left(M_{\rm ZAMS}/20\right)^{1.27}$ which can 
be combined with the previous relation in \Sect{stars} to give 
$L_{\rm preSN}\approx6.5\times10^{38}\ (M_{\rm ZAMS}/20\ \Msun)^{1.92}\
\rm ergs\ s^{-1}$. Since presupernova core compactness is chiefly a
function of helium core mass \citep{Suk14}, these results suggest a
near universal dependence of presupernova core structure on initial
mass, provided mass loss does not remove the entire hydrogen envelope.

Iron core masses increase, on the average with increasing stellar mass
reaching maximum of about 2.0 \Msun\ for the most massive stars studied
($>$ 40 \Msun). Still larger iron cores, up to about 2.5 \Msun\ 
characterize more massive stars in the pulsational-pair instability range 
\citep[70 - 140 \Msun;][]{Woo17}. For stars below 23 \Msun, the iron 
core mass is markedly multi-valued for stars with nearly the same initial 
mass. This reflects the operation of multiple shells of carbon and 
oxygen burning as will be discussed further in \Sect{interpret}. The 
two major branches of iron core masses below 20 \Msun, which are most of 
the stars that leave neutron star remnants might result in bimodality in 
the neutron star mass function.

\begin{figure}
\begin{center}
\leavevmode
\includegraphics[width=0.48\textwidth]{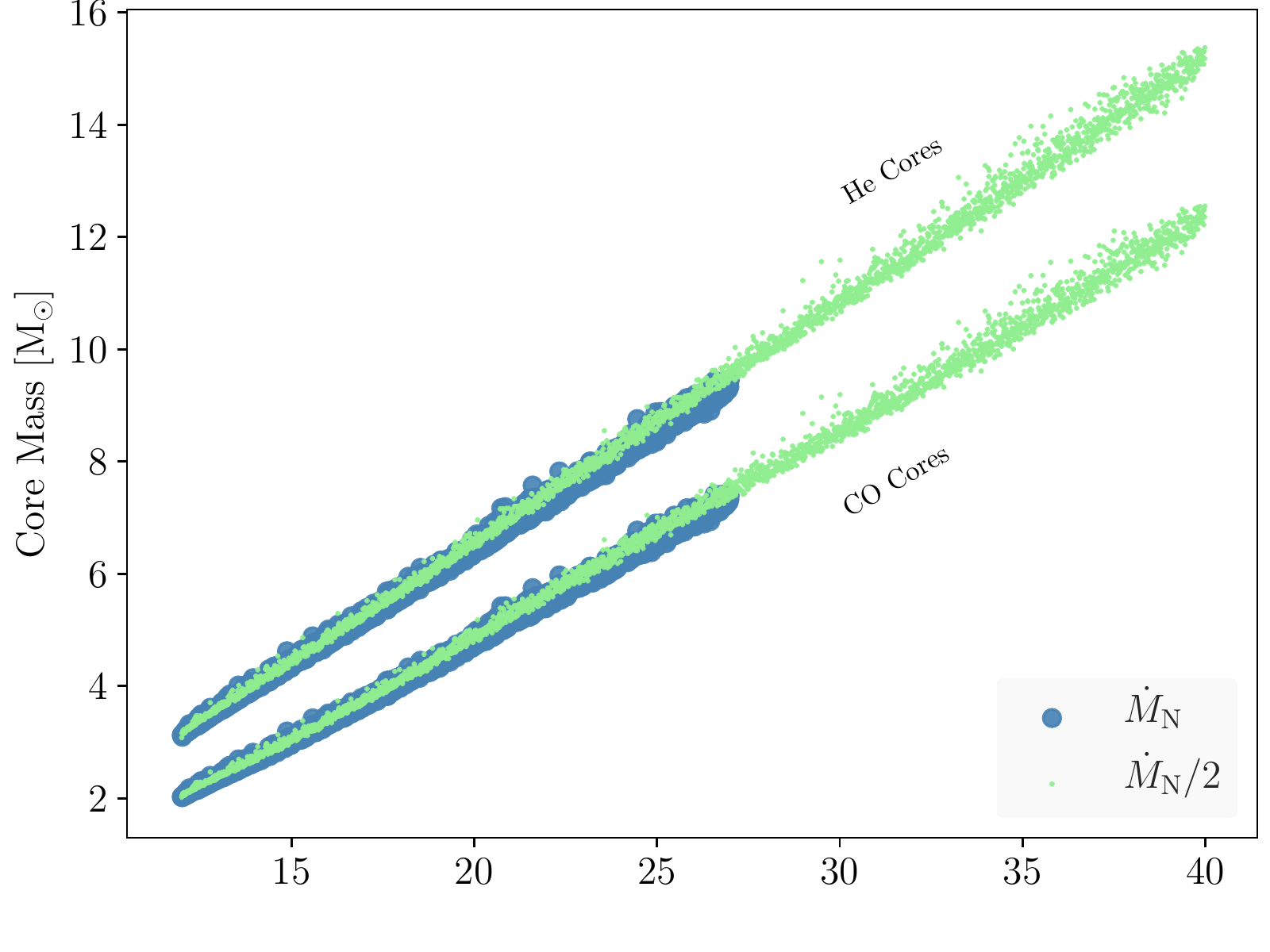}
\includegraphics[width=0.48\textwidth]{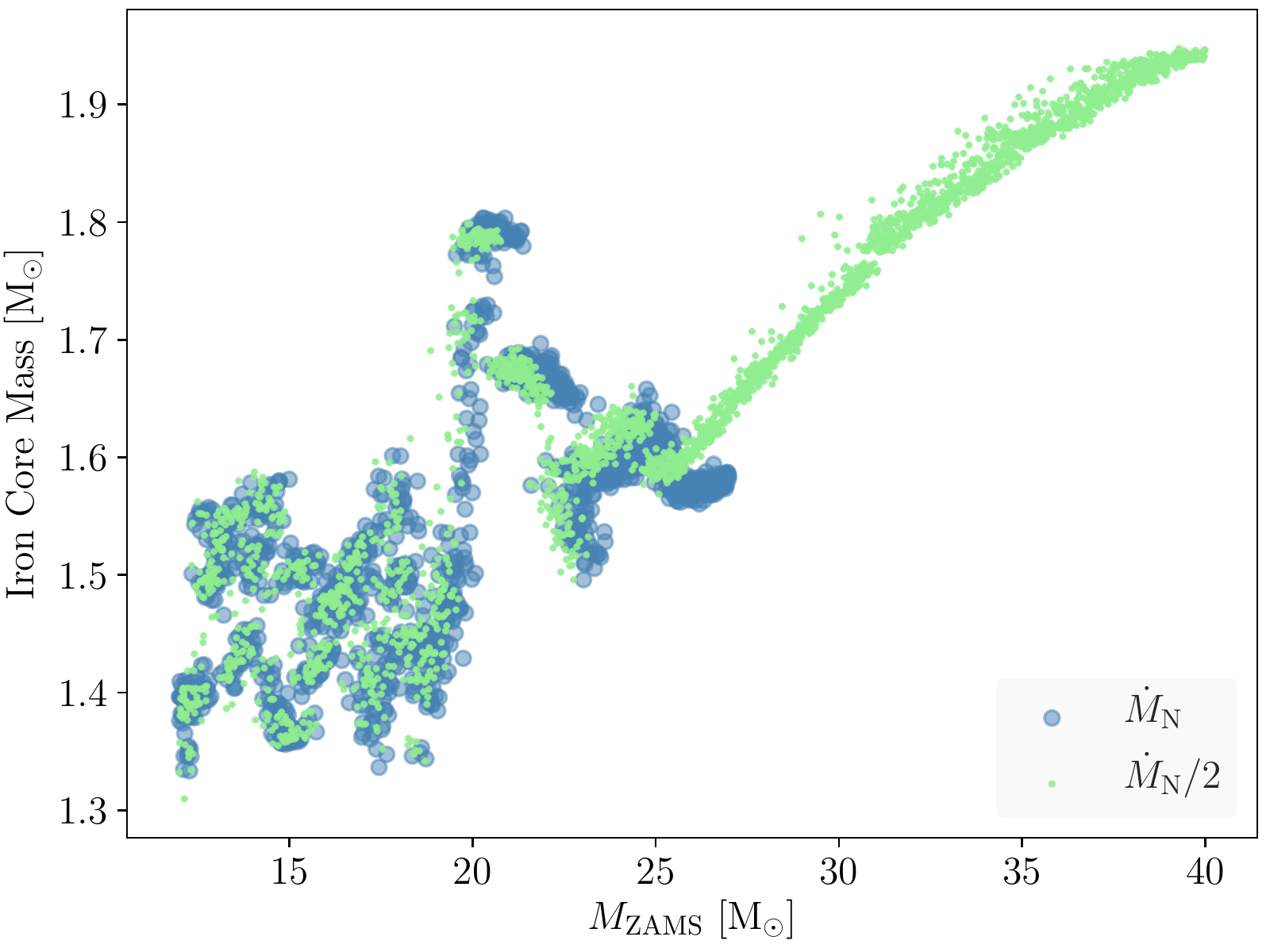}
\caption{Helium core, CO core, and Fe core masses for all presupernova
  stars from sets with $\dot{M}_{\rm N}$ and $\dot{M}_{\rm N}/2$. Despite 
  significant variations in mass loss (\Fig{mfin}), the final helium 
  and carbon-oxygen cores are well determined by the star's initial 
  mass and a standard choice of stellar physics. Note multiple branches 
  for the iron core mass below 19 \Msun. \lFig{cores}}
\end{center}
\end{figure}

\subsection{Core Structure}
\lSect{compact}

\subsubsection {Measures of ``Explodability''}

Early theoretical studies of supernovae noted a strong correlation of
a rapidly declining density outside the iron core with the degree of
difficulty encountered in trying to explode the star using the neutrino
energy transport \citep[e.g.,][]{Bur87,Fry99}. \citet{Oco11} introduced
a simple, single parameter measure of this density decline called the
``compactness parameter'':
\begin{equation}
\xi_{\rm M}=\frac{M/\Msun}{R(M_{\rm bary}=M)/1000\, {\rm km}}\Big|_{t_{\rm bounce}},
\lEq{compact}
\end{equation}
where $R(M_{\rm bary} = M)$ is the radius of the Lagrangian mass shell
enclosing mass $M$ in the presupernova star. The fiducial mass is
often chosen as the innermost 2.5 \Msun\ so that for a wide range of
initial masses it not only encloses the iron-core but samples enough
of the overlying shell material around it. Though the parameter is
defined to be evaluated at the time of bounce, it is more often
measured at the time of presupernova (when the collapse begins), since
the systematics are insensitive to slightly different fiducial choices
\citep{Suk14}. Subsequent studies by \citet{Ugl12}, \citet{Oco13}, and
\citet{Suk16} showed strong correlation between the ease with which
model stars exploded and the $\xi$ parameter, in the sense that stars
with small $\xi$, i.e., steep density gradients outside the iron core
were easier to explode using a standard, albeit approximate, set of
supernova physics. Although this is a useful starting point, a single
parameter conveys limited information about the structure of the core
and more physics-based representations followed.

\begin{figure*}
\includegraphics[width=0.5\textwidth]{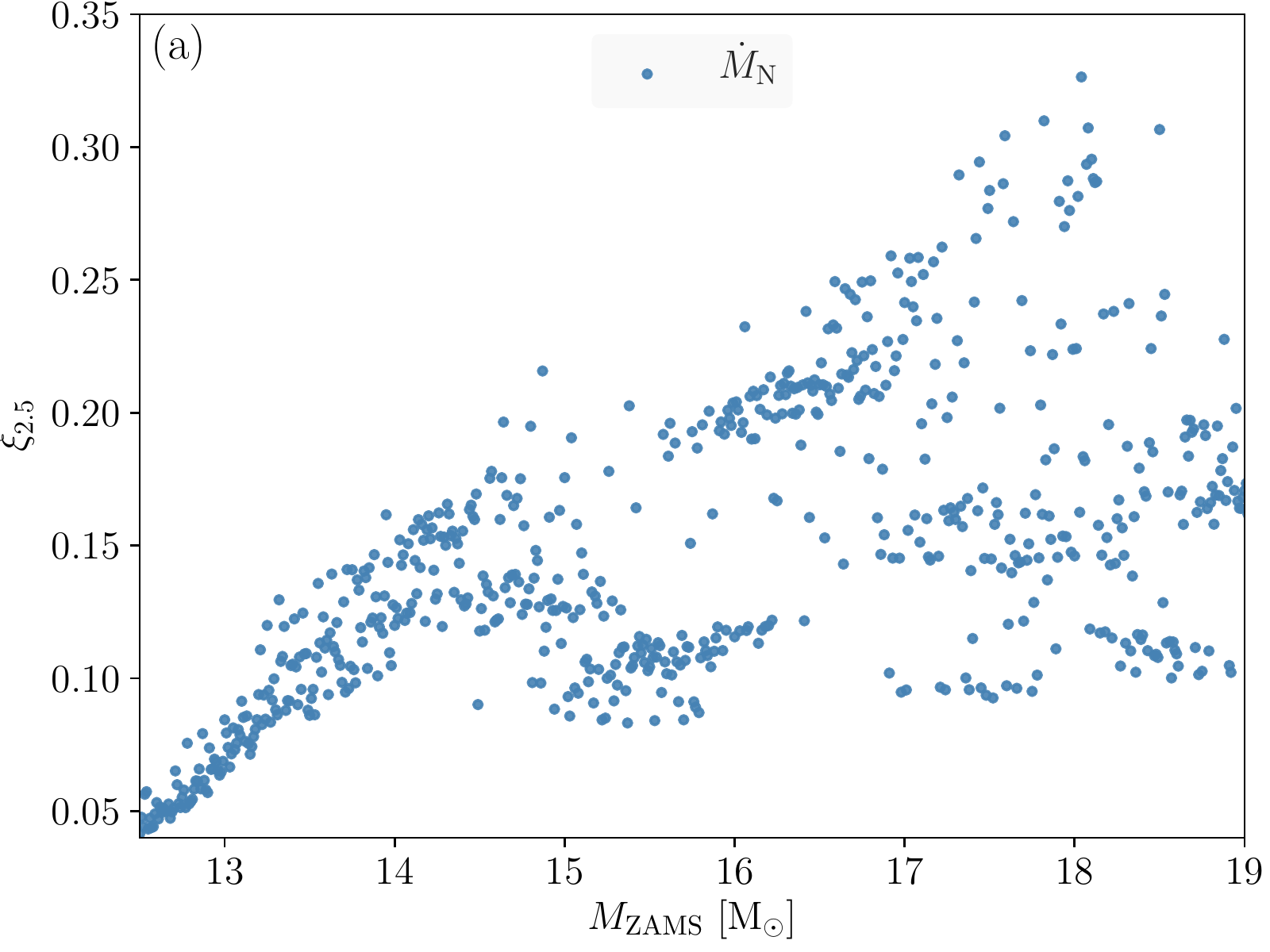}
\includegraphics[width=0.5\textwidth]{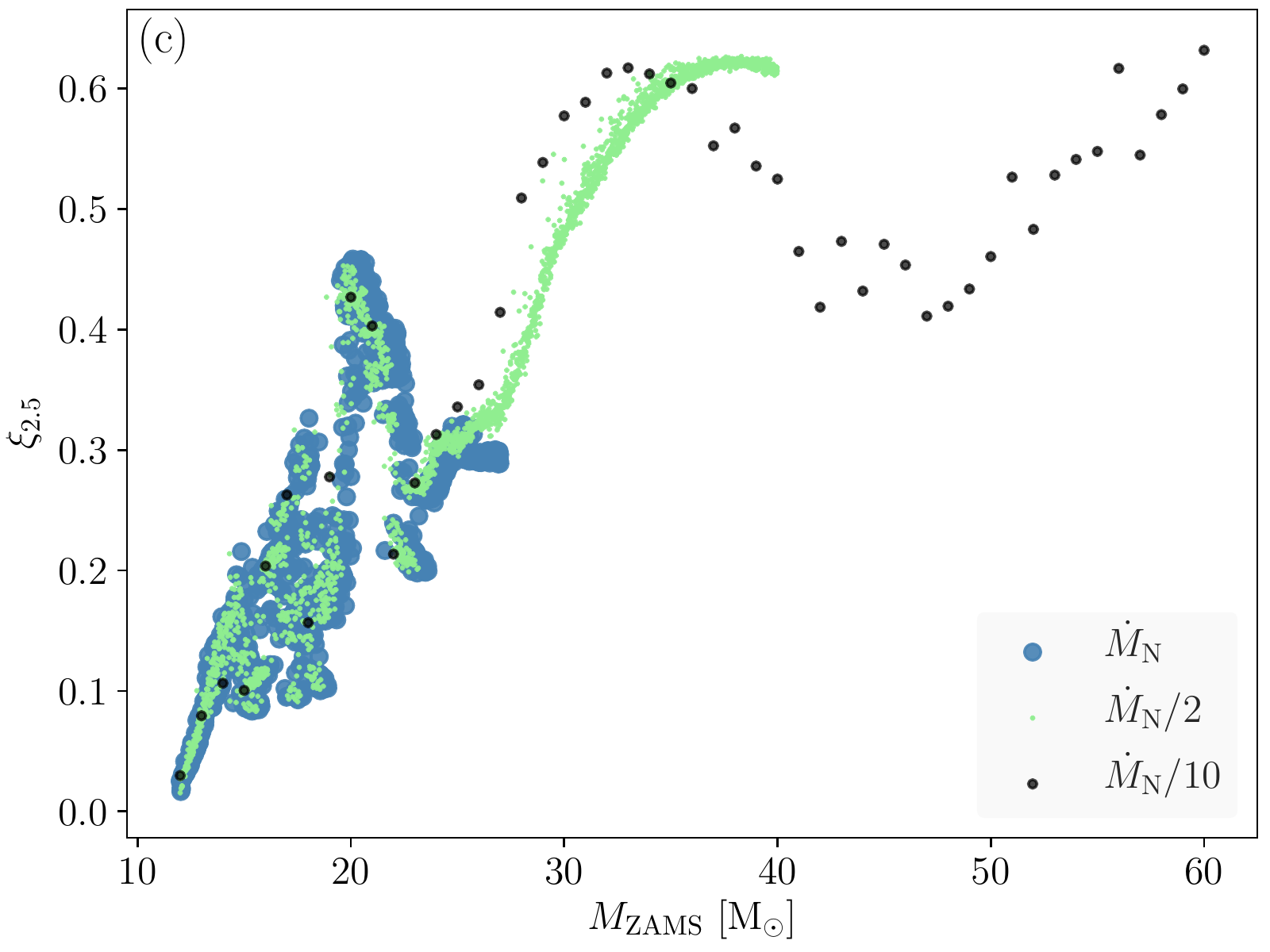}
\includegraphics[width=0.5\textwidth]{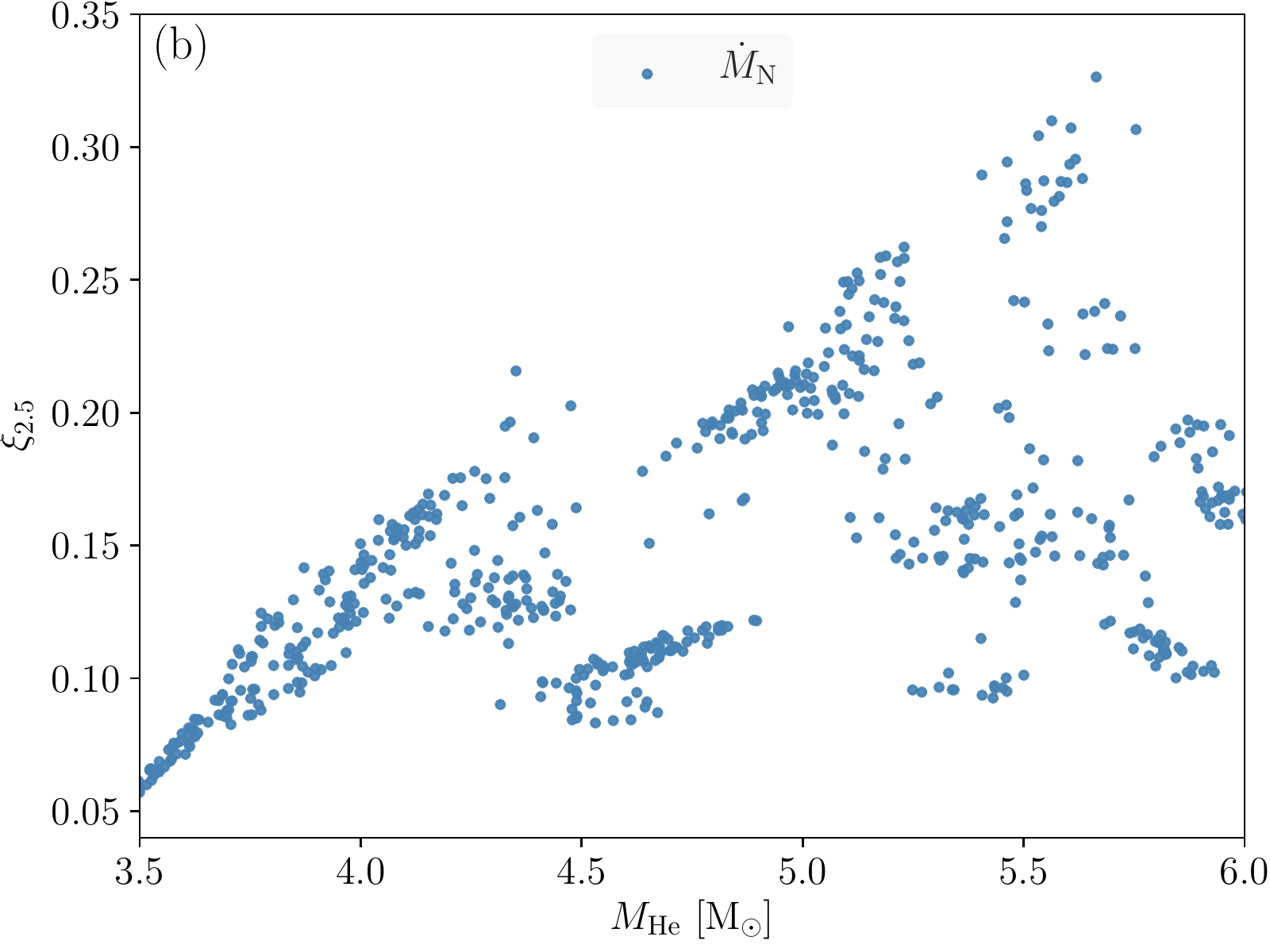}
\includegraphics[width=0.5\textwidth]{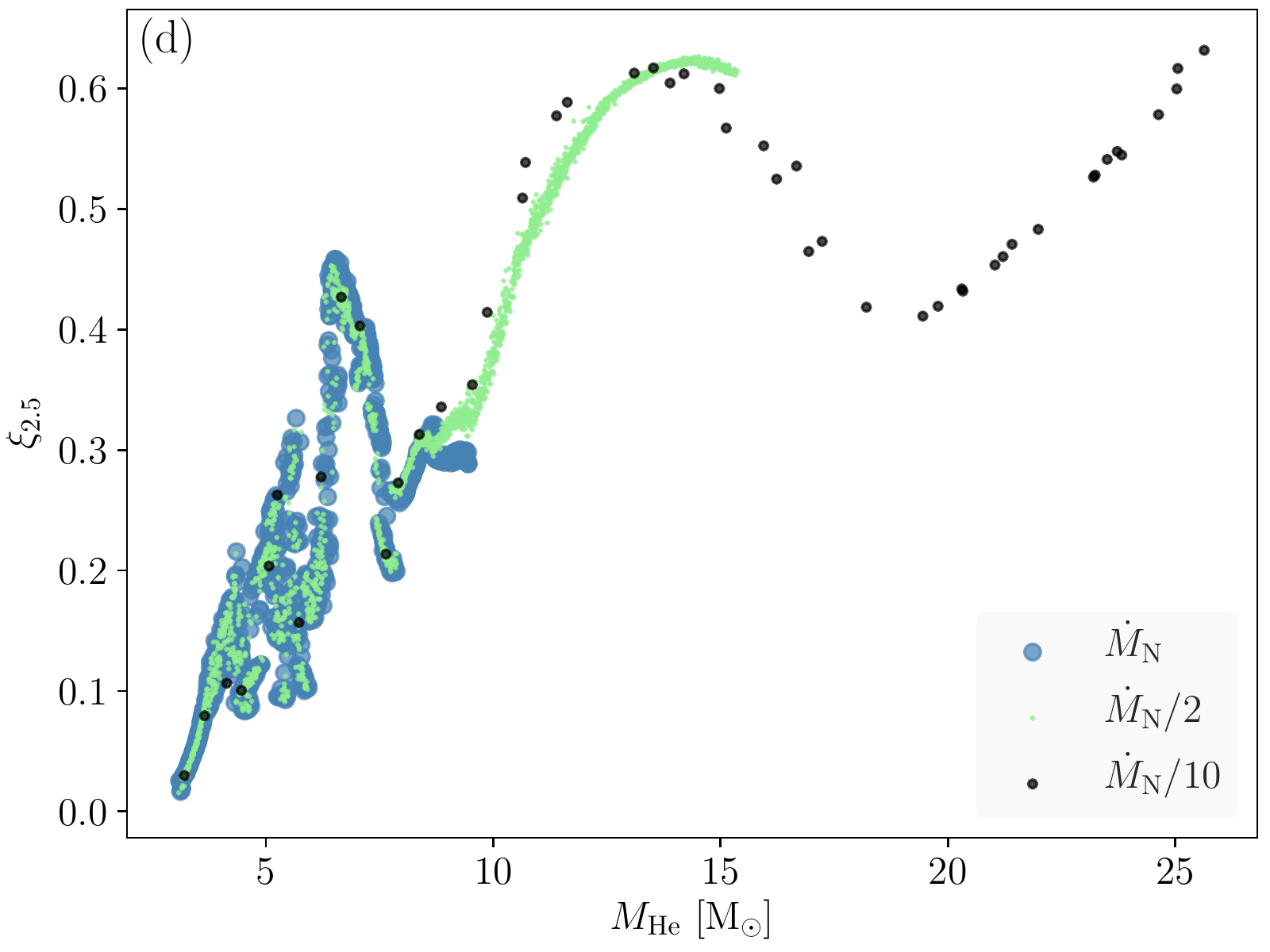}
\caption{(a): The compactness parameter for a fiducial mass of 2.5
  \Msun \ evaluated at the presupernova stage shown for models below
  $<19$\Msun\ as a function of initial mass for the $\dot{M}_{\rm N}$
  set. Even where rapid variations occur, the scatter is not wholly
  random. Points tend to concentrate in one of several solutions. (b):
  Same points as in (a) now shown as a function of helium core
  mass. The branching solutions are more clearly seen. The range of
  $\xi_{2.5}$ seen in this mass range includes stars that might
  explode or implode, and thus both outcomes are possible for stars
  with nearly identical initial mass.  (c): Compactness shown for all
  three values of mass loss. Slight shifts between the sets are due to
  the growth of the helium core by hydrogen shell burning, resulting
  in a slightly larger compactness for the models with reduced mass
  loss. (d): same points as in (c) are now shown as a function of
  helium core mass. Much of the shift seen in (c) is gone since the
  helium core mass is a chief determinant of the core structure.
  \lFig{cp}}
\end{figure*}

In particular, \citet{Ert16} suggested an alternative two-parameter 
characterisation based upon $M_4$, the mass coordinate, in solar 
masses, where the entropy per baryon
reaches $4.0 \, k_B$, and the radial gradient, $\mu_4$, of the density
at that point. In practice, $\mu_4$, was obtained by evaluating the
change in mass over the change in radius between several mass shells
separated by 0.3 \Msun\ in the vicinity of $M_4$, i.e.,
\begin{equation}
\mu_4 \ = \ \frac{dm/\Msun}{dr/1000 \, {\rm km}}
\end{equation}
where $dm=0.3$ \Msun. Smaller values of $\mu_4$ thus imply steeper
density gradients (less change in enclosed mass when the radius
changes). The quantity $M_4$ has long been used to locate the the
steep density decline often associated with a strong oxygen-burning
shell in the presupernova star \citep[e.g.,][]{Woo07}. Where the
entropy per baryon abruptly rises at nearly constant temperature,
the density declines.  When the core collapses, the arrival at the
neutrinosphere of lower density matter reduces the ``ram pressure''
and facilitates the launch of an outgoing shock. Besides its location,
it is helpful to characterise the rate of the density change, which is
the role played by $\mu_4$.

It is reasonable that higher neutrino luminosities and smaller
accretion rates on the proto-neutron star should favor explosion.
\citet{Ert16} made the case that $\mu_4$, multiplied by its radius
(i.e., the radius at which $M_4$ is found) and divided by a collapse 
time scale, is a surrogate for the accretion rate at the time of 
explosion. The product $\mu_4 M_4$ is similarly a surrogate for the 
accretion luminosity. This assumes, as the models suggest, that the 
radius where the neutrino luminosity is generated does not vary 
greatly with mass.  Explosion is thus favored by small $\mu_4$ 
(accretion rate) and large $\mu_4 M_4$ (luminosity). \citet{Ert16} 
determined a simple condition expressed as a straight line
\begin{equation}
\mu_4 \ = \ k_1 (\mu_4 M_4) \ + k_2
\end{equation}
such that models that lay below the line (lower $\mu_4$) were more
likely to explode than those above. The values $k_1$ and $k_2$ were
determined from several large scale simulations and leading models for
SN 1987A. For one representative set, $k_1$ = 0.194 (\Msun)$^{-1}$ and
$k_2$ = 0.058 (``N20'' model in their Table 2).

More recently, \citet{Mue16} adopted a different approach using a
semi-analytic model for the explosion based upon presupernova
properties. By using a fuller desription of the presupernova star than
afforded by just one or two parameters, they were able to
approximately estimate not only the success or failure of the
explosion, but also the explosion energy, ejected nickel mass, and
compact remnant mass. This approach does not include, in its present
form, several important pieces of explosions physics, such as
proto-neutron star cooling, self-consistent fallback and explosive
nuclear burning, and thus it is not as powerful as the parameterised
simulations such as \citet{Suk16}. Its main power lies is its ability
to rapidly calculate general trends in explosion properties that
depend on presupernova structure. Their approach uses five physically
motivated parameters to determine the outcome. Here we analyze our new
presupernova models using their standard values - $\rm
\beta_{expl}=4,\ \zeta=0.7,\ \alpha_{out} =
0.5,\ \alpha_{turb}=1.18,\ and\ \tau_{1.5}=1.2$ (see their Table
1). For the definitions of these parameters and further technical
details see \citet{Mue16}.

\subsubsection{Results for the Compactness Parameter}

\Fig{cp} shows the compactness parameter, $\xi_{2.5}$, i.e., $\xi$ for
our three model sets measured at a fiducial mass of 2.5 \Msun \ at the
time the presupernova collapse speed reaches 900 km s$^{-1}$. One 
notable feature seen in \citet{Suk14} is the relatively large
``scatter'' of compactness parameter between 18 and 22 \Msun, which 
are now shifted to roughly 14 and 19 \Msun\ (Panel a) with updated 
neutrino-losses (\Sect{physics}). As has long been noted \citep[e.g.][]{Tim96}, 
$\sim$20 \Msun\ marks the transition region between exoergic convective 
central carbon burning, at lower masses, to endoergic radiative burning 
at higher ones. That is, at high mass the energy produced by carbon 
fusion does not exceed neutrino losses in the star's center and it 
proceeds from helium depletion to oxygen ignition on a short 
Kelvin-Helmholtz time scale. As a result subsequent burning especially
in shells, where the net energy generation {\sl is} positive, proceeds
differently. The carbon convective shells move inwards (with
increasing initial mass) up until this transition point and are weaker
and more in number \citep{Bar94}.

Quite noticeable in the new results and also in the prior work of
\citet{Mue16} are concentrations of points in the compactness
parameter plot, which were not clearly seen in our earlier studies due
to much coarser increments in mass-space. Points are not randomly
scattered between some local maximum and minimum value. Note for
example, the existence of two, and possibly three discrete solutions
for helium core masses near 4.8\Msun \ (Panel b). Multiple branches
also exist at other masses but with less clarity. For a helium core
mass of 5.6 \Msun, the compactness parameter for stars with nearly
identical masses varies by a factor of three, from $\xi_{2.5} = 0.1$
to 0.3. Such a large variation spans the range of stars that might
explode or collapse to black holes, and suggests that both outcomes
are possible for relatively low mass stars of almost the same initial mass.

When all three new sets are plotted as a function of initial mass
(Panel c), the models with lower mass loss rate are slightly shifted
with respect to models with higher mass loss rate. The shifts are
hardly noticeable at lower initial mass ($<20$ \Msun), but they grow
with increasing mass and for our most massive models the $\dot{M}_{\rm
  N}/10$ set is shifted by a few \Msun\ with respect to $\dot{M}_{\rm
  N}/2$ set. The primary reason for this shift is that for a given
initial mass, models with lower mass loss rate attain slightly larger
He-cores and slightly different compositions. There is also a general
underlying scatter, similar to the one of total star mass shown in
\Fig{mfin}, some of which result from the slight change of He core
mass due to semiconvective effects discussed in \Sect{stars}.

Panel (d) of \Fig{cp} shows the compactness as a function of
He-core mass for all of our models. Since the initial mass scales
cleanly with He-core mass (\Fig{cores}), the structure of the curve
stays nearly identical, however, the above mentioned shifts and
scatters are mostly gone. The He-core mass and within it, the CO-core
mass, are the chief determinants of core structure, not the final
supernova mass which includes the hydrogen envelope. Uncertainties in
mass loss rates are thus not particularly relevant to the compactness,
except for determining the main sequence mass above which the entire
envelope is lost.

As was noted by \citet{Oco11} and studied many times since, the
compactness plotted against mass is non-monotonic and highly
structured with two distinct peaks in the vicinity of 21 and 35
\Msun\ \citep{Suk14}. A shallow dip, occurs near 50 \Msun\ followed by
a slow rise at still higher masses \citep{Suk14} until finally the
pulsational-pair domain is reached near 70 \Msun. As discussed in
\citet{Suk14} and \citet{SukT}, this structure is crafted out of an
overall gradually rising curve by the operation of an advanced stage
shell burning episode of one fuel modulating the core burning episode
of the next fuel. In particular, the structure near the first peak is
primarily driven by the effect of shell carbon burning on core
oxygen burning, while the structure near the second peak is driven by
the effect of shell oxygen burning on core silicon burning. At
very high mass, oxygen burning ultimately ignites a strong shell
outside of 2.5 \Msun, and therefore the compactness stays high.

\begin{figure}
\begin{center}
\leavevmode
\includegraphics[width=0.48\textwidth]{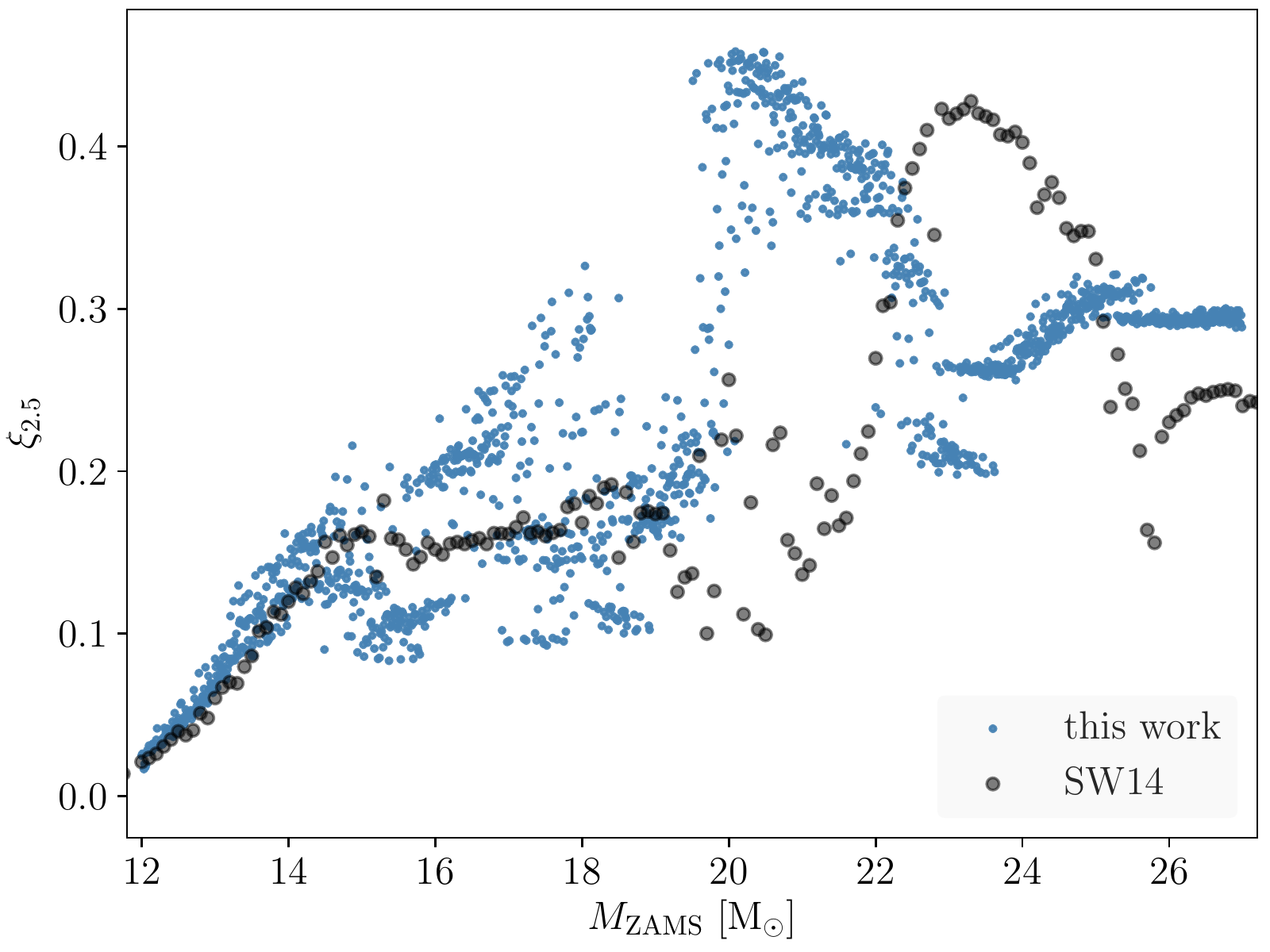}
\includegraphics[width=0.48\textwidth]{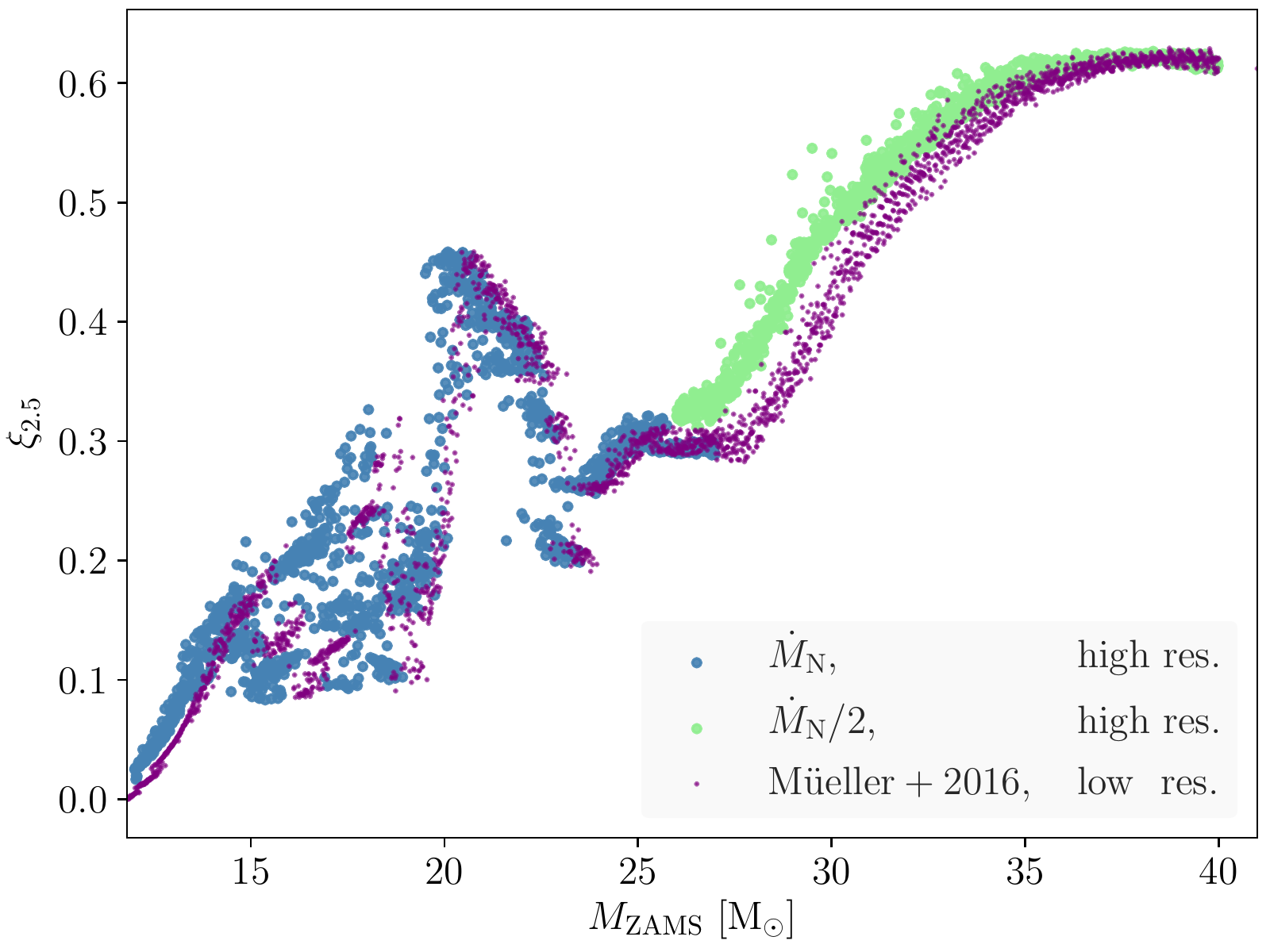}
\caption{Presupernova compactness compared with previous results for
  solar metallicity stars obtained by \citet{Suk14} (top frame) and
  \citet{Mue16} (bottom frame).  The new models differ little from
  \citet{Suk14} below 14 \Msun, but the peak in compactness formerly
  at 23 \Msun\ is shifted downwards to about 20.5 \Msun. The
  ``noise'' noted by \citet{Suk14} from 17 to 22 \Msun \ is also
  shifted downwards to 14 to 19 \Msun\ and now with finely
  incremented models it shows several concentrations of points
  suggestive of a multi-valued solution.  The lower panel shows
  comparison with the work of \citet{Mue16}. Since the $\dot{M}_{\rm N}$
  and $\dot{M}_{\rm N}/2$ sets are nearly identical at lower mass
  (\Fig{cp}), for clarity, we have plotted $\dot{M}_{\rm N}$ set for
  lower mass and $\dot{M}_{\rm N}/2$ set for higher mass. The agreement
  between these models with those from \citet{Mue16} is substantial,
  even though the current survey uses more models, carried more zones and
  timesteps per model. Reducing the mass loss (see text) shifts the
  results towards slightly lower main sequence mass.
\lFig{cpold}}
\end{center}
\end{figure}

\Fig{cpold} compares the new results with those of \citet{Suk14} and
\citet{Mue16}, and illustrates these points. Most notable is a shift
of the new models above $\sim14$ \Msun \ downwards in mass by about 2
to 3 \Msun \ compared with \citet{Suk14}. This is a consequence of
fixing the bug in the neutrino losses as discussed in \Sect{physics},
and not due to differing resolution. Models below 14 \Msun \ are
affected very little, but the more massive stars change
appreciably. The region of variability which was between 18 and 22
\Msun \ in \citet{Suk14} is now shifted down to 14 to 19 \Msun. If
anything that variability, is now greater with the finer resolution,
perhaps due to the larger number of models surveyed.

Indeed, the comparison with \citet{Mue16}, who used the corrected
neutrino loss rate and sampled many more models is excellent, both in
the location of structures and the range of $\xi_{2.5}$ within those
structures. This agreement persists despite the use of 4 to 10 times
more mass shells in each of the new models and a radical decrease in
the surface boundary pressure. The latter affected the mass lost by
the star, but not appreciably the helium core masses. There is no
reason to believe that still finer zoning, smaller time steps, or a
different reaction network will greatly alter these results, unless
the code physics itself (reaction rates, semiconvection, rotation,
etc.) is changed. The studies of \citet{Mue16} and the present work
are mutually confirming.

\subsubsection{New Results and the Ertl Parametrization}

An important subsidiary question is whether $\xi_{2.5}$ is really
the best measure for presupernova core structure. Might some of the
variability seen in \Fig{cp} be simply because of the choice of a
single arbitrary point in the star to sample its structure? Perhaps
other parametrizations might give less variability and more reliable
predictions? In particular, $\rm\xi_{2.5}$ is sensitive to recent shell
activity in the vicinity of 2.5 \Msun\ that might not always describe
well what went on deeper inside.

The Ertl parametrization of our results is given in \Fig{ertl}.  As
previously discussed (\Sect{compact}), points beneath the dashed line 
represent models that are more likely to explode in a simple neutrino 
transport scheme. Multi-valued solutions are clearly seen, especially 
for stars with $\mu_4 M_4$ less than 0.25. In some ranges, the 
difference between solutions is enough to significantly affect the 
probable outcome of the explosion. Stars with $\mu_4 M_4$ less than 
0.25 typically have masses less than 20 \Msun. For more massive stars, 
and, in particular, for $\mu_4 M_4$ greater than 0.25, explosion
seems quite unlikely. Reasons for the multi-valued solution are
discussed in \Sect{interpret}.

Also striking is the much tighter clustering of points in the Ertl
representation of our new models (\Fig{ertl}) in contrast with an
equivalent plot of ``compactness'' (\Fig{cp}). The results are less
noisy and also show less overall variability. This is due to 
the obvious correlation between two Ertl parameters on one hand, 
but it also presumably reflects
the better representation of core structure by a two-parameter,
physics-based representation than a single parameter anchored to a
single point in the core. It also suggests that much of the ``noise''
in \Fig{cp}, especially for initial masses 14 -- 19 \Msun\ may not be 
so much a consequence of ``non-convergence'' of the models, but as a poor
representation of the results.

\begin{figure}
\includegraphics[width=0.48\textwidth]{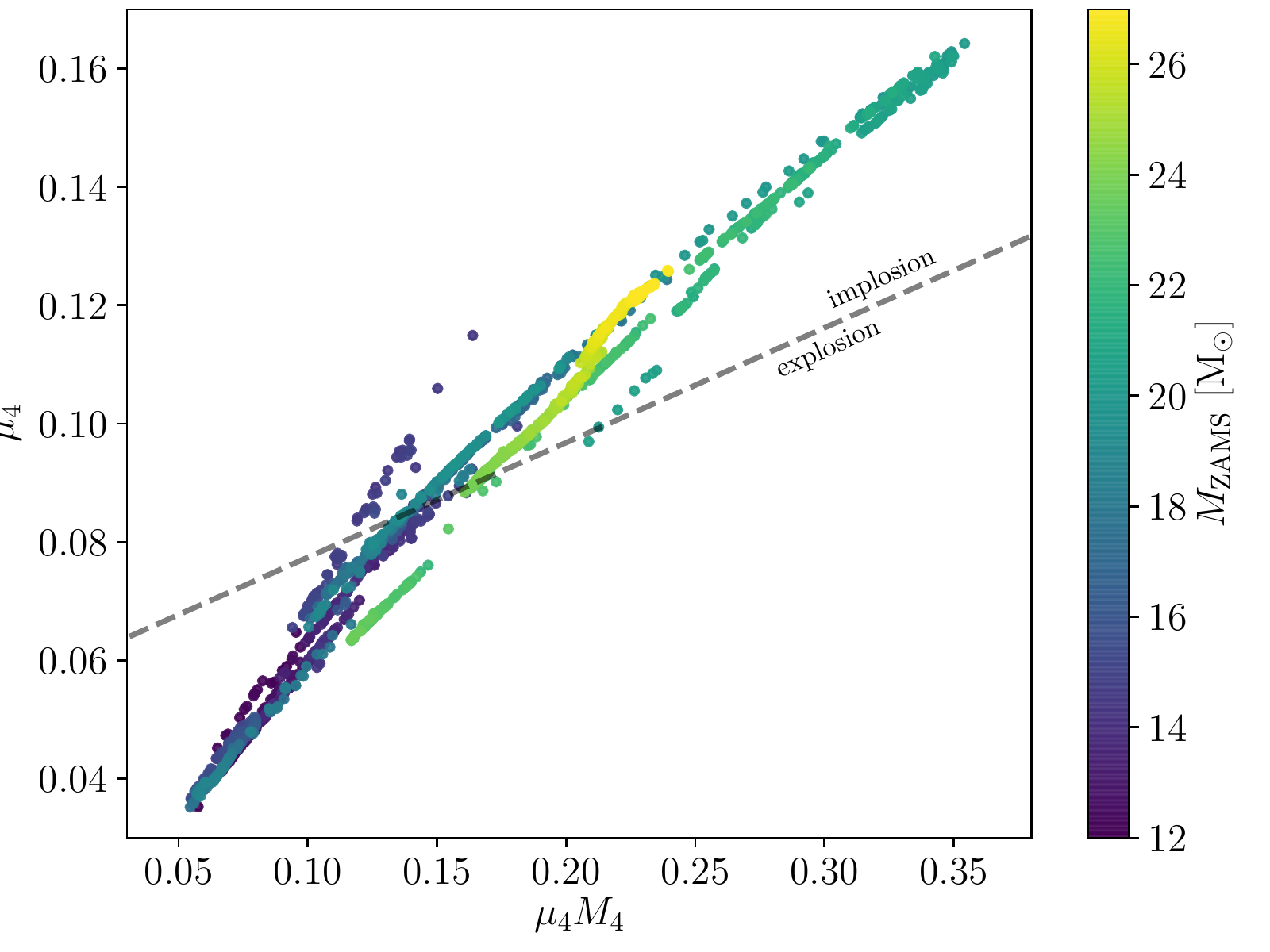}
\caption{``Explodability'' according to sample ``engine'' model from
  \citet{Ert16}.  The Ertl parameters, $\mu_4$ and $\mu_4M_4$, are 
  shown for the new models with standard mass loss rates. See text 
  for definitions. Points above the dashed line are more likely to 
  explode than those below. Multiple solutions are clearly visible 
  for $\mu_4M_4$ below 0.25 (\Msun)$^{-1}$
The scatter in this figure is a lot less than \Fig{cpold}.\lFig{ertl}}
\end{figure}

\section{Interpretation}
\lSect{interpret}

Why is the core structure of presupernova stars non-monotonic and
multi-valued for some ranges of mass? The short answer is that the
advanced stage evolution of massive stars involves two to three carbon
burning shells (plus central core burning for models below $\sim$19
\Msun) and one or two oxygen shells (plus core burning).  Combinations
of these shells lead to variable outcomes, but not a continuum of all
possibilities, because shells have finite sizes and their number is an
integer. The transitions are abrupt and small changes upstream in the
strength or extent of one or more shells can send a star down one path
or another. This is especially true for stars below 19 \Msun\ where
there are more carbon burning shells.

Globally, $\xi_{2.5}$, $\mu_4$, and $M_4$ all increase gradually with
mass.  Higher mass stars have greater entropy in their middles and are
less degenerate in their final stages. Greater degeneracy leads to an
increased central concentration of the mass and ``core convergence.''
For lighter stars, the presupernova structure resembles more a white
dwarf embedded in a low density envelope, where the density declines
rapidly at the edge of the central ``white dwarf'. This effect is
clearly at work in the lightest stars surveyed. Below 13 \Msun,
compactness defined at a mass of 2.5 \Msun \ has little meaning as a
measure of explodability, since the fiducial point is not even inside
of the helium burning shell. For stars in this mass range, the helium
burning shell always lies outside $2 \times 10^9$ cm, thus
guaranteeing a small compactness parameter.  For a 10 \Msun
\ presupernova star, 2.5 \Msun \ is even outside the helium core and
in the hydrogen envelope. For all parametrizations of core structure
considered in this paper, these light stars will always explode and
need no further discussion here.

\begin{figure}
\includegraphics[width=0.48\textwidth]{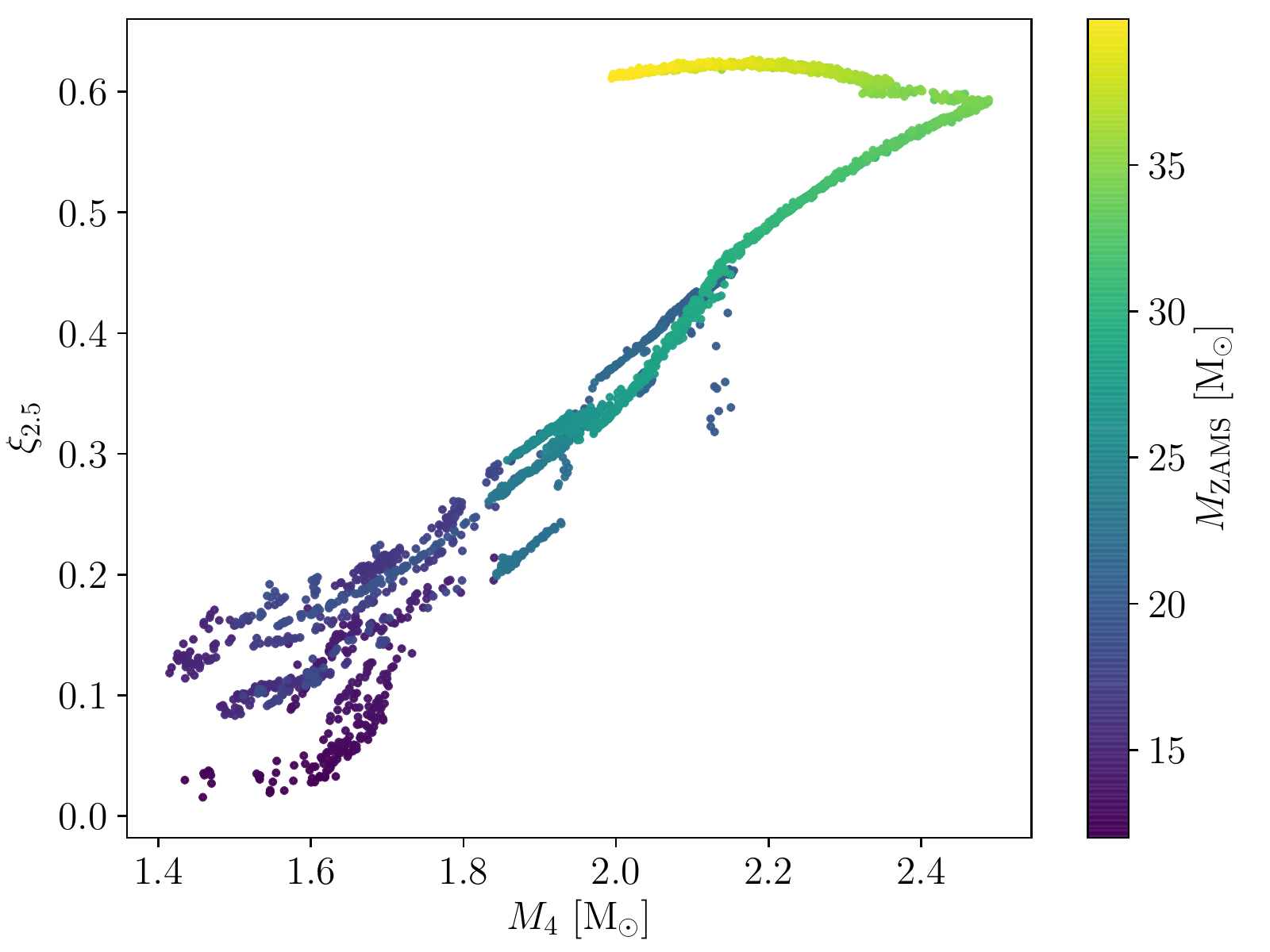}
\includegraphics[width=0.48\textwidth]{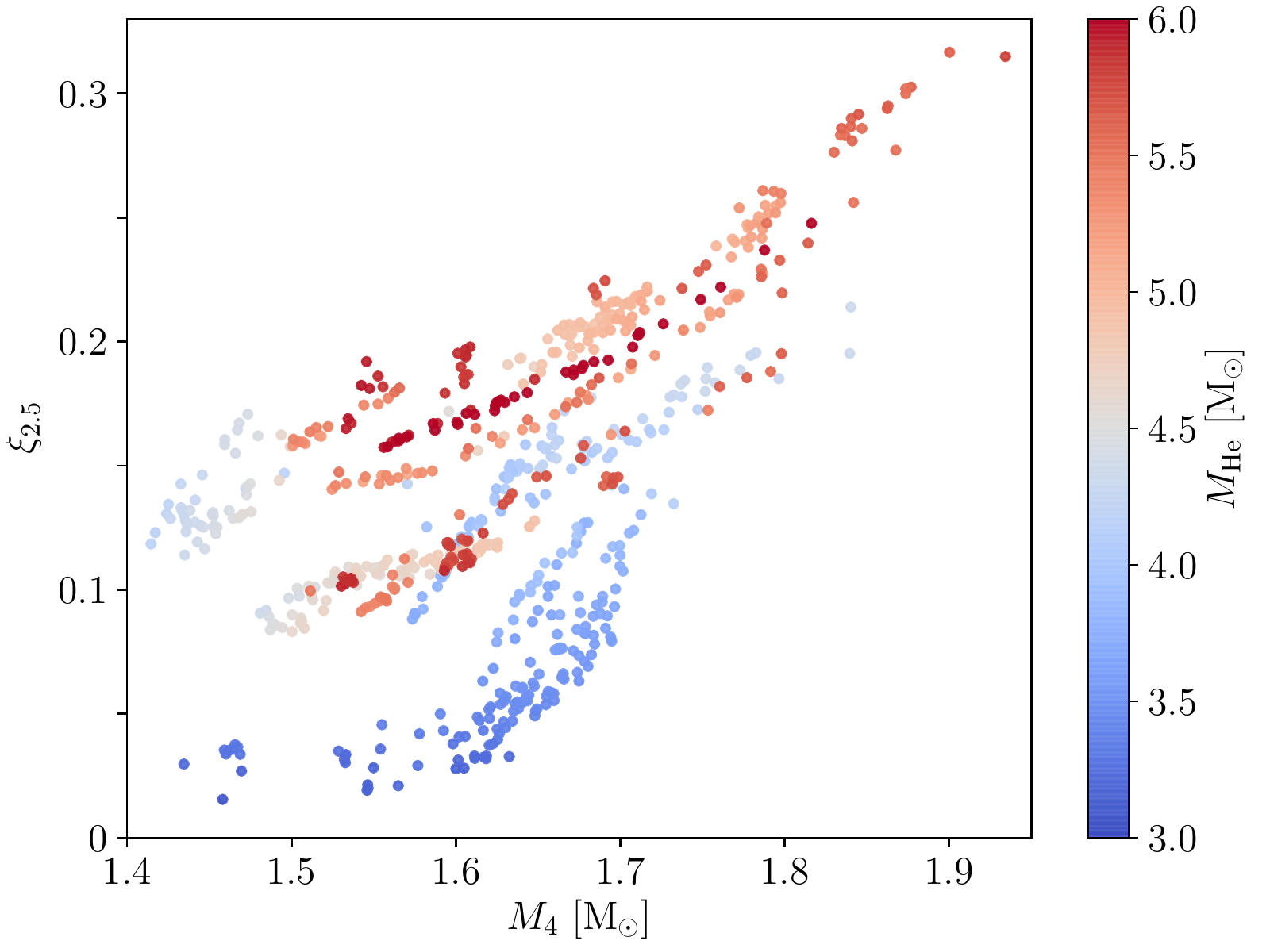}
\caption{Compactness enclosing innermost 2.5 \Msun\ at the time of
  presupernova for $\dot{M}_{\rm N}/2$ and $\dot{M}_{\rm N}/10$ set of
  models are plotted as a function of $M_4$, the Lagrangian coordinate
  of the first mass shell where the entropy per baryon exceeds $4.0
  k_B$. For lower mass models three major branches are clearly
  apparent below $M_4\approx$ 1.75 \Msun.  The lower panel shows a
  zoomed in version of the top panel, characterizing models with
  $M_{\rm ZAMS}<20$ \Msun, and with color contours denoting their
  corresponding helium core masses. \lFig{M4}}
\end{figure}

Degeneracy remains an important consideration though, even for the
larger stars. A 12 \Msun \ presupernova star has a core that is
degenerate (degeneracy parameter $\eta\equiv\mu/kT > 0$) out to 
1.46 \Msun. For 30 \Msun, degeneracy extends out to 1.87 \Msun. Even 
for 100 \Msun, the iron core is degenerate out to 2.1 \Msun. Most of 
the increase in degeneracy occurs between carbon depletion and 
oxygen depletion in the core when the core is efficiently cooled by 
the neutrinos \citep{Suk14}. Degeneracy affects the core structure 
making the (thermally adjusted) Chandrasekhar mass relevant for all 
stars that are likely to explode as common supernovae. On the other 
hand, the shells outside what will be the iron core in the presupernova 
star, and in particular the carbon and oxygen burning shells, are always 
non-degenerate. Strong burning thus leads to expansion that affects 
both $\xi_{2.5}$ and $\rm\mu_4$.

The upper panel of \Fig{M4} shows a generally strong correlation among
the compactness parameter, $\xi_{2.5}$, and $M_4$, except at very high
and low masses. A correlation is expected since a sharp density
decline close to the iron core requires a large radius to enclose 2.5
\Msun, thus small $\xi_{2.5}$. The sharp density decline implies a
deeper location where the entropy per baryon exceeds $4.0 \, k_B$ and
thus a smaller $M_4$ and a steeper mass gradient at that point as
well. The reversal of the plot for the most massive models is
a consequence of both the increasing central entropy and the outward
migration of the first oxygen shell. This migration causes a
non-monotonic dependence on mass for both $M_4$ and $\xi_{2.5}$ as
shown in \Fig{cp}, but $M_4$ starts to decline at a slightly lower 
initial mass due to increasing entropy. Therefore when the location of 
$M_4$ recedes from $\sim$2.5 to 2 \Msun, the $\xi$ stays roughly 
constant, and results in the reversal near $\sim$35 \Msun.

The lower panel of \Fig{M4} shows the behavior of compactness
  and $M_4$ for stars lighter than about 20 \Msun \ is clearly
multi-valued.  Three distinct branches are apparent for the lighter
stars. There is no reason to believe that more dense grid of
calculations would randomly fill out the spaces between the branches.

\Fig{M4He} delves deeper into this correlation between compactness and
$M_4$ and helps to understand why both are multi-valued relations of
mass for moderate mass stars (13\ $<M_{\rm ZAMS}<$19 \Msun). The first
panel shows $M_4$ in the presupernova star as a function of helium
core mass. A cleaner pattern results from using the helium core mass
instead of the main sequence mass since it eliminates the variations
due to envelope mass loss (\Fig{mfin}). As the figure shows, $M_4$ is
usually pegged to the ``oxygen shell'', by which we mean the location,
in the presupernova star, where the energy generation from oxygen
fusion (excluding neutrino losses) is a maximum. This is frequently,
though not always interior to the boundary of the ``silicon core'',
inside of which the silicon mass fraction is greater than the oxygen
fraction. Though one might naively assume that the oxygen burning
shell is at the edge of the silicon core, this is not generally
true. There are often gradients on the silicon to oxygen ratio left
behind by receding convection or radiative burning.  In these cases,
the oxygen shell lies inside the silicon core. On the other extreme,
in some stars the oxygen shell can be so active as to merge with the
carbon and neon burning shells producing one large convective shell
where all three are burning vigorously and the silicon core and oxygen
shell are coincident.  When this occurs, the compactness is usually
small \citep[Fig. 16 of][]{Suk14}. Many of the red points in the top
panel of \Fig{M4He} are of this sort.

It is not surprising that the oxygen burning shell, the strongest
burning shell during the late stages of stellar evolution, is usually
the location of a jump in entropy. The oxygen fusion rate is very
temperature sensitive, so oxygen burns at a nearly constant
temperature.  The overlying star expands, decreasing the local density
and thus increasing the entropy. The quantity $\mu_4$ in the Ertl
parametrization is a measure of the strength of this burning.

\begin{figure}
\begin{center}
\leavevmode
\includegraphics[width=0.48\textwidth]{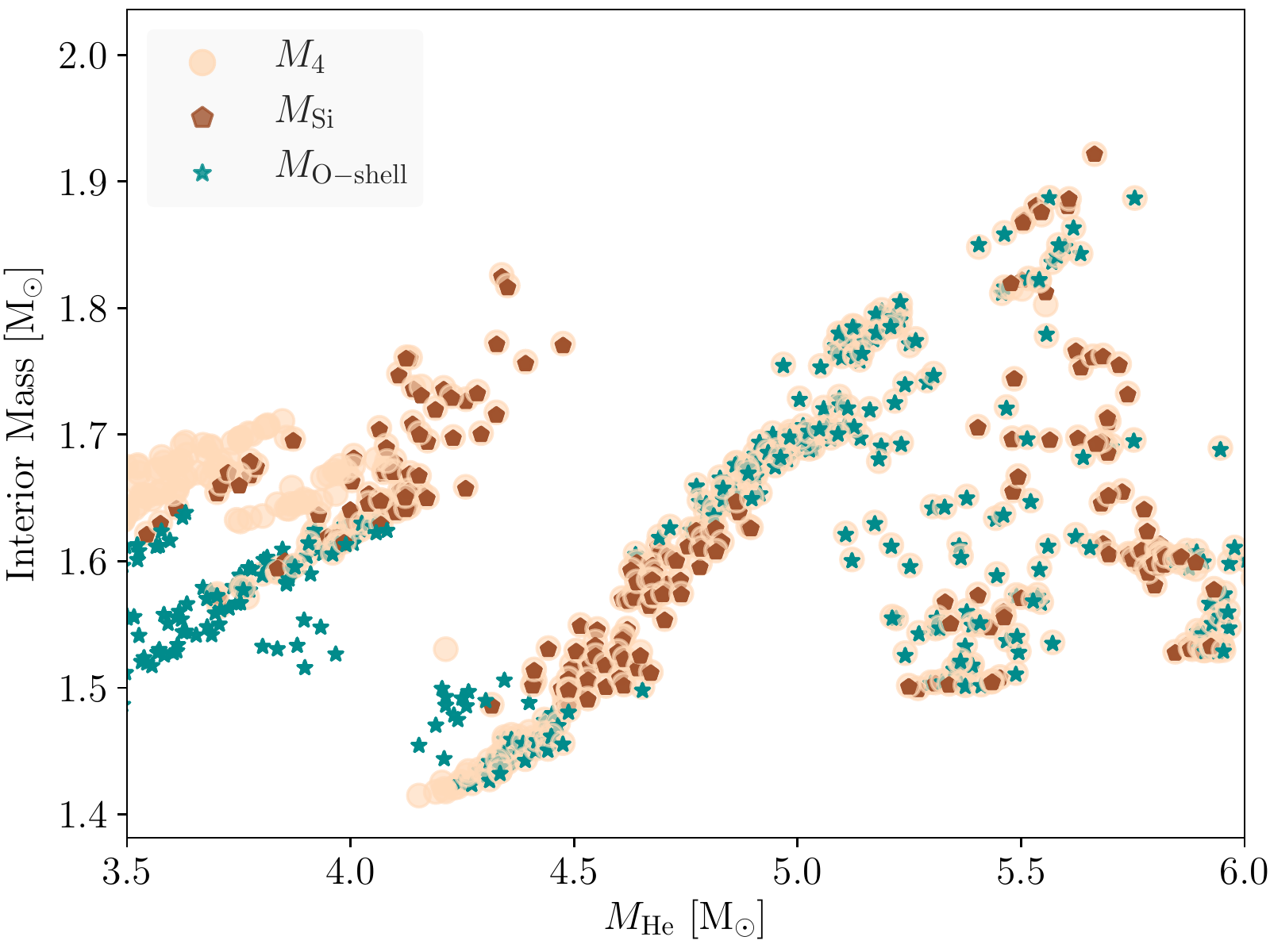}
\includegraphics[width=0.48\textwidth]{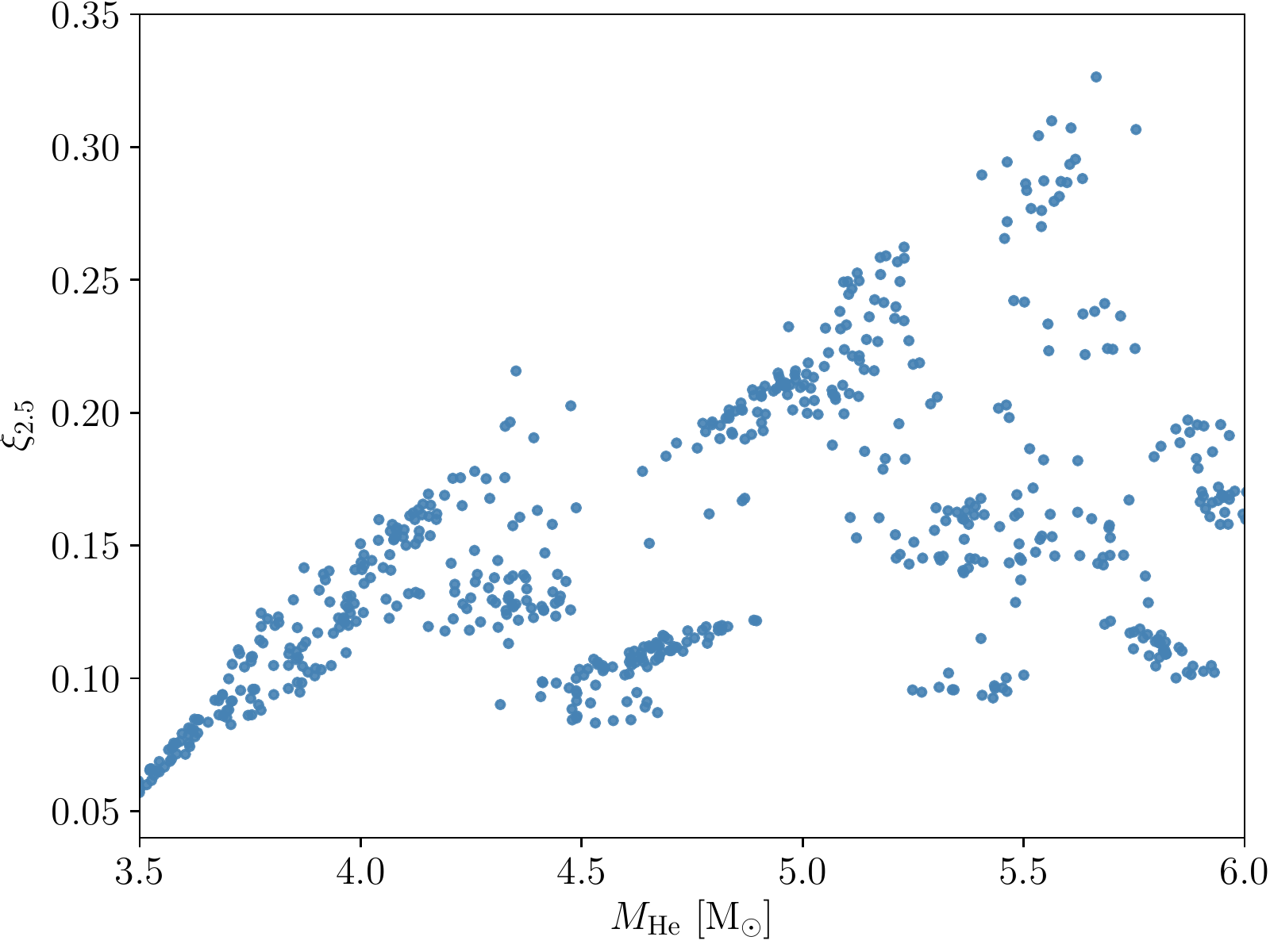}
\includegraphics[width=0.48\textwidth]{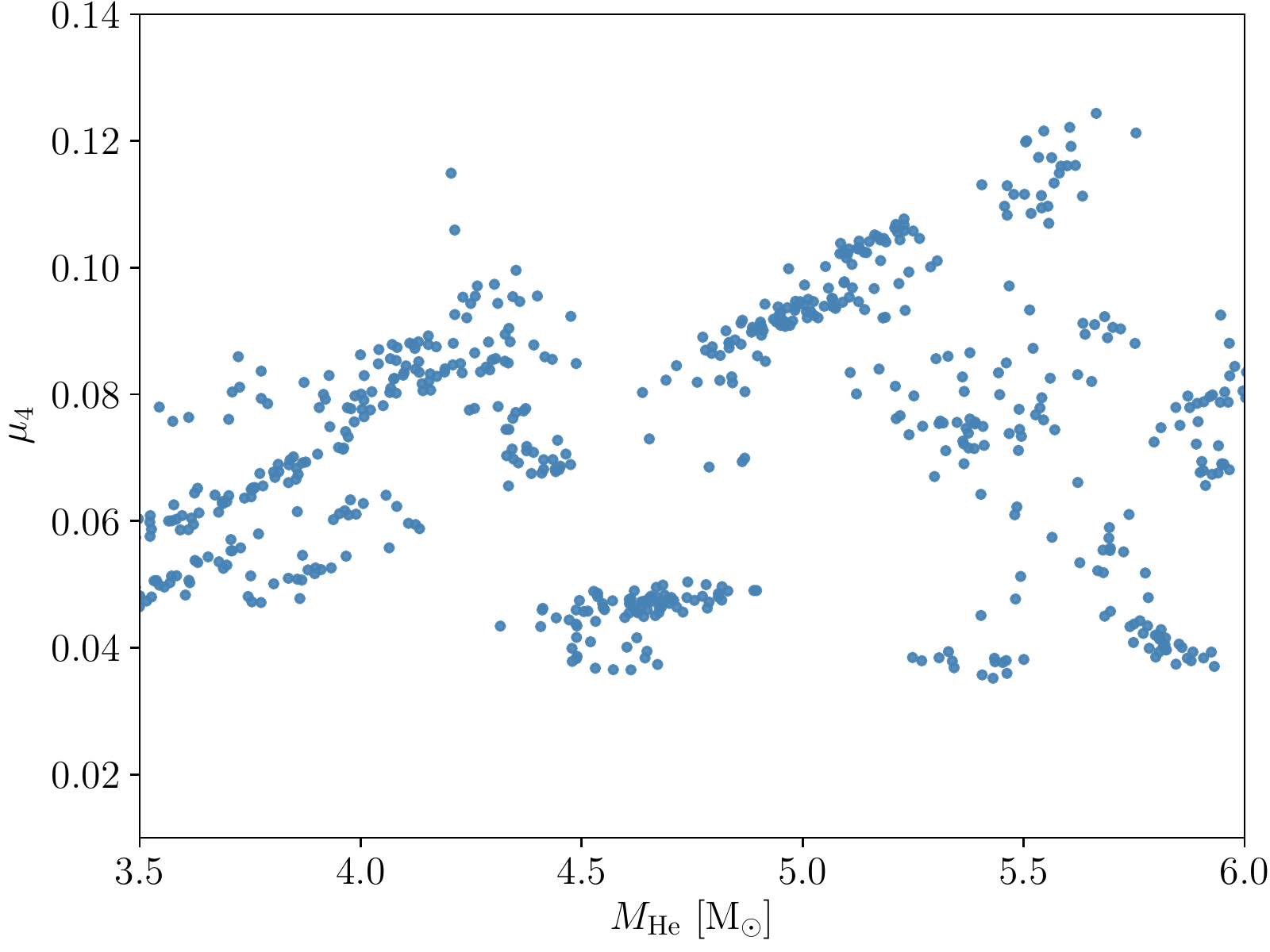}
\caption{(Top:) $M_4$ vs Helium core mass for presupernova stars with
  main sequence masses between 13 and 19 \Msun. Light pink points
  which usually underly green stars or red pentagon points indicate
  $M_4$.  Green stars mark the location of the most active oxygen
  burning zone and red pentagons are where the silicon mass fraction
  first exceeds the oxygen mass fraction, going inward. For clarity,
  red and green points not closely affiliated to a $M_4$ within
  0.01 \Msun\ have not been plotted. Usually the $M_4$ point is
  found at or near the maximum oxygen burning zone.  Several clusters
  of points sometimes lying in nearly straight lines are apparent.
  (Middle:) The compactness parameter, $\xi_{2.5}$ plotted for the
  same helium cores, equivalent to Panel b of \Fig{cp}. Note also a
  clustering of points here and a high degree of correlation with 
  $M_4$ in the top panel. (Bottom:) $\mu_4$ as a function of helium 
  core mass. Note the almost completely congruent pattern with 
  $\xi_{2.5}$, since both measure the most active burning shell at 
  the time of presupernova. \lFig{M4He}}
\end{center}
\end{figure}

The middle panel of \Fig{M4He} shows that the behavior of $\xi_{2.5}$
in this same mass range (equivalent to Panel b of \Fig{cp}) correlates
reasonably well with that of $M_4$, including, approximately, the
location and extent of clusters of solutions. Within a cluster, there
is usually a quasi-linear relation with a well-defined slope. A larger
value of $M_4$ implies a larger value of $\xi_{2.5}$. The farther out
in the star the strong burning shell, the less centrally concentrated
is its density. This correlation is not perfect though. Sometimes long
lines in the $M_4$ plot, e.g. the long string of points from $M_{\rm
  He}$ = 4.1 \Msun\ to 5.2 \Msun, break into several segments with
different $\xi_{2.5}$. This is partly due to the arbitrary pinning of
$\xi_{2.5}$ to a single mass shell, but also because specifying $M_4$
alone does not measure the strength of the burning there.

Instead, as might be expected, the bottom panel of \Fig{M4He} shows
that $\xi_{2.5}$ correlates better with $\mu_4$, the mass gradient
at $M_4$. This correlation is very strong since both quantities are
sensitive to the strength of the most active burning shell in the last
days of the star's life.  Why though are $M_4$ and $\mu_4$ not
randomly scattered between their extrema? Why the patterns? The first
panel of \Fig{M4He} suggests a reason. $M_4$ traces the location of
the strongest oxygen burning shell. Oxygen burns there because it is
at the deepest location that has not already depleted oxygen by prior
shell burning. That location is, in turn, set by the extent of oxygen
burning shell(s) during the previous evolution, which in turn depends
upon the entropy structure set up during carbon shell burning.

\citet{Suk14} pointed out this correlation between the core
compactness and the extent of the first oxygen burning convective
shell (their Fig. 14). \Fig{oshell} shows a related
quantity for the new model set. The figure shows the mass of the
``silicon core'', the mass interior to which oxygen has burned out, at
the time silicon burning ignites in the star's center. Here a specific
central temperature, $3.0 \times 10^9$ K, was chosen in order to make
the plot, but the conditions here reflect what the oxygen burning
shell (or shells) have accomplished prior to silicon ignition.  Visual
inspection of the convective history of these 700 models shows that
the silicon core mass at silicon ignition is very nearly equal to the
maximum extent of the second convective oxygen burning shell for stars
below $\sim14.6$ \Msun \ (helium core mass 4.2 \Msun) and of the first
oxygen convective shell for more massive stars up to at least 20
\Msun. In this plot, we see the clearest evidence yet for regular, but
non-monotonic and occasionally multi-valued behavior. Two helium cores
of very nearly 4.2 \Msun \ can give rise to presupernova cores with
structures in one of two well-defined states. Slight shifts in mass,
composition, or even numerical approach (including zoning) can send the
star one way or another.

Further analysis of the convective histories reveals the systematics
behind this behavior - at least in the 1D code, if not in nature:

\begin{itemize}

\item The little jump at $M_{\rm He}$ = 3.7 \Msun \ ( $M_{\rm ZAMS}$ =
  13.2 \Msun) from $M_{\rm Si}$ = 1.7 \Msun \ to 1.6 \Msun, reflects
  the operation of the third convective carbon burning shell. Below
  this mass, the third shell ignites inside the former full extent of
  the second shell and outside the effective Chandrasekhar mass. Thus
  the core oxygen burning start to burn in a smaller extent while this
  outermost carbon shell operates. For masses above this value, the
  third shell ignites at the outer boundary of the previous shell and
  within the Chandrasekhar mass. Core oxygen burning now has to wait
  until this third shell is complete and therefore the base of this
  third shell sets (approximately) the extent of the first oxygen
  convective shell, and thus the base of the second one (Panel 13.45 
  \Msun\ of \Fig{cnv}).

\item The much larger decrease in $M_{\rm Si}$ from 1.7 \Msun \ to 1.4
  \Msun \ at $M_{\rm He}$ = 4.2 \Msun \ ($M_{\rm ZAMS}$ = 14.6 \Msun)
  is due to the diminished significance of the second
  oxygen burning shell. The silicon core shrinks to the extent of the
  first oxygen burning shell though the second shell continues to be
  sporadically important for a time (Panel 15.01 \Msun\ of \Fig{cnv}). 

\item The jump and wild variations starting at $M_{\rm He}$ = 5.2
  \Msun \ and extending up to 5.7 \Msun \ ($M_{\rm ZAMS} $ = 17.2
  \Msun \ to 19.0 \Msun) mark the transition from convective carbon
  burning to radiative burning. After the transition near $\sim$19
  \Msun, carbon no longer burns exoergically at the center of the
  star, and there are only two convective shells before oxygen burning
  rather than three or more. During this transition, carbon core and
  shell burning vary greatly in location and extent. The large spread
  in $M_{\rm Si}$, and ultimately in compactness, reflects the
  irregularity of this transition (Panels 17.90 and 18.81 \Msun\ of 
  \Fig{cnv}).

\end{itemize}

Other shell interactions result in the weaker multiple branches seen
in higher mass stars in \Fig{cp}. Despite the specific masses given in
the list above, none of these transitions are abrupt and, given slight
nudges, the star may oscillate from one to the other solution when it
is close to boundaries. This leads to the ``multivalued'' behavior.
Although complicated and probably sensitive to the one-dimensional
treatment of the problem, the conclusion is that the presupernova
structure in stars from 13 to 19 \Msun \ results from an interplay of
convective carbon and oxygen burning shells after carbon ignition.

Putting these various factors together, a deterministic picture
emerges. Multiple carbon burning stages - core burning below 19 \Msun,
and two or more episodes of shell burning - act to sculpt an entropy
distribution in the core so that oxygen shell burning, when it occurs,
ignites at and extends to various mass shells.  Whether the number of
carbon shells is two or three or four and the number of oxygen shells
one or two strongly affects the core structure of a presupernova star
in the mass range where a significant number of events are
observed. These changes in the extent of convective shells, which
amplify small differences in the earlier evolution, may end up
determining whether the star explodes or makes a black hole.

\begin{figure}
\includegraphics[width=0.48\textwidth]{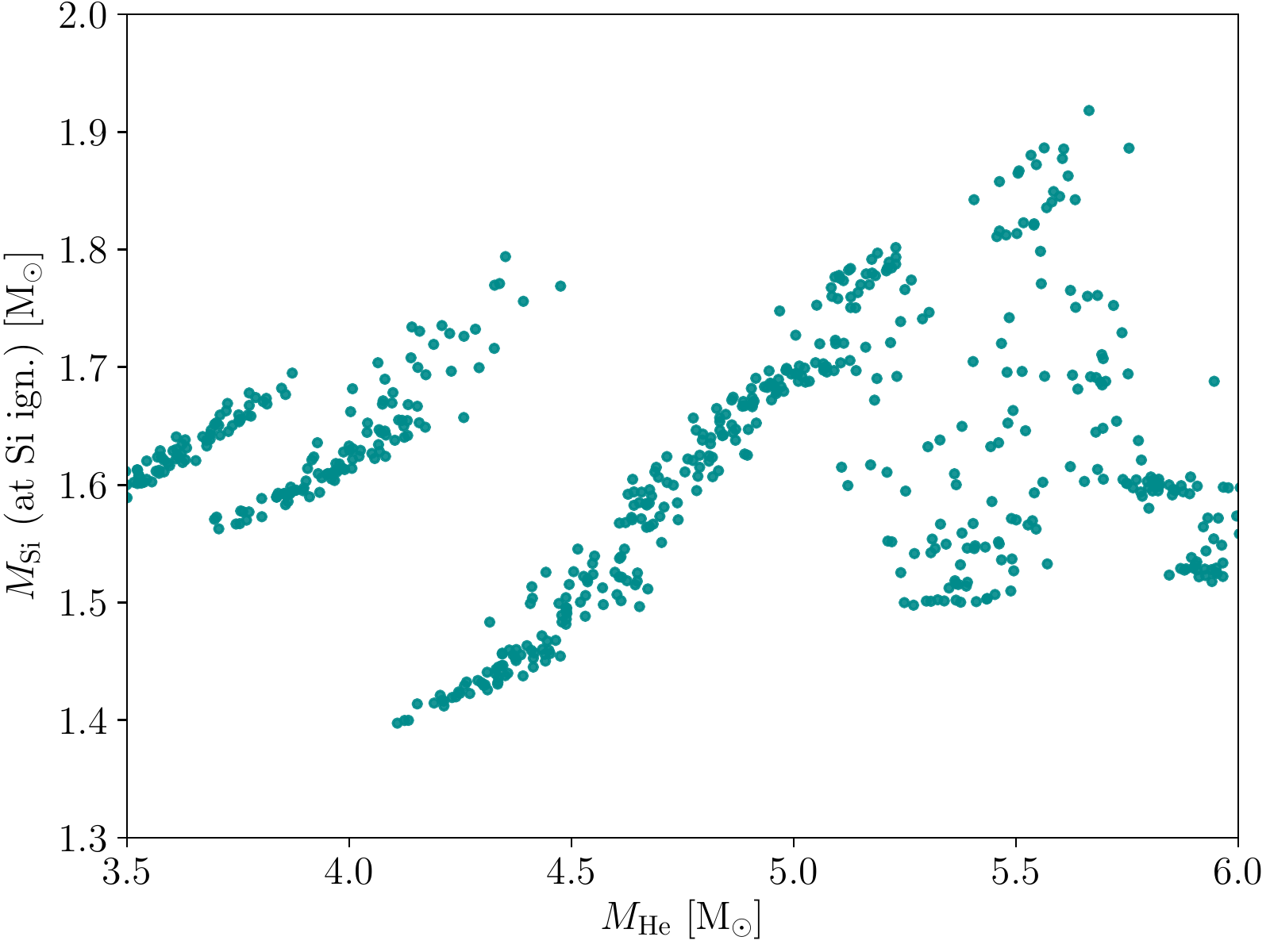}
\caption{The size of the oxygen depleted core (oxygen less than
  silicon or iron) as a function of helium core mass at silicon ignition
  (when the central temperature first reaches $3.0 \times 10^9$ K).
  This is approximately the maximum extent of the first oxygen
  convective shell. The function has multiple trajectories that
  correlate with the major features in the plot of $ M_4$ vs helium
  core mass in the top panel of \Fig{M4He}. \lFig{oshell}}
\end{figure}

\begin{figure*}
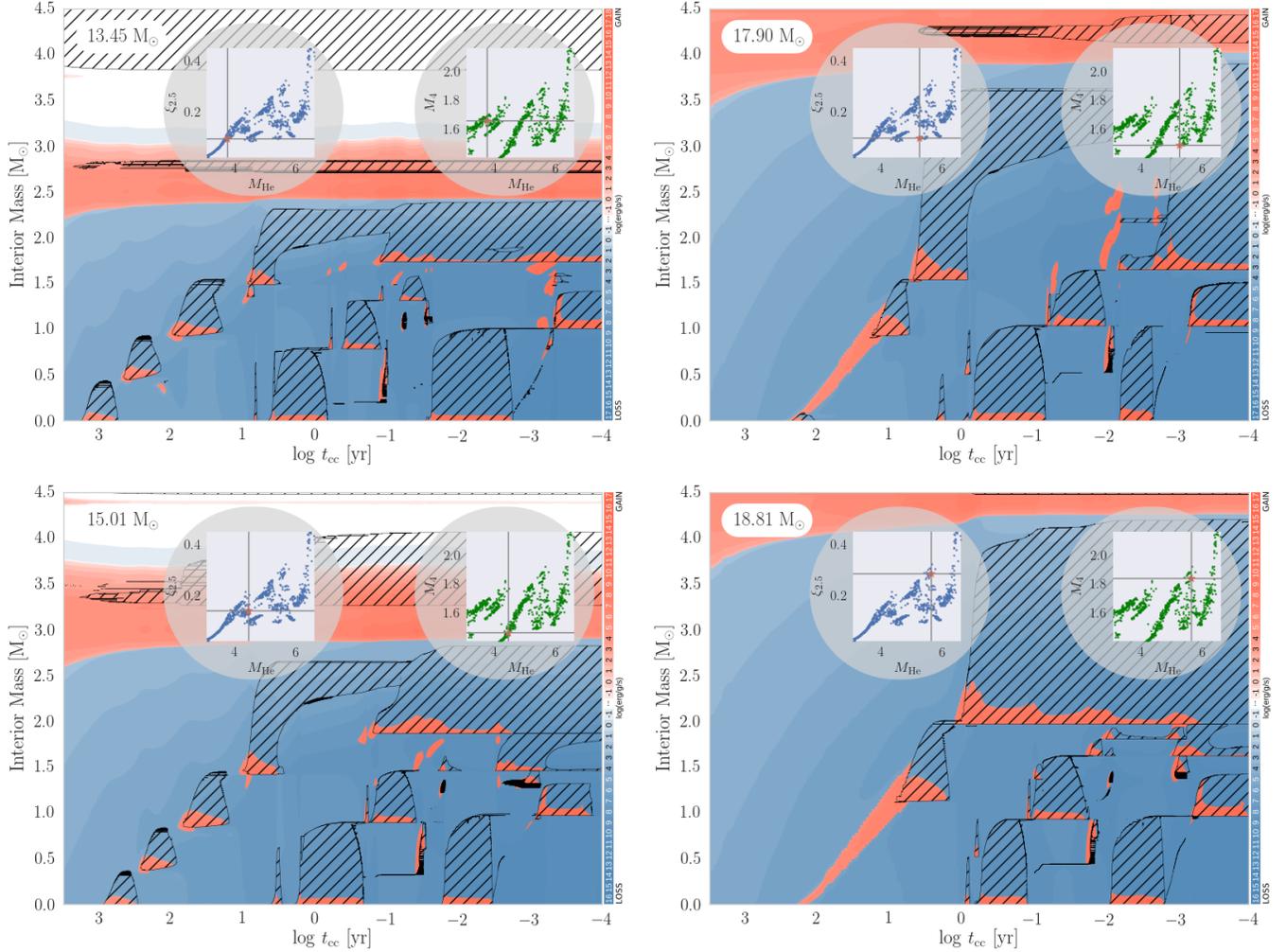

\includegraphics[width=0.5\textwidth]{fig14a.pdf}
\includegraphics[width=0.5\textwidth]{fig14c.pdf}
\includegraphics[width=0.5\textwidth]{fig14b.pdf}
\includegraphics[width=0.5\textwidth]{fig14d.pdf}
\caption{Convection histories of four sample models from the range of
  initial masses from 14 to 19 \Msun, representing key structural
  changes that are responsible for the significant variations in the
  final presupernova core properties. Each panel shows the evolution
  of the innermost 4.5 \Msun\ material roughly during the last
  thousand years of its life, i.e., from core carbon burning until
  presupernova. Colored shades denote energy generation (red) and
  energy loss (blue) gradients. Hatched black regions are convective
  episodes. The $x$-axis is shown as the log of time until core
  collapse, $t_{\rm cc}$. The initial mass of each model is denoted on
  the top left corner of panels. Two inserted mini-plots illustrate
  the compactness parameter (purple) and the lagrangian location of
  entropy per baryon equal $4.0 \, k_B$ point (green) respectively,
  both as a function of helium core mass corresponding to the above
  mentioned initial mass range. The crosses inside each mini-plots
  denote the values corresponding to the model. (13.45 \Msun:) The
  lowest mass stars have 4 or more convective carbon burning episodes
  followed by 3 oxygen burning episodes (including central core
  burning episode). The shell helium burning lies within the 2.5
  \Msun\ location, the point on which the compactness parameter,
  $\xi_{2.5}$, is measured, and thus these models have small
  compactness and small $M_4$. (15.01 \Msun:) Central carbon burning
  weakens with increasing mass, and as a result the shell carbon
  burning episodes gradually ``migrate'' inwards (with increasing
  initial mass). Once the third shell carbon burning ignites within
  the effective Chandrasekhar mass, the core oxygen burning has to
  wait until the overlying carbon burning episode is finished, which
  leads to significantly weakened or, in some cases absent, immediate
  second oxygen shell burning. In these models oxygen eventually burns
  later when silicon is already ignited in the core. This marks the
  transition of $M_4$ to the ``second'' branch.  Due to its
  sensitivity to the last major shell burning episode, the compactness
  parameter has multiple solutions depending on the final
  configuration of oxygen, neon and carbon burning (i.e. together burn
  vigorously in one shell, or separately). (17.90 \Msun:) Around
  $\sim$ 18 \Msun\ it becomes increasingly harder for carbon to burn
  convectively near the center. The tiny convective central burning
  episode is followed by a long lasting radiative flame, and as a
  result, with increasing initial mass, the base of the convective
  carbon burning shells start to migrate outwards. (18.81 \Msun:)
  Eventually at high enough initial mass all central carbon burning is
  radiative. Models between 17 and 19 \Msun\ have rapidly changing
  core structures, and thus rapidly varying $\xi_{2.5}$ and $M_4$.
  Same format plots for all 800 models between 12.00 and 19.99
  \Msun\ are available online (see Footnote 9). \lFig{cnv}}
\end{figure*}

\section{Compact Remnants and Explosions}
\lSect{remnants}

Since the new models differ from those of \citet{Suk14} and
  \citet{Suk16}, and to a lesser extent, from \citet{Mue16}, it is
  worth revisiting some of the conclusions of those papers regarding
  compact remnants using the new models.

\begin{figure}
\includegraphics[width=0.48\textwidth]{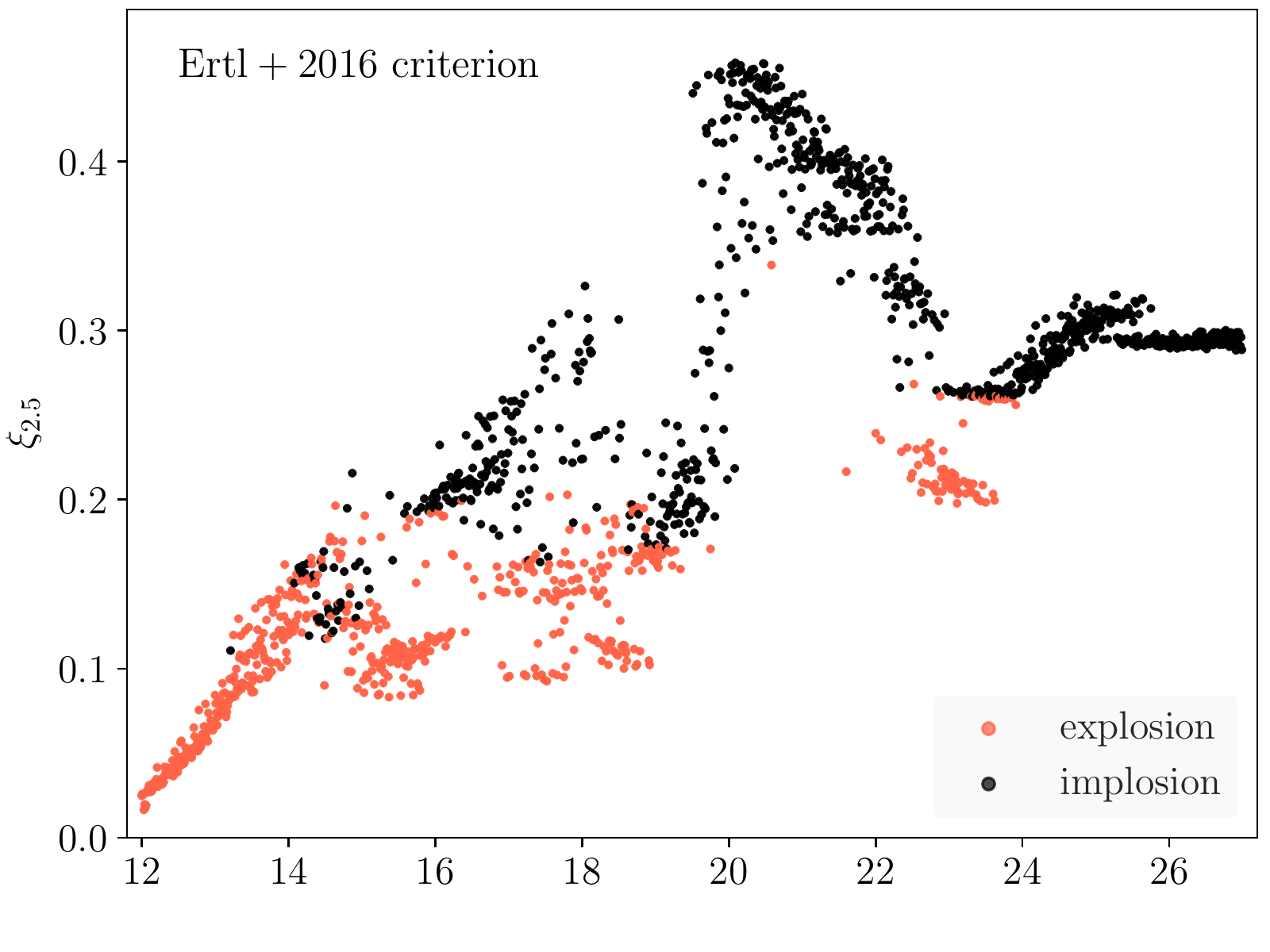}
\includegraphics[width=0.48\textwidth]{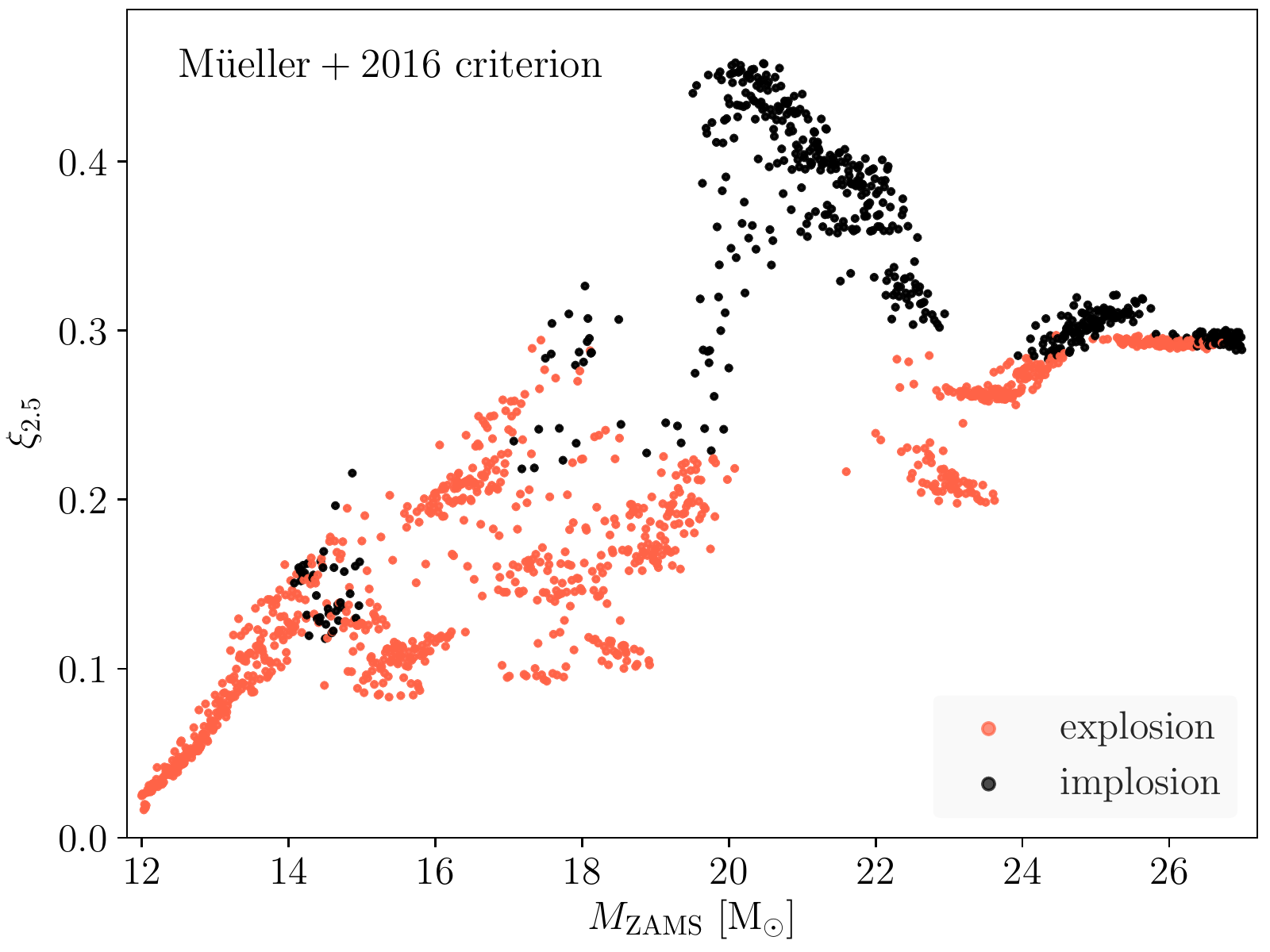}
\caption{The compactness parameter shown in \Fig{cp} is plotted again
  twice, color coded as to the success or failure of the explosion
  based upon the parameterization of \citet{Ert16} (top) and the
  semi-analytical method of \citet{Mue16} (bottom). Red symbols denote
  successfull explosions by these criteria, and black symbols,
  failures. The very good agreement between these two different
  approches suggests that both are good representations of core
  structure, though the criterion of \citet{Mue16} favors slightly
  more explosions for the parameters chosen.  \lFig{explode}}
\end{figure}

\Fig{explode} shows the models now expected to explode based on the
criteria of \citet{Ert16} and \citet{Mue16}. These figures can be
compared with Fig. 4 of Ertl et al. and Fig. 6 of M\"uller et al.,
which they very closely resemble. The ``N20'' engine parameterization 
was used to make the comparison with \citet{Ert16} and the ``standard'' 
choices of five parameters were used for \citet{Mue16} method.

The most significant difference with the Ertl criterion is the shift
of the pattern for models above about 14 \Msun\ to slightly ($\sim$10
\%) lower masses.  Changes in average quantities like the remnant
masses are small, however. \Fig{BH} shows a comparison of IMF-weighted
remnant mass distributions of imploding models assuming a
\citet{Sal55} initial mass function with a power of -2.35. The usual
two cases are considered: a) collapse of the helium core, but ejection
of the hydrogen envelope; and b) collapse of the entire presupernova
star. Case a) is more appropriate for stars where the envelope
  is lost to a binary companion or, for very massive stars, to
  winds. The envelope might also be lost to a very weak explosion that
  did not unbind the helium core \citep{Qua12,Ful17,Lov13}.  Case b)
  is for more robust explosions. In each case there is an element of
  uncertainty because the final mass depends on the mass loss rate,
  e.g., for red supergiants in Case b) and Wolf-Rayet stars in Case
  a). In making this \Fig{BH}, but not in computing the averages,
only the mass range covered by our $\dot{M}_{\rm N}$ mass loss survey,
12 - 27 \Msun, was included.

Astronomers, of course, observe black holes coming from all masses of
stars, not just 12 - 27 \Msun. To compute the averages for solar
metallicity stars (and the answer will be sensitive to metallicity),
the new results were supplemented with the prior models of
\citet{Suk16} for stars more massive than 27 \Msun. For just the
limited mass range 12 - 27 \Msun, the average black hole masses from
\citet{Suk16} were previously 6.52 and 15.3 \Msun\ respectively for
the helium core and full star assumptions. The corresponding new
numbers are 6.60 and 14.2 \Msun.  Considering the entire mass range of
stars that experience iron core collapse, and using the N20 parameters
in \citet{Suk16}, the previous averages were 9.25 \Msun \ and 13.7
\Msun. The equivaluent new numbers are 8.61 \Msun \ and 13.5 \Msun.  A
less than 1\% adjustment has been applied to Table 4 of \citet{Suk16}
based upon a slightly different way of interpolating the grid.

Similar corrections can be estimated for the neutron star
gravitational masses. Lacking an explosion model, it is assumed, based
on prior experience \citep[i.e.,][]{Ert16}, that the baryonic mass of
the resulting neutron star is usually equal to $M_4$. After
appropriately correcting for neutrino mass loss 
\citep[same as in][]{Mue16}, the new average neutron star gravitational 
mass in the range 12 - 27 \Msun\ is 1.45 \Msun. Explosions below 12 
\Msun\ were calculated by \citet{Suk16} using presupernova models in 
which the neutrino bug had been fixed. They can thus be combined with 
the current set. Assuming neutron star production above 27 \Msun \ is 
negligibly small, the new global average neutron star mass is 1.38 \Msun.
In Table 4 of \citet{Suk16} the corresponding global average was 1.41
\Msun.

\begin{figure}
\includegraphics[width=0.48\textwidth]{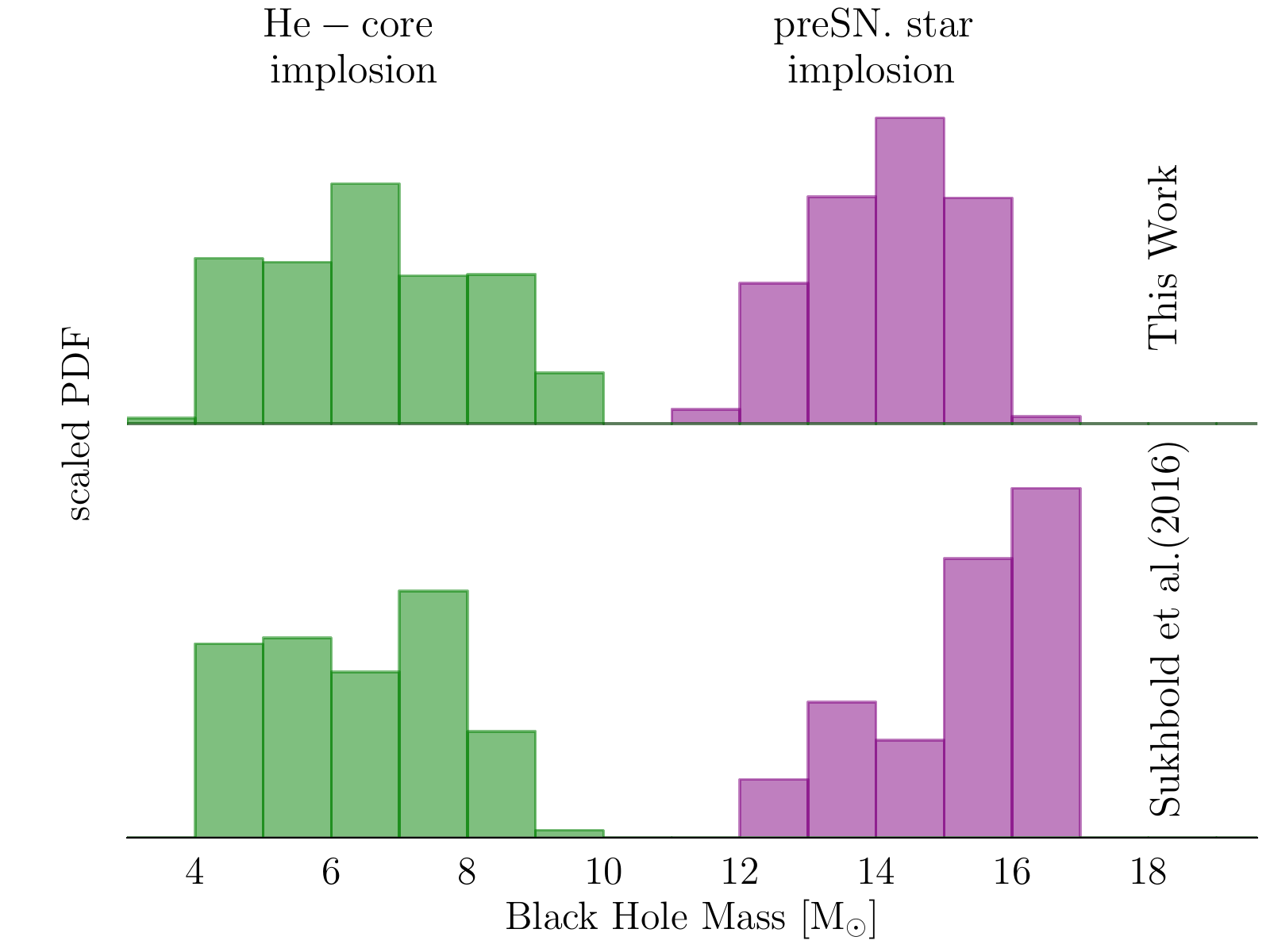}
\caption{The IMF-weighted black hole mass distributions for the new
  $\dot{M}_{\rm N}$ set is shown in comparison with the results from
  \citet{Suk16} over the same initial mass range between 12 and 27
  \Msun. The differences are small and well within the range of
  variation seen for different explosion models. The average black
  hole masses for the older study is 15.3 and 6.52 \Msun\ for two
  scenarios respectively, while the new models yield 14.2 and 6.60
  \Msun. \lFig{BH}}
\end{figure}

The agreement of \Fig{explode} with the earlier work \citep[Fig. 6 of
][]{Mue16} is excellent, and it is thus expected that the explosion
outcomes will also be very similar. Indeed, plots of neutron star mass
(\Fig{nstar}), black hole mass, and explosion energy as a function of
initial mass (not shown) are virtually indistinguishable from the
panels in their Figure 2. This suggests that the difference in
envelope structure and zoning between the series of \citet{Mue16} and 
the present paper did not have much influence on the core structure 
and statistical explosion properties.

\Fig{explode} shows that a significantly larger fraction of stars in
the interesting mass range 15 to 20 \Msun \ explode using the
\citet{Mue16} formalism, even though ''N20'' is one of the more
energetic formulations of the \citet{Ert16} model. Conservatively,
this could be regarded as an uncertainty in outcome until more
realisic simulations of the actual explosion can be done.  The various
energies in the \citet{Ert16} model came, however, from calibrating a
central engine with 1D neutrino transport, a shrinking protoneutron
star, and fallback to SN 1987A using various presupernova models and
might, for now, be considered the more realistic of the two. This
calibration to SN 1987A has its own uncertainties though, especially
since the structure of the presupernova star could be
different in more realistic binary merger models \citep[e.g.,][and 
references therein]{Men17}.

\Fig{nstar} gives the expected neutron star masses that result when
our new presupernova models are analyzed using the formalism and
standard parameters of \citet{Mue16}. Fallback is neglected in their
analysis, so the baryonic mass of the proto-neutron star is equal to
the mass coordinate where the neutrino-driven engine shuts off a few
hundred ms after shock revival. As mentioned before, this correlates
strongly with $M_4$ (top panel). There are branches of neutron star
masses above the $M_4$ line, which primarily originate from less
massive progenitors ($M_{\rm ZAMS}<15$ \Msun), reflecting the
operation of a prominent second oxygen burning shell and the resulting
shallow entropy profile. The most massive neutron stars have a
baryonic mass that is bounded by the base of the convective carbon
burning shell, which is located much further out than $M_4$ in the
presupernova star. The second panel of \Fig{nstar} shows the expected
neutron star (gravitational) mass as a function of main sequence mass.
These results agree quite well with those of \citet[][see their
  Fig.~2c]{Mue16} including the existence of very massive neutron
stars produced for main sequence stars 14 - 15 \Msun \ and the highly
variable nature of the solution between 14 and 19 \Msun. For
the majority of models, the resulting neutron star mass is tightly
correlated with $M_4$ (\Fig{M4He}) and oxygen depleted core at the
time of silicon ignition (\Fig{oshell}), and in certain initial mass
ranges the explosion of models with nearly identical mass (or slightly
different input physics) can result in very different neutron stars.

\begin{figure}
\includegraphics[width=0.48\textwidth]{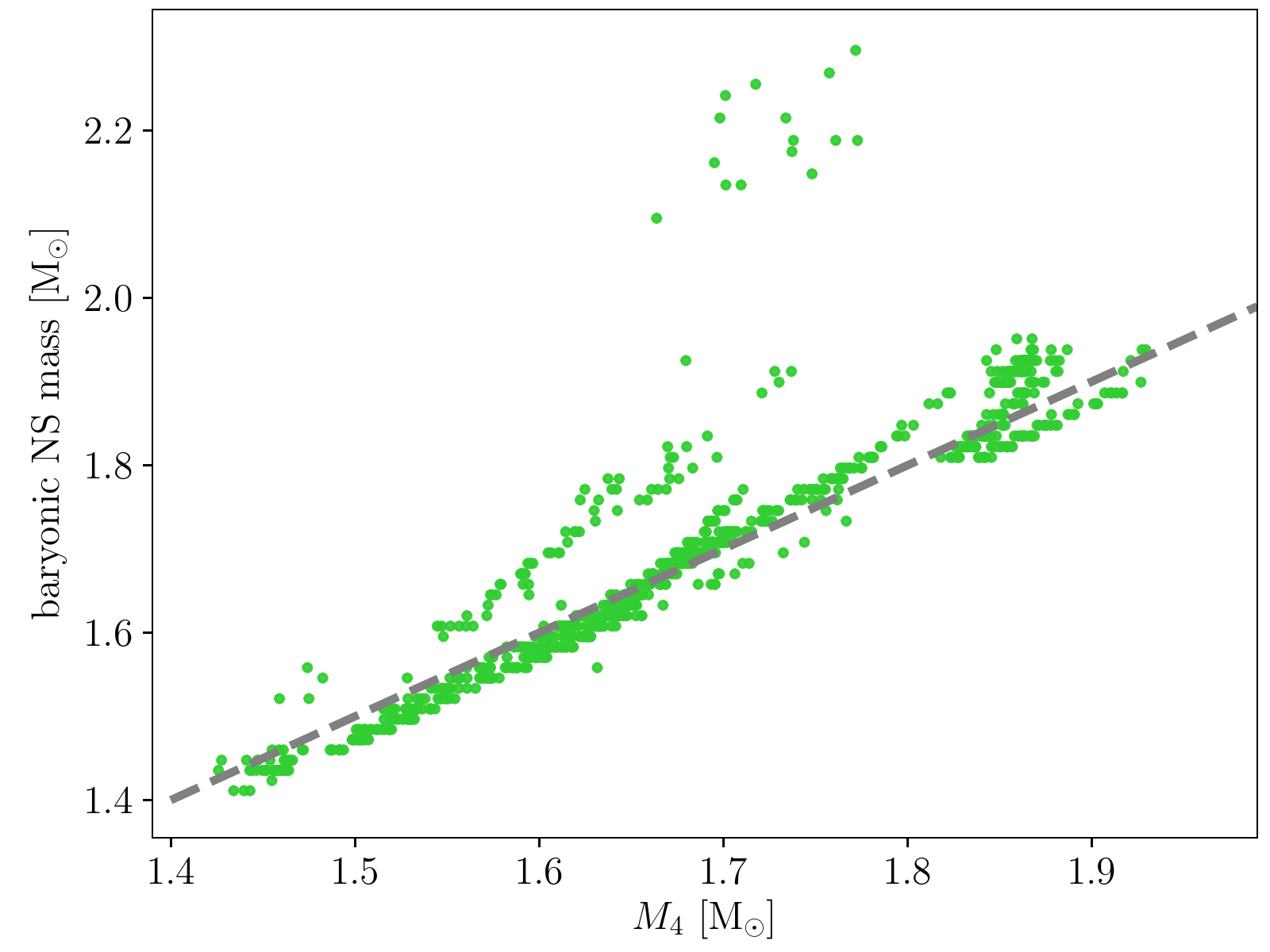}
\includegraphics[width=0.48\textwidth]{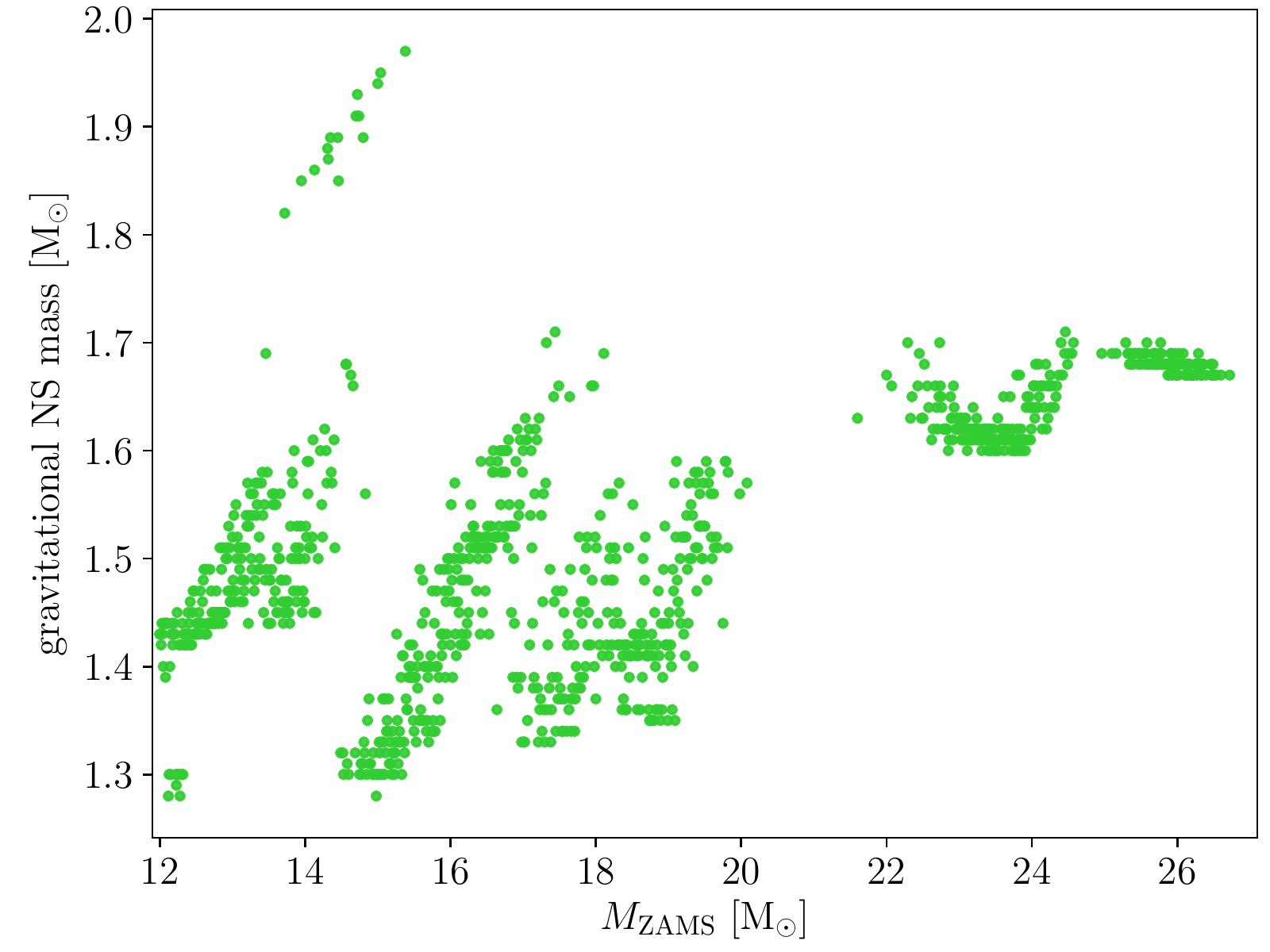}
\caption{Neutron star mass distribution resulting from analyzing the
  new model set with normal mass loss ($\dot{M}_{\rm N}$) for stars in
  the mass range 12 to 27 \Msun\ using the approach of
  \citet{Mue16}. (Top:) The {\sl baryonic mass} of the expected
  neutron star is plotted as a function of $M_4$, the mass where the
  entropy per baryon reaches $4.0 k_B$ in the presupernova star
  (\Sect{interpret}). Neutron stars on the grey dashed line have a
  mass, before neutrino losses, equal to $M_4$. Many points are found
  here because $M_4$ is usually the ``mass cut'' in a successful
  explosion, unless the final entropy profile is significantly more
  shallow.  (Bottom:) The gravitational mass \citep[adjusted from the
    baryonic mass in the same way as in][]{Mue16} of the expected
  neutron star as a function of main sequence mass. More massive
  neutron stars are typically made by more massive main sequence stars
  with a greater $\xi_{2.5}$ and $\mu_4$, but the most massive points
  on this plot come from models with $M_{\rm ZAMS}\sim14-15\ \Msun$,
  the most massive models with a significant second oxygen shell
  burning. The existence of multiple branches for these quantities
  (\Fig{M4He} and \Fig{oshell}) results in a large range of neutron
  star masses being acessible by stars of nearly the same main
  sequence mass for some initial masses. Note also the cluster of
  neutron stars with mass near 1.65 \Msun \ coming from the most
  massive supernova progenitors. \lFig{nstar}}
\end{figure}

It is not expected that the nucleosynthesis and light curves
calculated by \citet{Suk16} will be significantly altered by using the
new models, though further exploration is certainly encouraged.  In
particular, the deficiency of light \textsl{s}-process elements seen
in \citet{Suk16} will persist in the new model set, since most of the
production of these elements is due to the most massive stars
\citep{Bro13} which still fail to explode. 
The fraction of solar metallicity stars above 9 \Msun \ exploding as
supernovae was 74\% based on the N20 parameterization \citep[Table 4
  of][]{Suk16} and is now 65 \%. Above 18 \Msun, the fraction of SN
reduces to only 8\%, which actually is in a slightly better accord
with the results from \citet{Sma09,Sma15}, as compared to the previous
study.  Since the envelope masses have not changed appreciably and the
explosion energies are also expected to be unchanged, the light curves
will be unaltered.  The fractions given in their Table 4 for
supernovae above 12, 20, and 30 \Msun \ will also probably not change
within the variations already seen for the different central engine
characteristics.

\vskip 0.2 in
\section{Conclusions}
\lSect{conclude}

The full evolution of over 4,000 massive stars of solar metallicity in
the mass range 12 to 60 \Msun\ has been studied using unprecedented
resolution, both in zones per star and number of stars within a given
mass range. The mass loss rate was varied, and an important bug in the
neutrino loss routine was repaired that caused a significant variation
in the presupernova properties from those calculated by \citet{Suk14}
and prior works. Our chief conclusions are:

\begin{itemize}

\item The pattern of core compactness seen in previous studies
  \citep[e.g.,][]{Suk14,Mue16} is robust. The ``noise'' in these
  studies was not a consequence of inadequate zoning, but reflects
  real variability. The range of variation and the location of peaks
  (\Fig{cp}) in the new study are virtually identical to that
  seen by \citet{Mue16}, even though the new models use 4 to 10 times
  the zoning and a much smaller surface boundary pressure. The
  results are also qualitatively similar to \citet{Suk14}, but
  important peaks in the compactness plot are shifted downwards by
  about 10 \% in initial mass (\Fig{cpold}) due to the corrected
  neutrino loss rate. For the stellar physics used and spherically
  symmetric nature of the calculation, the variations and peaks seen
  in \Fig{cp} are now well determined.

\item The large variation in core structure seen for stars between 14
  and 19 \Msun\ by \citet{Suk14} is not completely random. For a
  larger set of models with finer spacing in initial mass, several
  branches of solutions emerge. This behavior was also seen by
  \citet{Mue16}. The branches apparently result from variations in the
  location of the oxygen burning shells in the presupernova star,
  which in turn ``remember'' the location of several carbon burning
  shells of variable extent. Some noise is introduced by the fact that
  the carbon abundance and size of the carbon-oxygen core are not
  precisely monotonic with mass (\Fig{cores}).  This, in turn,
  reflects the operation of semiconvection at the boundary of the
  hydrogen and helium convective core. In the mass ranges 14 to 19
  \Msun\ and 22 to 24 \Msun, the presupernova core structure is
  more sensitive to events in the last year of a star's life (and
  sometimes the last hours) than to the star's initial
  mass. Presupernova structure in these mass ranges does not
  necessarily follow the Vogt-Russell theorem \citep{Vog26,Rus27}.  It
  may be more comparable to weather on earth.

\item Even accounting for these systematics, the branches of solutions
  are still noisy. The level of noise is reduced if one characterizes
  the presupernova star by its helium core mass, not its starting mass
  or final total mass. The scatter is also reduced in the Ertl
  two-parameter characterization of core-structure rather than the
  O'Connor-Ott compactness parameter.

\item Given the large number of models, it is possible to give
  statistically meaningful results for the radius, luminosity and
  effective temperature of supernova progenitors (\Sect{stars}). 

\item The mass distributions of neutron stars and black holes
  resulting from supernovae in the mass range studied are not greatly
  altered from our earlier surveys (\Sect{remnants}). The average
  gravitational mass of neutron stars, including all masses of
  supernovae, is now 1.38 \Msun.  The average black hole mass, if only
  the helium core implodes, is 8.61 \Msun. If the entire presueprnova
  star collapses, the average black hole mass is 13.5 \Msun.  The
  fraction of stars above 9 \Msun \ that explode rather than
  collapsing is estimated to be 65\%. The fraction above 18 \Msun \ is
  8\%. Nucleosynthesis and light curves are basically unchanged.

\end{itemize}

  Binary mass exchange and rotation undoubtedly play key roles,
  but have been neglected in this study. To first order, stars that
  end up with the same final helium core mass will have similar
  presupernova compactness and fates. Similar systematic variations
  and multi-valued solutions are expected to persist. The statistical
  averages of compact remnant masses may vary, however, and certainly
  the light curves will be different. Our models are publically
  available to those wanting to estimate outcomes based upon their own
  distribution of helium core masses, carbon-oxygen core masses,
  etc. In the future, we will consider rotating models, but this was
  principally a study of how resolution affects the solution to a
  well-defined, frequently studied problem.

  Our results suggest a slightly different strategy to the study
  of presupernova evolution and supernova modeling than sometimes used
  in the past. Given the variation in outcome for stars of nearly the
  same mass, or the same mass with different codes, a statistically
  meaningful sample of models must be calculated before drawing strong
  conclusions about supernova mass ranges, remnant mass distributions,
  nucleosynthesis, etc. Historically, researchers have sometimes
  focused on the calculation of just a few masses, e.g., 15, 20, 25
  \Msun, and sought to test the senitivity to changes in physics in
  just those cases. Here we see that calculating a statistically
  meaningful sample may be just as important as getting the physics
  precisely right. The size of such a sample depends upon the need to
  resolve regions of rapid variabilility found with a given code and
  physics, but a minimum initial mass grid of 0.1 \Msun \ is
  reasonable within such regions.

Some of the models calculated here had merged carbon and and oxygen
convective shells at the end. Many did not. Others were only separated
by a single thin zone from being coupled. Similar to the numerical
artifact seen when helium burning is calculated with too little
semiconvection, these mathematical bifurcations may not be real and
might be overcome, or at least smoothed out \citep{Ale04}, by
increasing the convective overshoot or doing a multi-dimensional
calculation  \citep[e.g.,][]{Mea07,Via13,Arn16,Jon17,Cri17}.  In
general, linked convective shells give a more compact core structure
and favor explosions \citep[see, e.g.,][]{Col17}.  Further study with
other representations of convective overshoot mixing and
  multi-dimensional codes are thus encouraged.

Semiconvection and overshoot mixing remain major uncertainties
  in studies of this sort and also play important roles in determining
  the helium and carbon-oxygen core masses. How \KEPLER\ treats
  semiconvection is described in \citet{Wea93}.It is expected that
  other studies using different treatments will find results that
  differ in important detail from those presented here. The overall
pattern, multi-peaked structure and range of variability of the
compactness parameter and other measures of core structure should
persist, however. Clusters of solutions due to variable numbers and
extents of the carbon and oxygen shells should also be a common
feature. Further exploration is again encouraged.

All of the new presupernova models presented here are available online. 
Also online are plots of the convective histories of 800 models between 
12 and 20 \Msun\ of the $\dot{M}$ set (see Footnote 9). Other auxiliary 
prespernova models (e.g., those presented in \Sect{ultra} and 
\Sect{network}) are available upon request to the authors.

\section*{Acknowledgments}

We thank Raphael Hirschi for a careful review of this work and
for providing many useful suggestions. We also thank Bernhard
M\"uller for helping us to process our new progenitor results through
his semi-analytical explosion modelling and feedback on this
manuscript. Thomas Janka and Thomas Ertl provided valuable insight
into the supernova explosion models and their dependence on progenitor
properties. We thank Todd Thompson and John Beacom for useful comments
on the manuscript. We also appreciate the efforts of Gang Gao in
discovering the neutrino bug in the \KEPLER\ code. All numerical
\KEPLER\ calculations presented in this work were performed on the
\texttt{RUBY} cluster at the Ohio Supercomputer Center \citep{osc}.
TS is partly supported by NSF PHY-1404311 to John Beacom. SW is
supported by NASA NNX14AH34G.  AH is supported by an ARC Future
Fellowship, FT120100363, and, in part, by the National Science
Foundation under Grant No. PHY-1430152 (JINA Center for the Evolution
of the Elements).

\end{document}